# Elliptification of Rectangular Imagery

Chamberlain Fong
spectralfft@yahoo.com
Joint Mathematics Meetings 2019, SIGMAA-ARTS

Abstract – We present and discuss different algorithms for converting rectangular imagery into elliptical regions. We will focus primarily on methods that use mathematical mappings with explicit and invertible equations. The key idea is to start with invertible mappings between the square & the circular disc; and then extend it to handle rectangles and ellipses. This extension can be done by simply removing the eccentricity and reintroducing it back after using a chosen square-to-disc mapping.

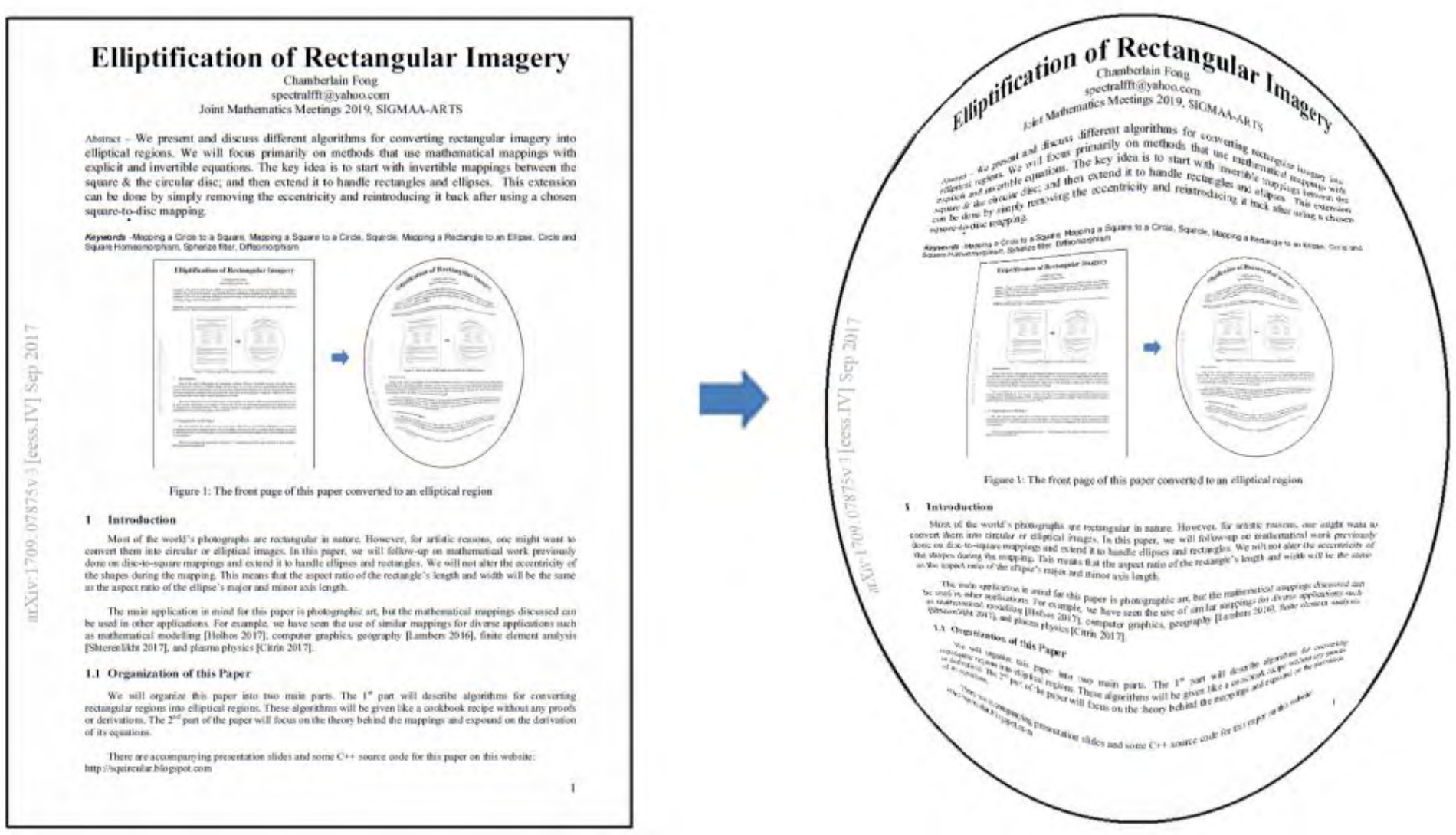

Elliptification of Rectangular Imagery

Chamberlain Fong
spectralfft@yahoo.com
Joint Mathematics Meetings 2019, SIGMAA-ARTS

Abstract – We present and discuss different algorithms for converting rectangular imagery into elliptical regions. We will focus primarily on methods that use mathematical mappings with explicit and invertible equations. The key idea is to start with invertible mappings between the square & the circular disc; and then extend it to handle rectangles and ellipses. This extension can be done by simply removing the eccentricity and reintroducing it back after using a chosen square-to-disc mapping.

Figure 1: The front page of this paper converted to an elliptical region

1 Introduction

Most of the world's photographs are rectangular in nature. However, for artistic reasons, one might want to convert them into circular or elliptical images. In this paper, we will follow-up on mathematical work previously done on disc-to-square mappings and extend it to handle ellipses and rectangles. We will not alter the eccentricity of the shapes during the mapping. This means that the aspect ratio of the rectangle's length and width will be the same as the aspect ratio of the ellipse's major and minor axis length.

The main application in mind for this paper is photographic art, but the mathematical mappings discussed can be used in other applications. For example, we have seen the use of similar mappings for diverse applications such as mathematical modelling [Holhos 2017], computer graphics, geography [Lambers 2016], finite element analysis [Shterenlikht 2017], and plasma physics [Citrin 2017].

1.1 Organization of this Paper

We will organize this paper into two main parts. The 1st part will describe algorithms for converting rectangular regions into elliptical regions. These algorithms will be given like a cookbook recipe without any proofs or derivations. The 2nd part of the paper will focus on the theory behind the mappings and expound on the derivation of its equations.

There are accompanying presentation slides and some C++ source code for this paper on this website:
http://squircular.blogspot.com

Figure 1: The front page of this paper converted to an elliptical region

## 1 Introduction

Most of the world's photographs are rectangular in nature. However, for artistic reasons, one might want to convert them into circular or elliptical images. In this paper, we will follow-up on mathematical work previously done on disc-to-square mappings and extend it to handle ellipses and rectangles. We will not alter the eccentricity of the shapes during the mapping. This means that the aspect ratio of the rectangle's length and width will be the same as the aspect ratio of the ellipse's major and minor axis length.

The main application in mind for this paper is photographic art, but the mathematical mappings discussed can be used in other applications. For example, we have seen the use of similar mappings for diverse applications such as mathematical modelling [Holhos 2017], computer graphics [Shirley 1997], geography [Lambers 2016], finite element analysis [Shterenlikht 2017], and plasma physics [Citrin 2017].

### 1.1 Organization of this Paper

We will organize this paper into two main parts. The 1st part will describe algorithms for converting rectangular regions into elliptical regions. These algorithms will be given like a cookbook recipe without any proofs or derivations. The 2nd part of the paper will focus on the theory behind the mappings and expound on the derivation of its equations.

There are accompanying presentation slides and some C++ source code for this paper on this website:
http://squircular.blogspot.com

## 1.2 Square to Disc Mapping Space

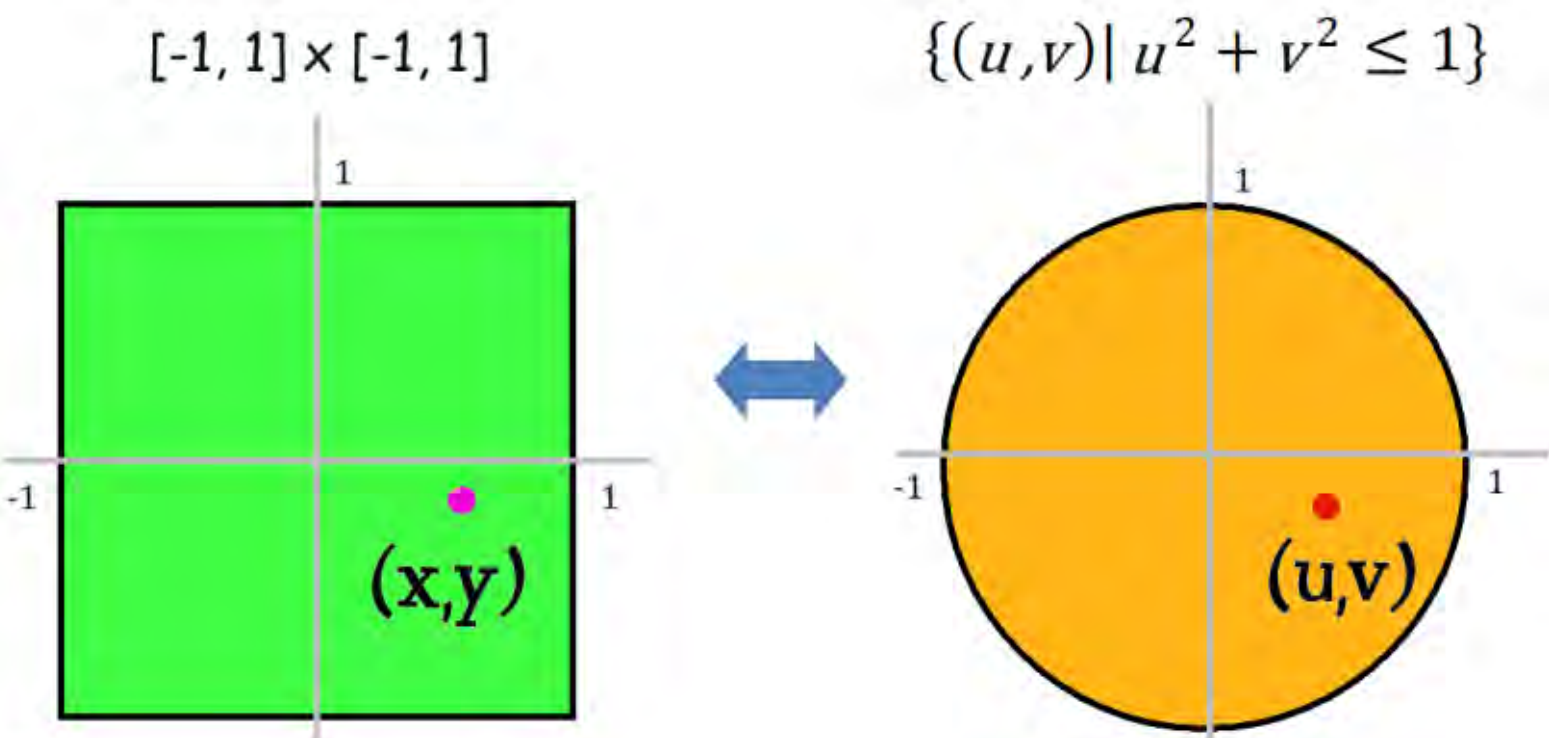

Figure 2: Canonical mapping space from square to disc

Before we discuss mapping between the rectangle to the ellipse (and vice versa), we have to first discuss the simpler case of mapping between the square and the circle. The author has written a paper about this titled "*Analytical Methods for Squaring the Disc*" [Fong 2014] which covers the topic extensively. This paper is a follow-up paper and will require the previous paper as a prerequisite. We shall reuse many of the concepts, equations, and notation from that paper in our exposition here.

The canonical mapping space between the square and disc is shown in Figure 2. It consists of a unit disc centered at the origin with a square circumscribing it. The square is defined as the set $\mathcal{S} = [-1,1] \times [-1,1]$. In contrast, the unit disc is defined as the set $\mathcal{D} = \{(u,v) \mid u^2 + v^2 \leq 1\}$. We shall denote (x,y) as the point in the interior of the square and (u,v) as the corresponding point in the interior of the disc after the mapping.

## 1.3 Open Variation of the Mapping Space

An important variation of our canonical mapping space arises when we do not include the boundary of the shapes in the mapping. We shall denote this as the *open mapping space*. We will introduce several mappings in this paper that only work in the open mapping space. These mappings do not include the points on the rims of the square or the circle. The open mapping space uses the open square as its domain and the open circular disc as its range.

The open square is defined as the set $\mathcal{S} = (-1,1) \times (-1,1)$. Similarly, the open circular disc is defined as the set $\mathcal{D} = \{(u,v) \mid u^2 + v^2 < 1\}$. Both of these sets do not include the boundary of the shape.

## 1.4 The Signum Function

Many of the mappings introduced in this paper will use the signum function, so we need to review its definition.

$$\mathrm{sgn}(x) = \frac{|x|}{x} = \begin{cases} -1 & \text{if } x < 0, \\ 0 & \text{if } x = 0, \\ 1 & \text{if } x > 0. \end{cases}$$

## 1.5 Rectangle to Ellipse Extension

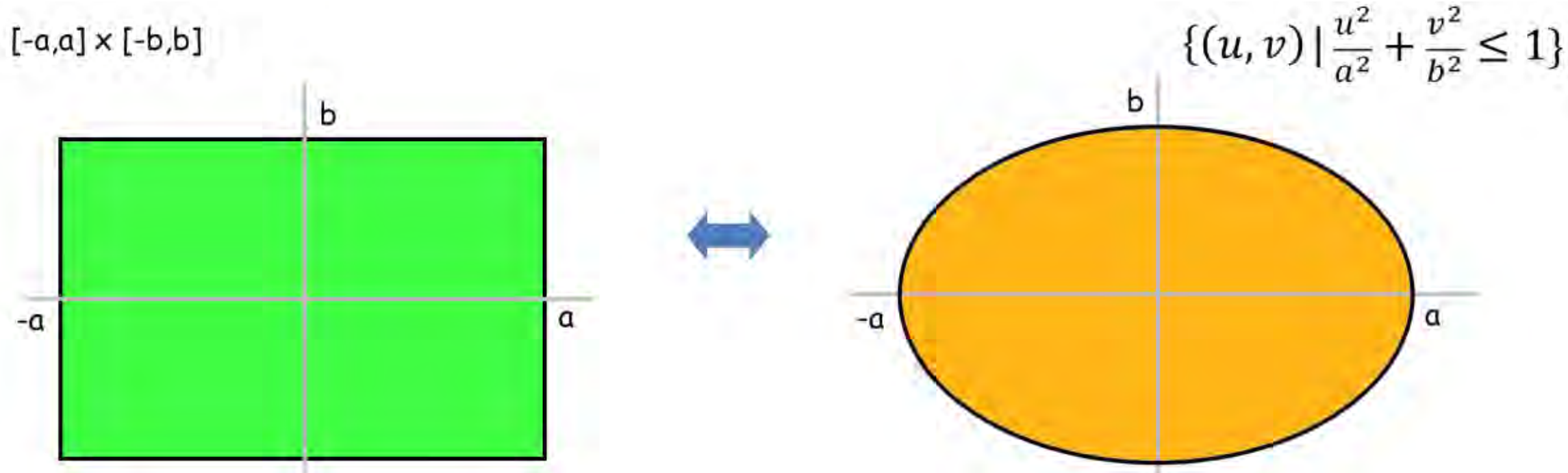

Figure 3: Canonical mapping space from rectangle to ellipse

The canonical mapping space between the rectangular and elliptical regions is shown in Figure 3. It consists of an ellipse centered at the origin with a rectangle circumscribing it. Both shapes have the aspect ratio a/b. The rectangular region is defined as the set $\mathcal{R} = [-a, a] \times [-b, b]$. In contrast, the elliptical region is defined as the set $\mathcal{E} = \{(u,v)\,|\,\frac{u^2}{a^2}+\frac{v^2}{b^2}\leq 1\}$ . These sets apply for closed mappings. There are also analogous sets for open mappings which do not include the boundary curves on the rims of the shapes.

## 1.6 The Mappings

In the next few pages, we shall present several mappings for converting the circular disc to a square and vice versa. We include pictures of a radial grid inside the disc converted to a square; and a square grid converted to a circular disc. This is followed by forward and inverse equations for the mappings.

The first three mappings have been thoroughly covered in the paper "*Analytical Methods for Squaring the Disc*" [Fong 2014]. We only provide them here for reference since they will be used and expanded upon by subsequent mappings. In particular, we will present several variations to the FG-Squircular mapping and the Elliptical Grid mapping.

The subsequent four mappings covered are all variations of the FG-Squircular mapping. They are all radial and based on the Fernandez-Guasti squircle. These are the

- 2-Squircular mapping
- 3-Squircular mapping
- Cornerific Tapered2 mapping
- Tapered4 mapping

An additional three mappings covered are also variations of the FG-Squircular mapping. They are also radial but have a property called axial nonlinearity which will be discussed later. These three mappings are

- Non-axial 2 mapping
- Non-axial ½ mapping
- Sham Quartic mapping

The next three mappings covered are variations of the Elliptical Grid mapping. These are also related to the Fernandez-Guasti squircle, but they are not radial. Furthermore, these are open-type mappings that only work in the open mapping space. In other words, these mappings do not include the boundary curves on the rim of the shapes.

- Squelched Grid mapping
- Vertical squelch mapping
- Horizontal squelch mapping

The last five mappings are blended generalizations of the previous mappings. These mappings include an adjustable free parameter $\beta$ used for blending. These blended mappings are

- Blended Elliptical Grid mapping
- Blended Squelch
- Power2 Blend
- Power3 Blend
- Sham Quartic Blend

Note that for the sake of brevity, we have not singled out cases when there are divisions by zero in the mapping equations. These degenerate cases usually arise after getting an indeterminate form $\frac{0}{0}$ in the explicit expressions given. For these degenerate cases, just equate x=u, y=v and vice versa whenever there is an unwanted division by zero in the equations. This usually happens when u=0 or v=0 or both. Mathematically, the expanded mapping equations can be interpreted as

*Disc-to-square*

$$x = \begin{cases} \dots & \textit{when there is no division by zero in the expression} \\ u & \textit{where there is a division by zero in the expression} \end{cases}$$

$$y = \begin{cases} \dots & \textit{when there is no division by zero in the expression} \\ v & \textit{where there is a division by zero in the expression} \end{cases}$$

*Square-to-disc*

$$u = \begin{cases} \dots & \textit{when there is no division by zero in the expression} \\ x & \textit{where there is a division by zero in the expression} \end{cases}$$

$$v = \begin{cases} \dots & \textit{when there is no division by zero in the expression} \\ y & \textit{where there is a division by zero in the expression} \end{cases}$$

After we cover a bevy of square-to-disc mappings in the following pages, we will proceed to extend these mappings to work with rectangle-to-ellipse mappings. The key idea for this extension is very simple. Simply remove the eccentricity of the input rectangle and use any of the given square-to-disc mappings to convert it into a circular disc. Afterwards, reintroduce the eccentricity back to get an ellipse.

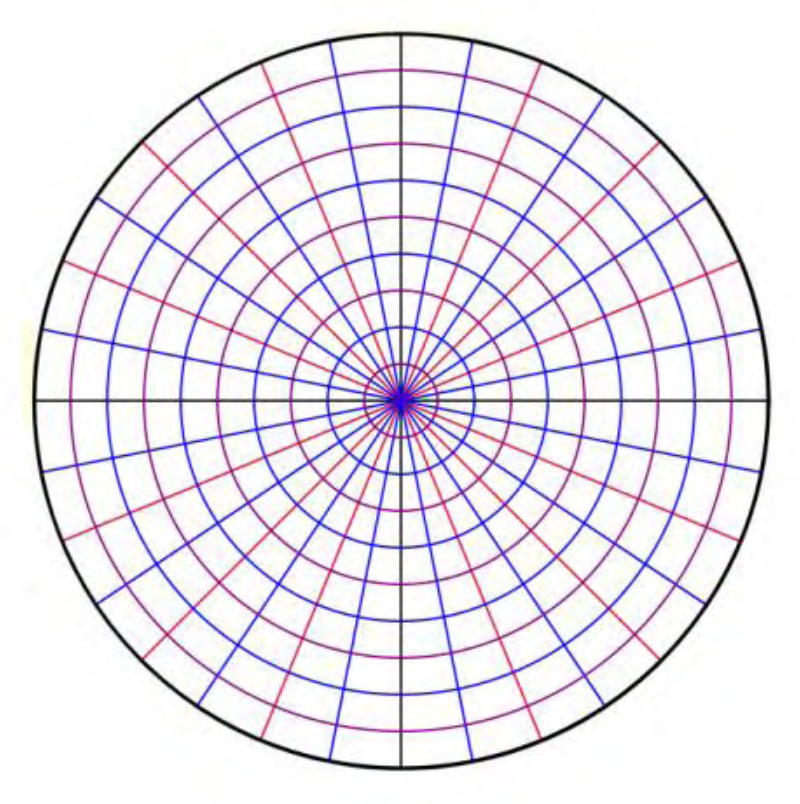

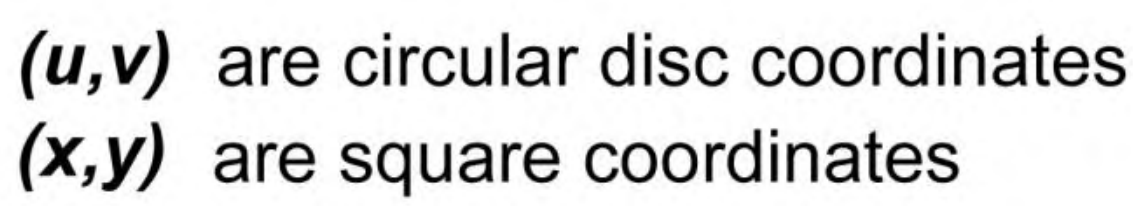
***(u,v)*** are circular disc coordinates

***(x,y)*** are square coordinates

***F*** is the Legendre elliptic integral of the 1st kind

***cn*** is a Jacobi elliptic function

$$K_e = F\left(\frac{\pi}{2}, \frac{1}{\sqrt{2}}\right) = \int_0^{\frac{\pi}{2}} \frac{dt}{\sqrt{1 - \frac{1}{2}\sin^2 t}} \approx 1.854$$

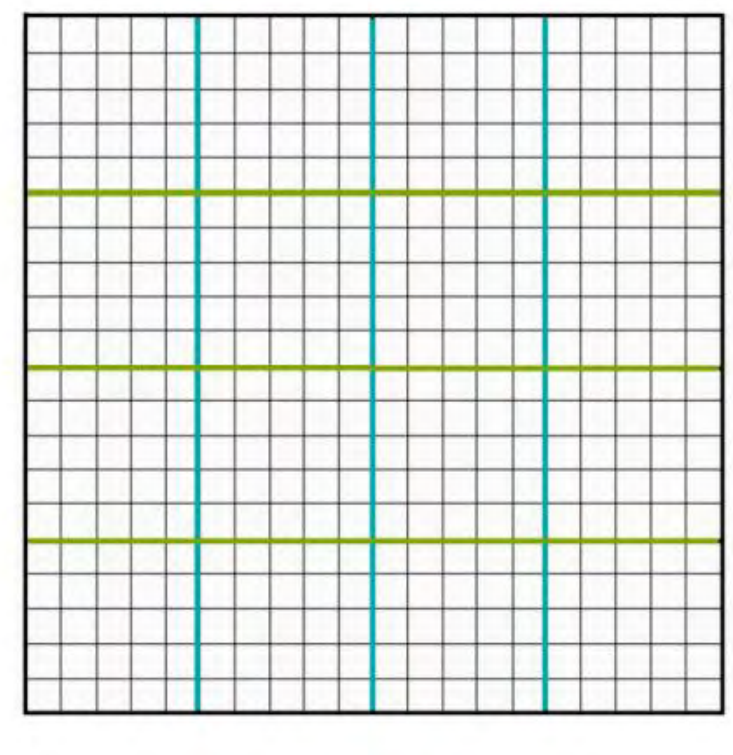

# Schwarz-Christoffel mapping

disc to square

$$x = Re\left(\frac{1-i}{-K_e} F\left(\cos^{-1}\left(\frac{1+i}{\sqrt{2}}(u + v\,i)\right) ,\frac{1}{\sqrt{2}}\right)\right) + 1$$

$$y = Im\left(\frac{1-i}{-K_e} F\left(\cos^{-1}\left(\frac{1+i}{\sqrt{2}}(u + v\,i)\right) ,\frac{1}{\sqrt{2}}\right)\right) - 1$$

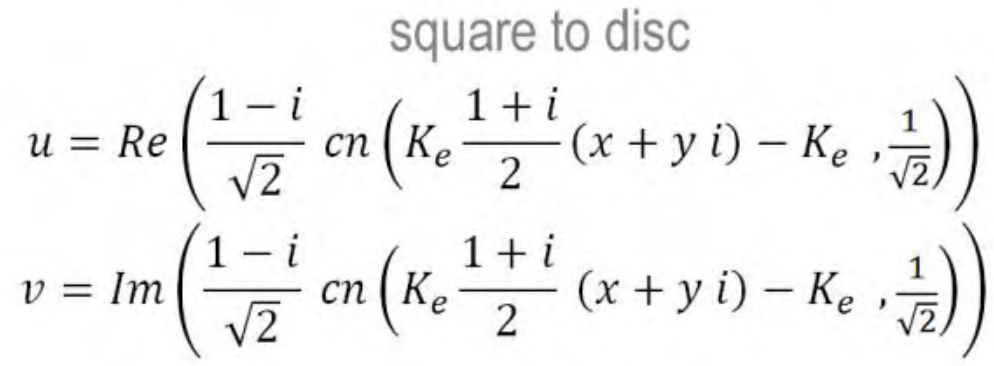
square to disc

$$u = Re\left(\frac{1-i}{\sqrt{2}}\ cn\left(K_e \frac{1+i}{2}(x + y\,i) - K_e\ ,\frac{1}{\sqrt{2}}\right)\right)$$

$$v = Im\left(\frac{1-i}{\sqrt{2}}\ cn\left(K_e \frac{1+i}{2}(x + y\,i) - K_e\ ,\frac{1}{\sqrt{2}}\right)\right)$$

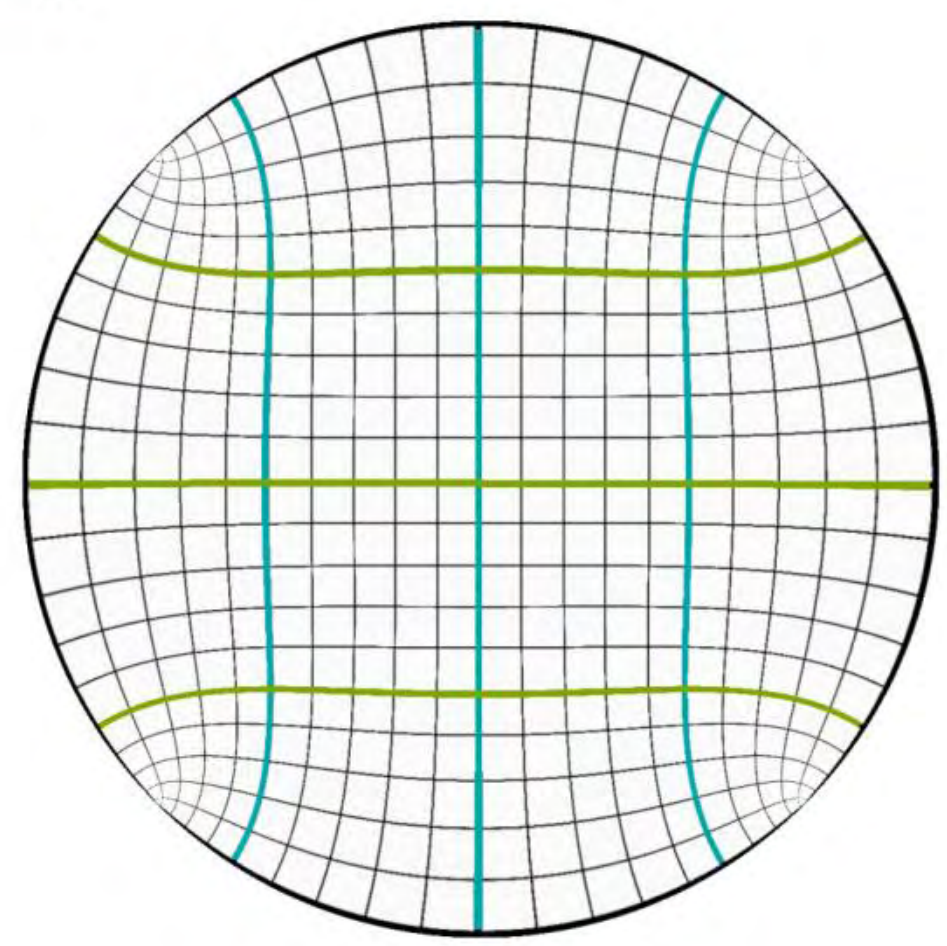

# FG-Squircular mapping

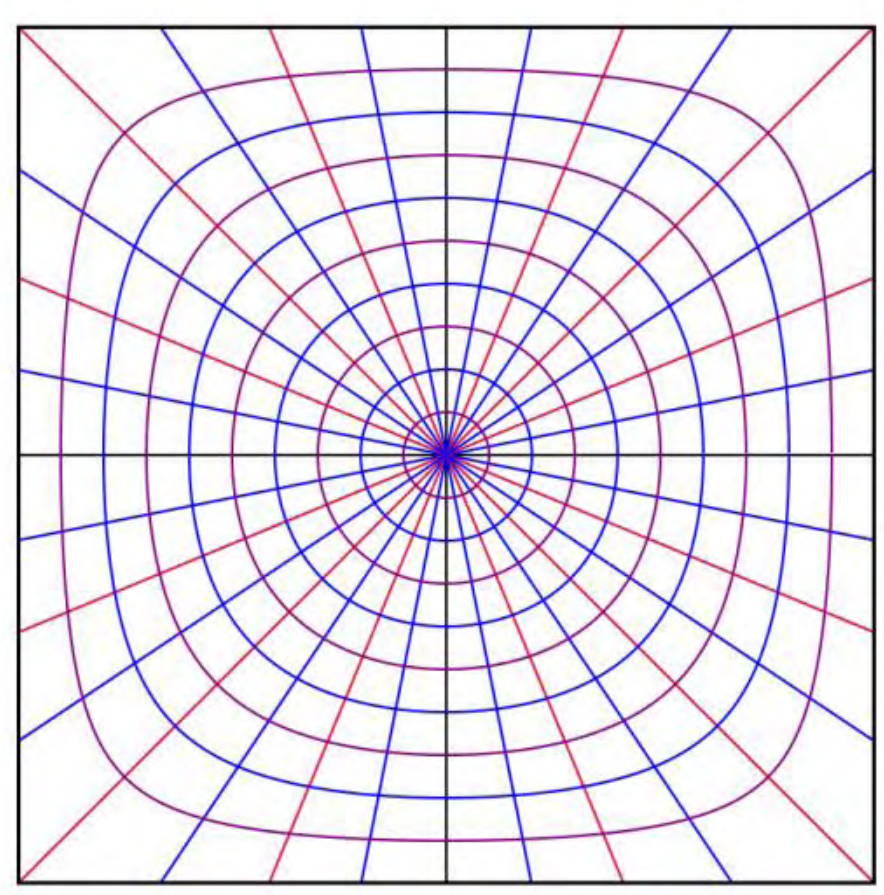

disc to square

$$x = \frac{sgn(uv)}{v\sqrt{2}} \sqrt{u^2 + v^2 - \sqrt{(u^2 + v^2)(u^2 + v^2 - 4u^2v^2)}}$$

$$y = \frac{sgn(uv)}{u\sqrt{2}} \sqrt{u^2 + v^2 - \sqrt{(u^2 + v^2)(u^2 + v^2 - 4u^2v^2)}}$$

square to disc

$$u = \frac{x\sqrt{x^2 + y^2 - x^2y^2}}{\sqrt{x^2 + y^2}} \qquad v = \frac{y\sqrt{x^2 + y^2 - x^2y^2}}{\sqrt{x^2 + y^2}}$$

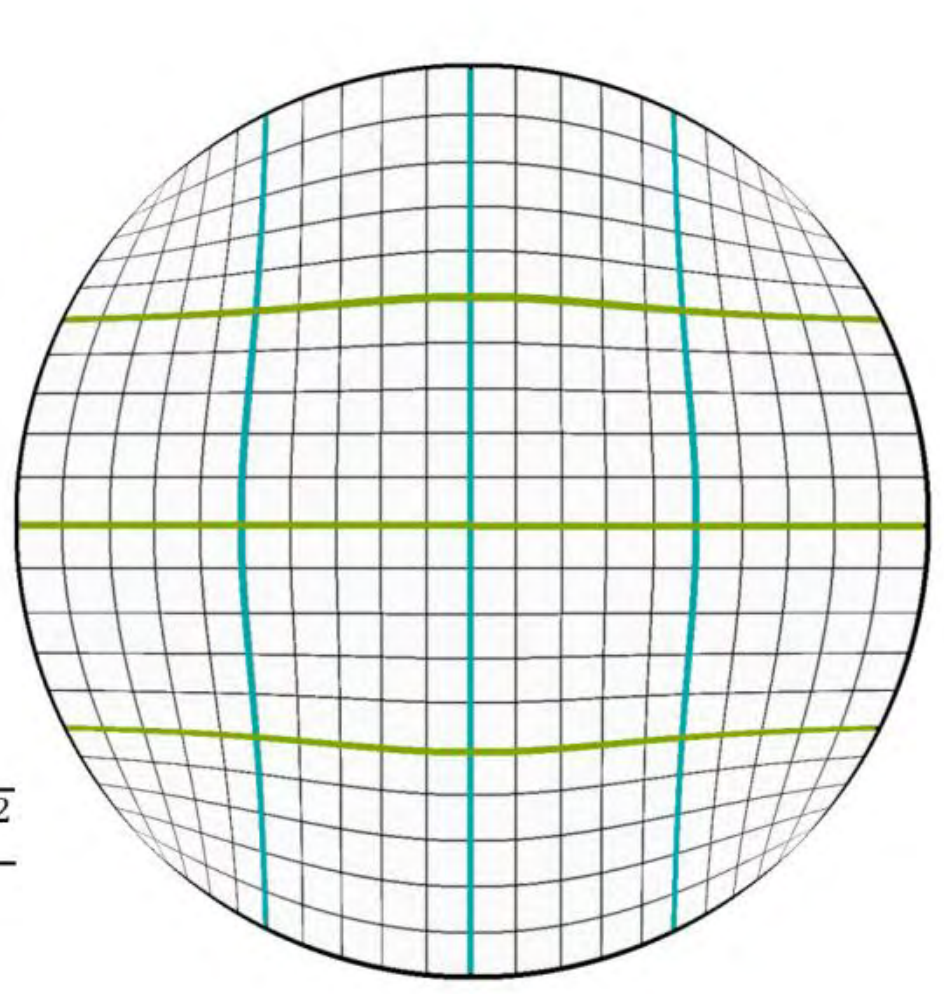

# Elliptical Grid mapping

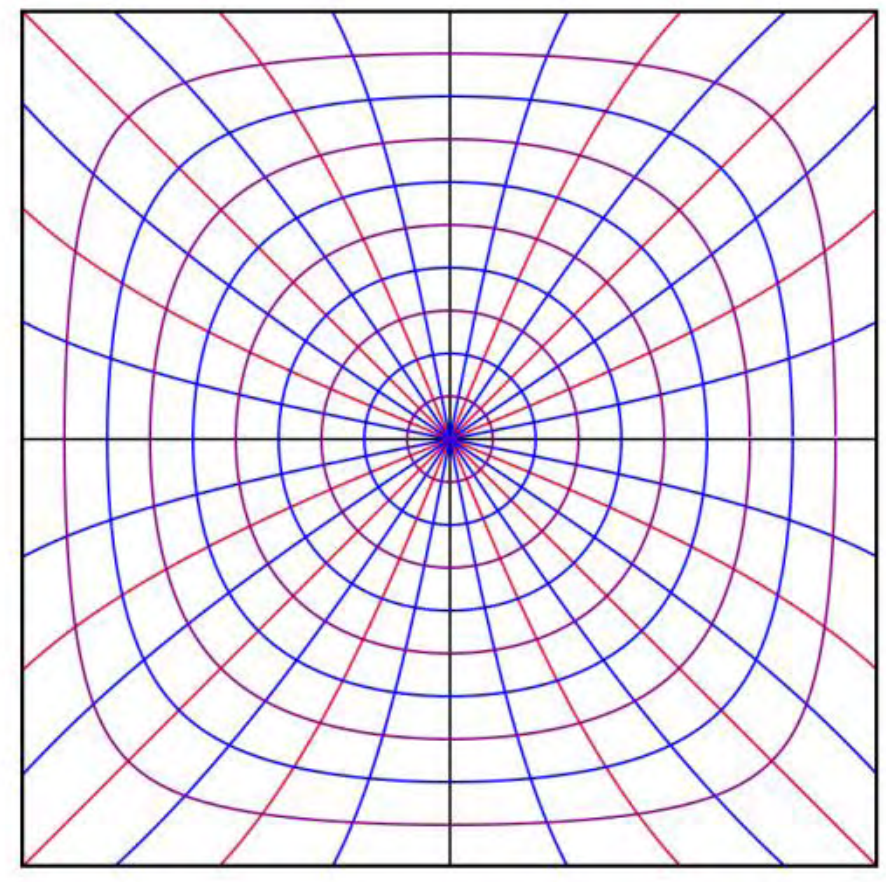

disc to square

$$x = \frac{1}{2}\sqrt{2 + u^2 - v^2 + 2\sqrt{2}\,u} - \frac{1}{2}\sqrt{2 + u^2 - v^2 - 2\sqrt{2}\,u}$$

$$y = \frac{1}{2}\sqrt{2 - u^2 + v^2 + 2\sqrt{2}\,v} - \frac{1}{2}\sqrt{2 - u^2 + v^2 - 2\sqrt{2}\,v}$$

square to disc

$$u = x\sqrt{1 - \frac{y^2}{2}} \qquad v = y\sqrt{1 - \frac{x^2}{2}}$$

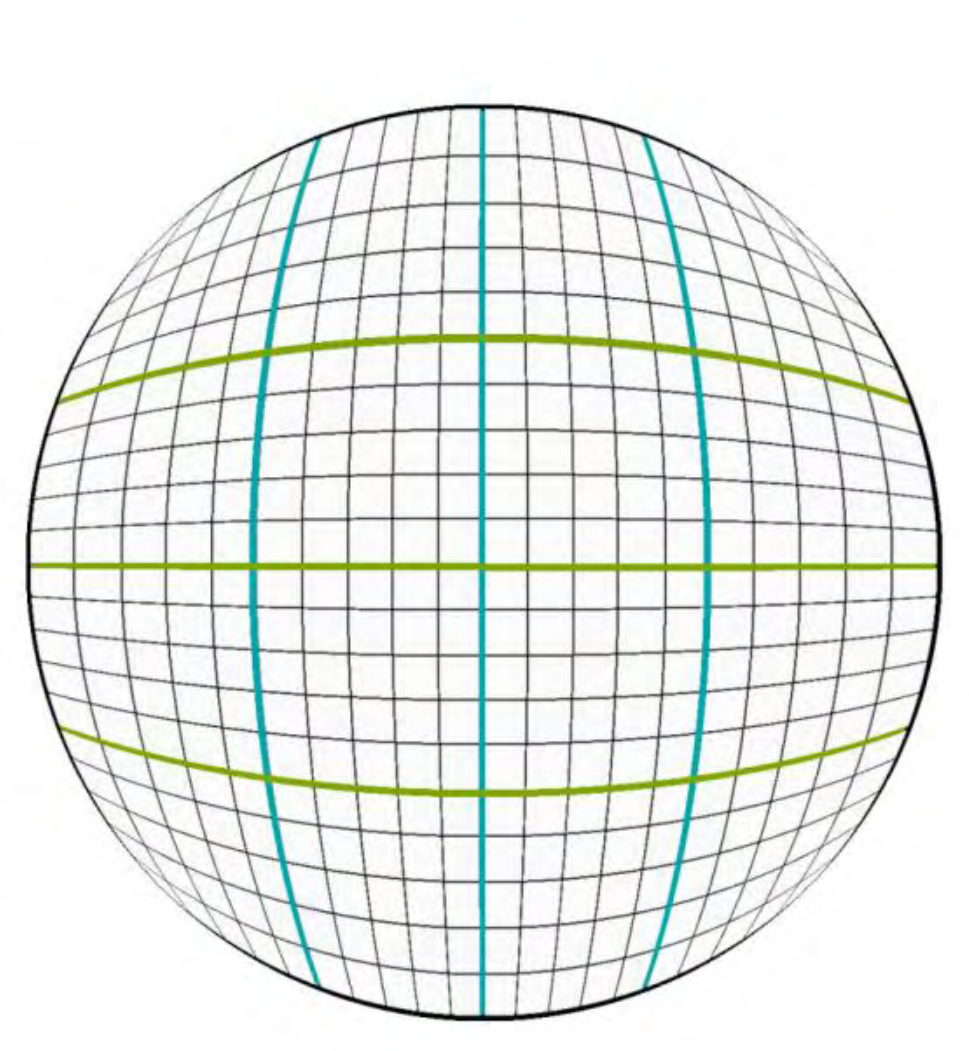

# 2-Squircular Mapping

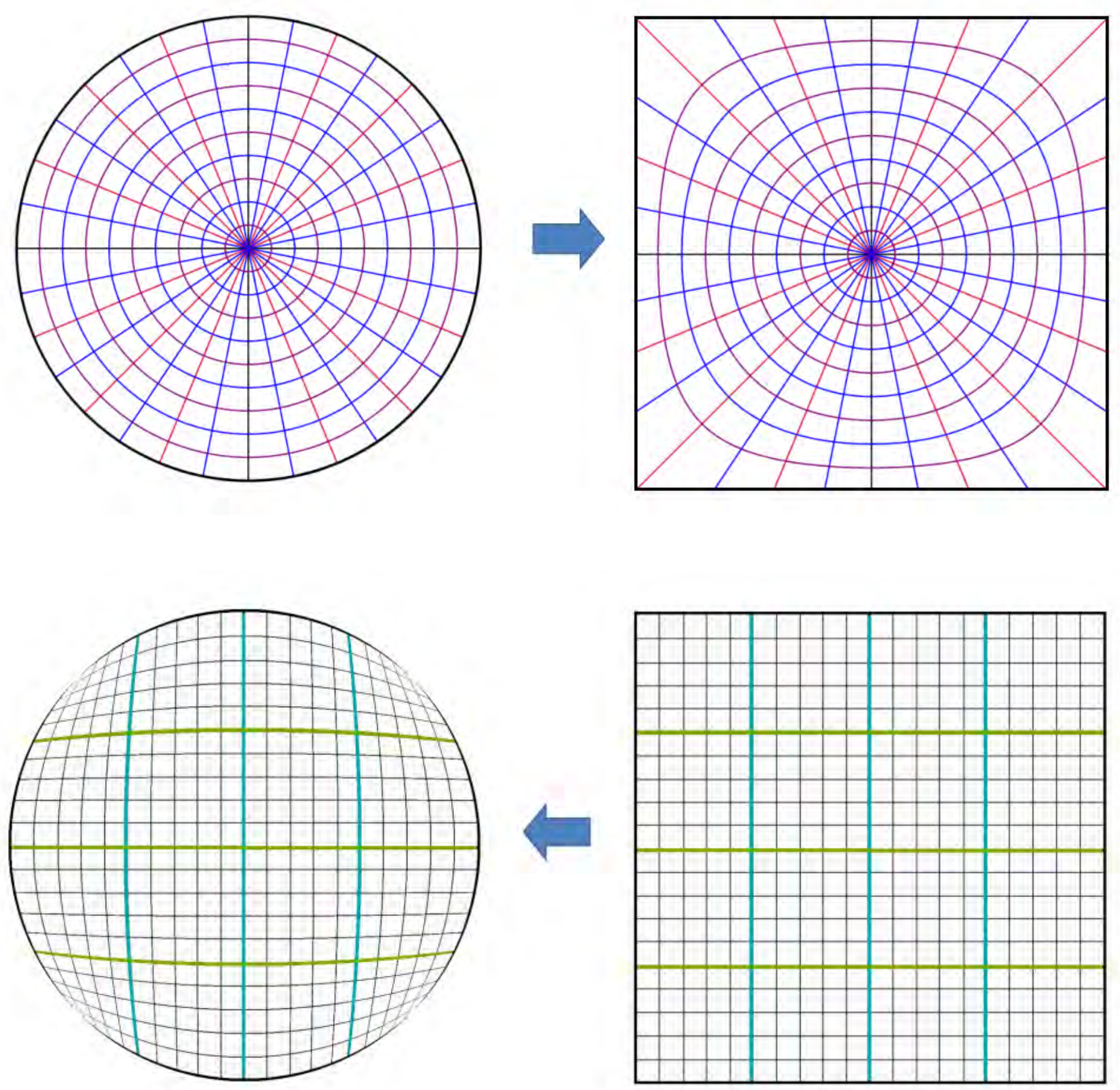

Disc to square mapping:

$$x = \frac{sgn(uv)}{v\sqrt{2}}\sqrt{1-\sqrt{1-4u^2v^2}}$$
$$y = \frac{sgn(uv)}{u\sqrt{2}}\sqrt{1-\sqrt{1-4u^2v^2}}$$

Square to disc mapping:

$$u = \frac{x}{\sqrt{1+x^2y^2}} \qquad v = \frac{y}{\sqrt{1+x^2y^2}}$$

# 3-Squircular Mapping

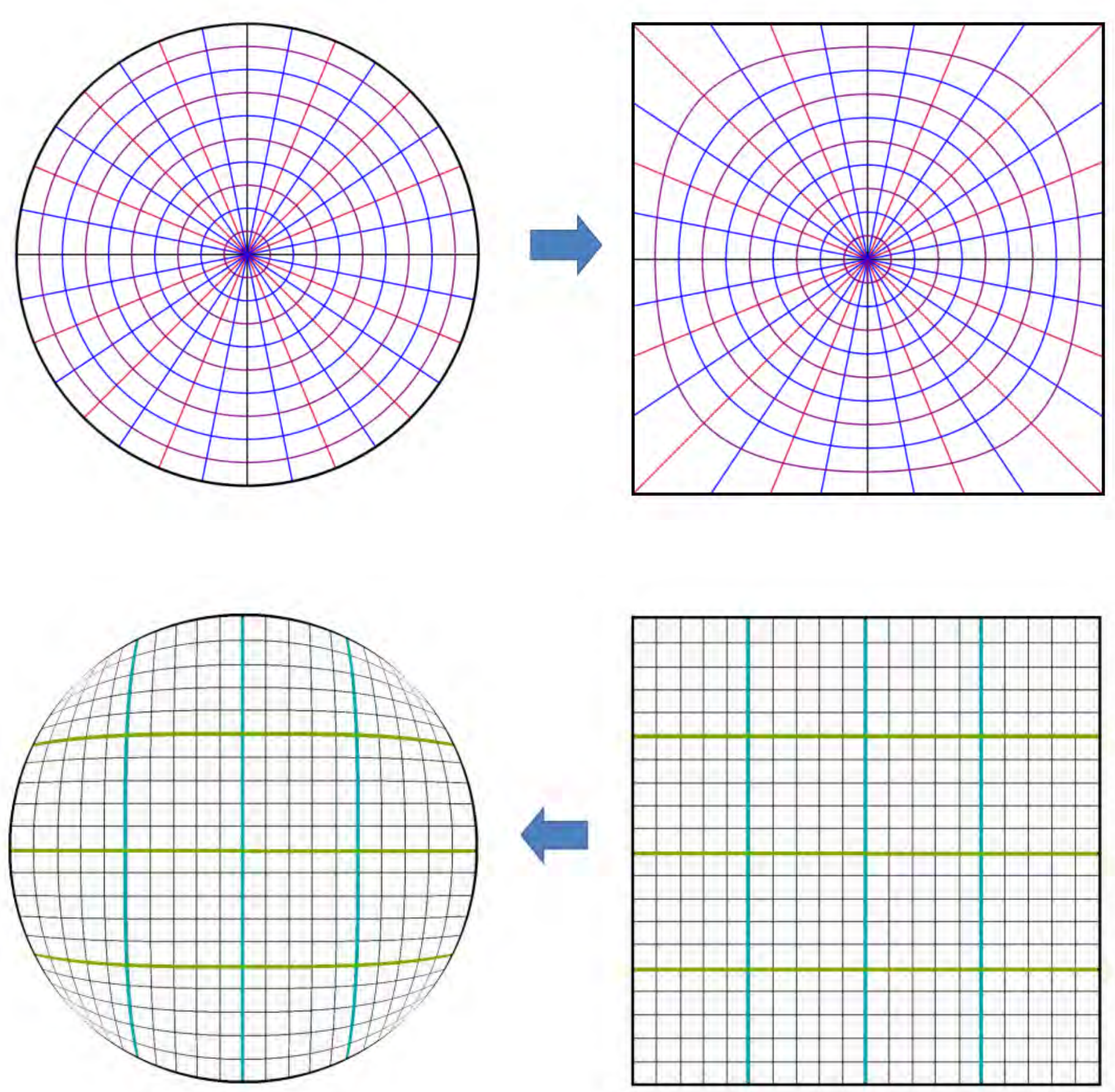

Disc to square mapping:

$$x = \frac{sgn(uv)}{v}\sqrt{\frac{1-\sqrt{1-4u^4v^2-4u^2v^4}}{2(u^2+v^2)}}$$

$$y = \frac{sgn(uv)}{u}\sqrt{\frac{1-\sqrt{1-4u^4v^2-4u^2v^4}}{2(u^2+v^2)}}$$

Square to disc mapping:

$$u = \frac{sgn(xy)}{y}\sqrt{\frac{-1+\sqrt{1+4x^4y^2+4x^2y^4}}{2(x^2+y^2)}}$$

$$v = \frac{sgn(xy)}{x}\sqrt{\frac{-1+\sqrt{1+4x^4y^2+4x^2y^4}}{2(x^2+y^2)}}$$

# Cornerific Tapered2 Mapping

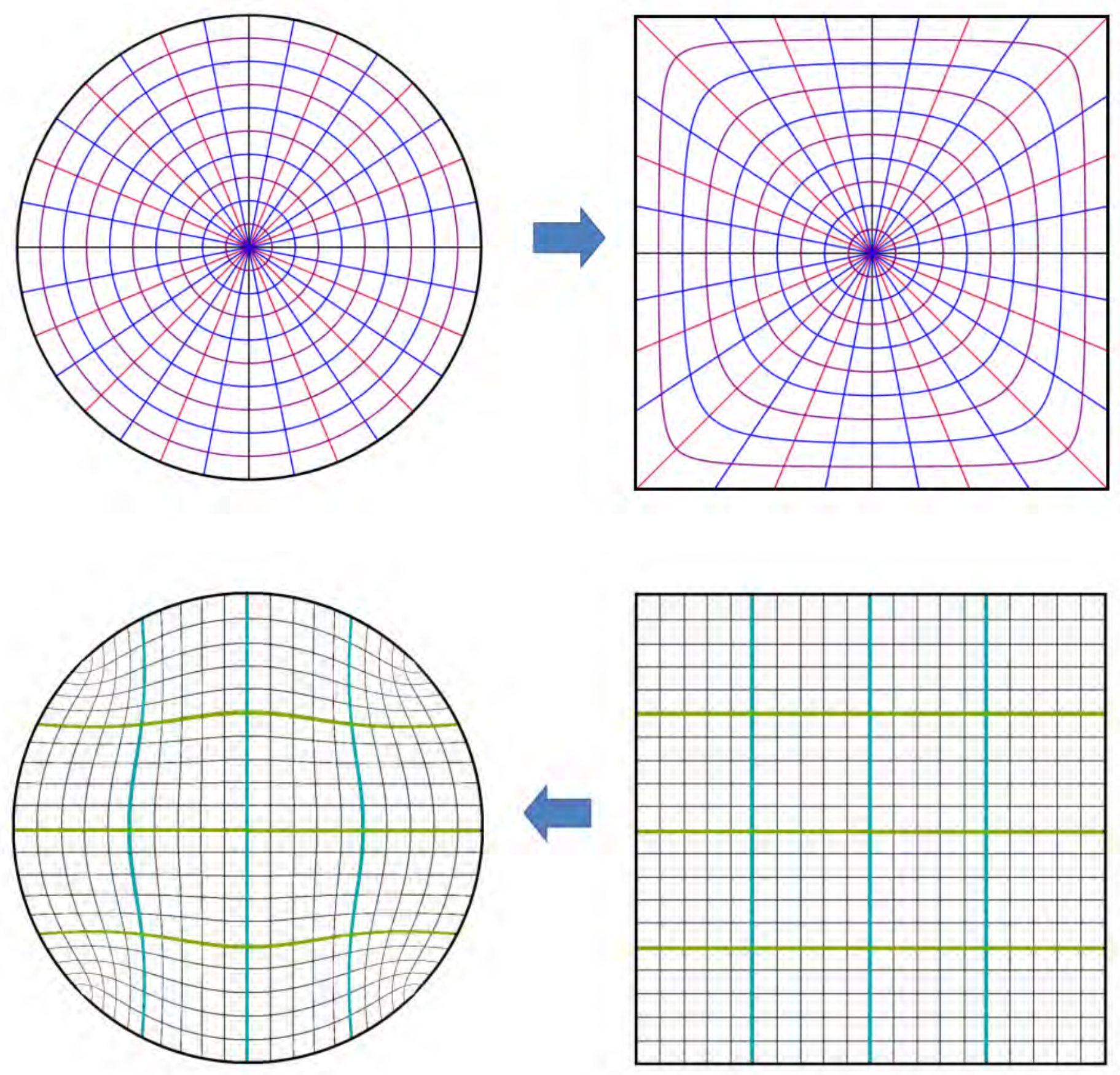

Disc to square mapping:

$$x = \frac{sgn(uv)}{v}\sqrt{\frac{u^2+v^2-\sqrt{(u^2+v^2)[u^2+v^2-4u^2v^2(2-u^2-v^2)]}}{2(2-u^2-v^2)}}$$

$$y = \frac{sgn(uv)}{u}\sqrt{\frac{u^2+v^2-\sqrt{(u^2+v^2)[u^2+v^2-4u^2v^2(2-u^2-v^2)]}}{2(2-u^2-v^2)}}$$

Square to disc mapping:

$$u = x\sqrt{\frac{x^2+y^2-2x^2y^2}{(x^2+y^2)(1-x^2y^2)}}$$

$$v = y\sqrt{\frac{x^2+y^2-2x^2y^2}{(x^2+y^2)(1-x^2y^2)}}$$

# Tapered4 Mapping

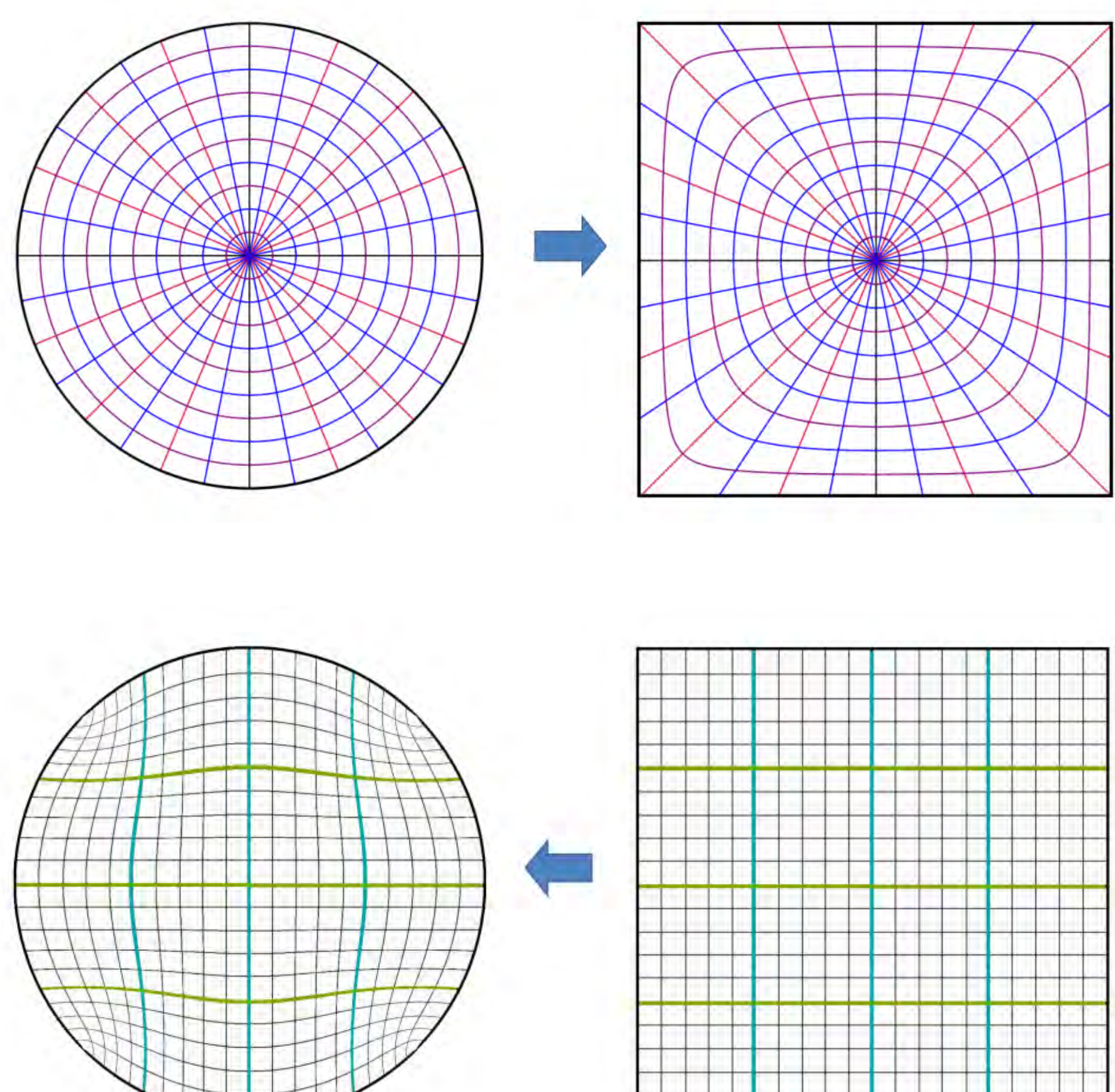

Disc to square mapping:

$$x = \frac{sgn(uv)}{v}\sqrt{\frac{u^2+v^2-\sqrt{(u^2+v^2)[u^2+v^2-2u^2v^2(3-u^4-2u^2v^2-v^4)]}}{3-u^4-2u^2v^2-v^4}}$$

$$y = \frac{sgn(uv)}{u}\sqrt{\frac{u^2+v^2-\sqrt{(u^2+v^2)[u^2+v^2-2u^2v^2(3-u^4-2u^2v^2-v^4)]}}{3-u^4-2u^2v^2-v^4}}$$

Square to disc mapping:

$$u = \frac{sgn(xy)}{y\sqrt{x^2+y^2}}\sqrt{1-\sqrt{1-2x^4y^2-2x^2y^4+3x^4y^4}}$$

$$v = \frac{sgn(xy)}{x\sqrt{x^2+y^2}}\sqrt{1-\sqrt{1-2x^4y^2-2x^2y^4+3x^4y^4}}$$

# Squelched Grid Open Mapping

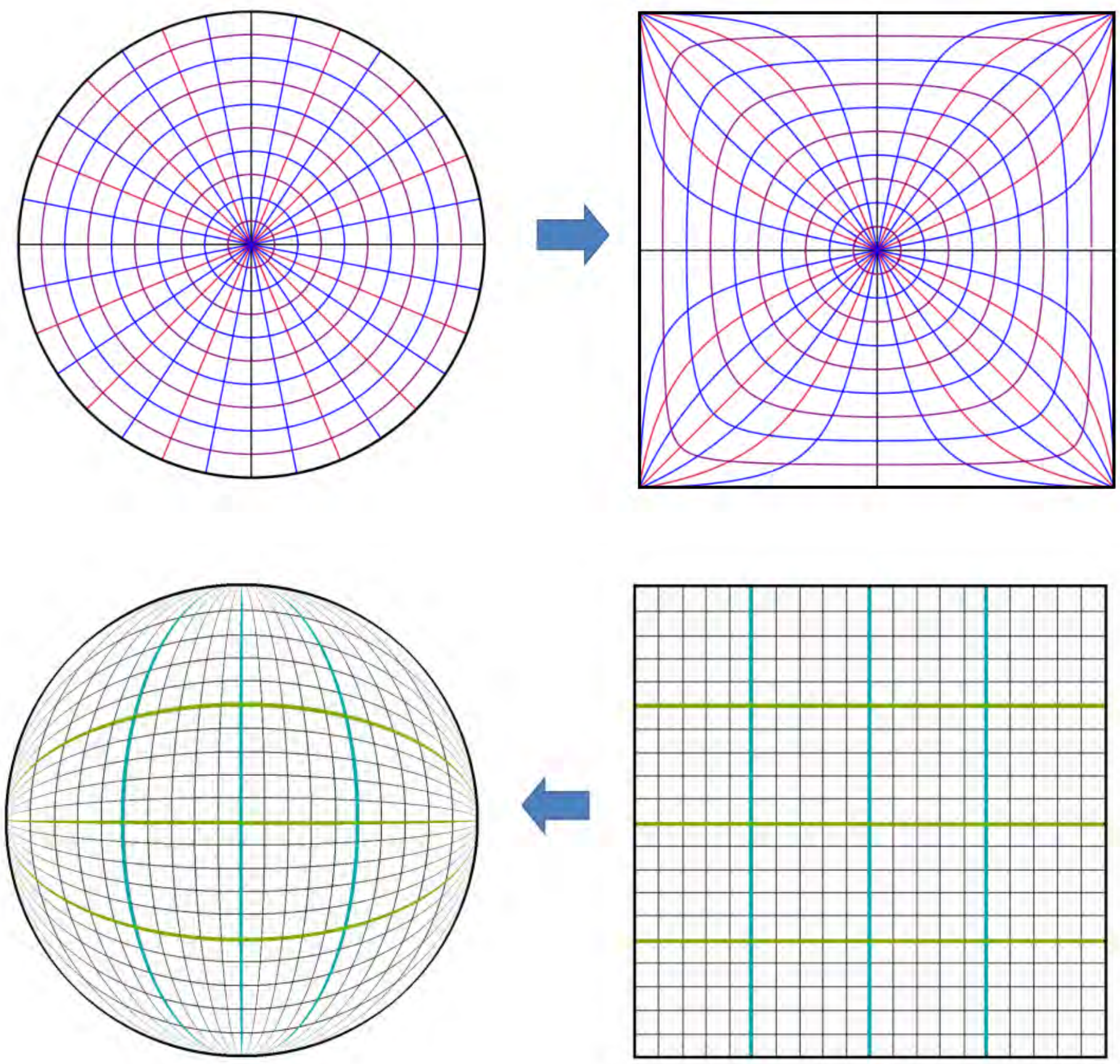

Disc to square mapping:

$$x = \frac{u}{\sqrt{1 - v^2}}$$

$$y = \frac{v}{\sqrt{1 - u^2}}$$

Square to disc mapping:

$$u = x\sqrt{\frac{1 - y^2}{1 - x^2y^2}}$$

$$v = y\sqrt{\frac{1 - x^2}{1 - x^2y^2}}$$

# Vertical Squelch Open Mapping

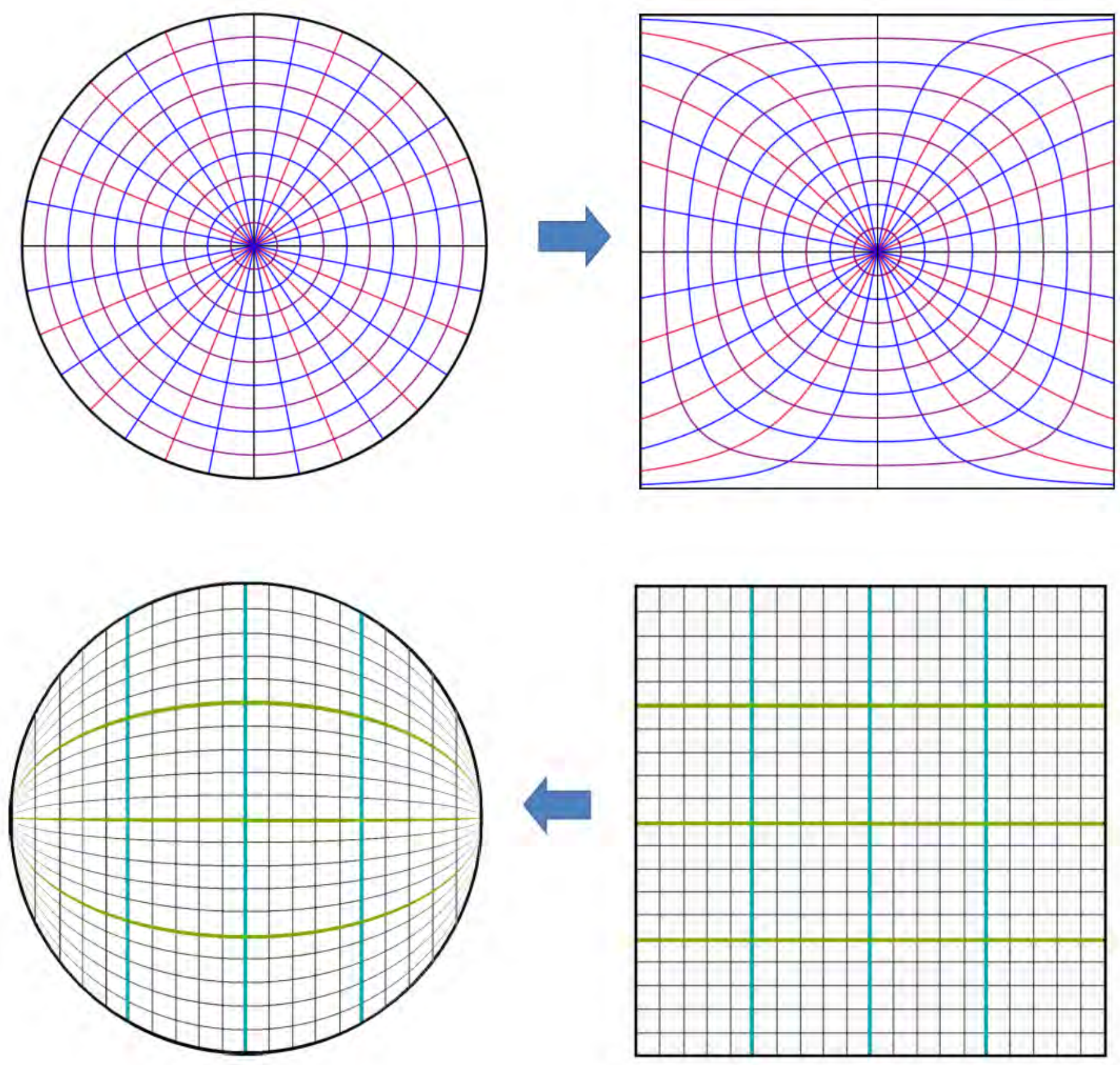

Disc to square mapping:

$$x = u$$

$$y = \frac{v}{\sqrt{1 - u^2}}$$

Square to disc mapping:

$$u = x$$

$$v = \ y\sqrt{1 - x^2}$$

# Horizontal Squelch Open Mapping

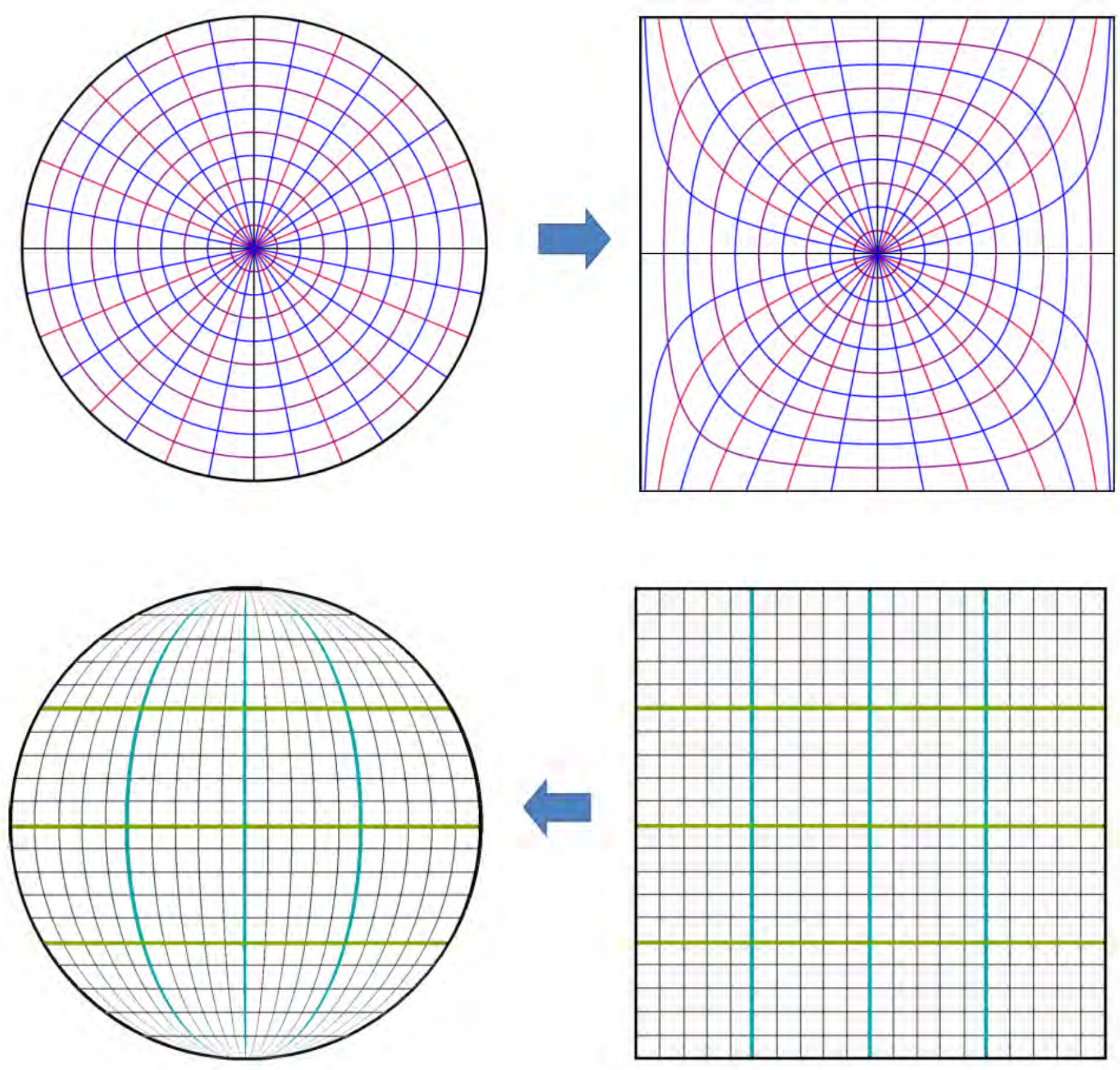

Disc to square mapping:

$$x = \frac{u}{\sqrt{1 - v^2}}$$

$$y = v$$

Square to disc mapping:

$$u = x\sqrt{1 - y^2}$$

$$v = y$$

# Non-axial 2-Pinch Mapping

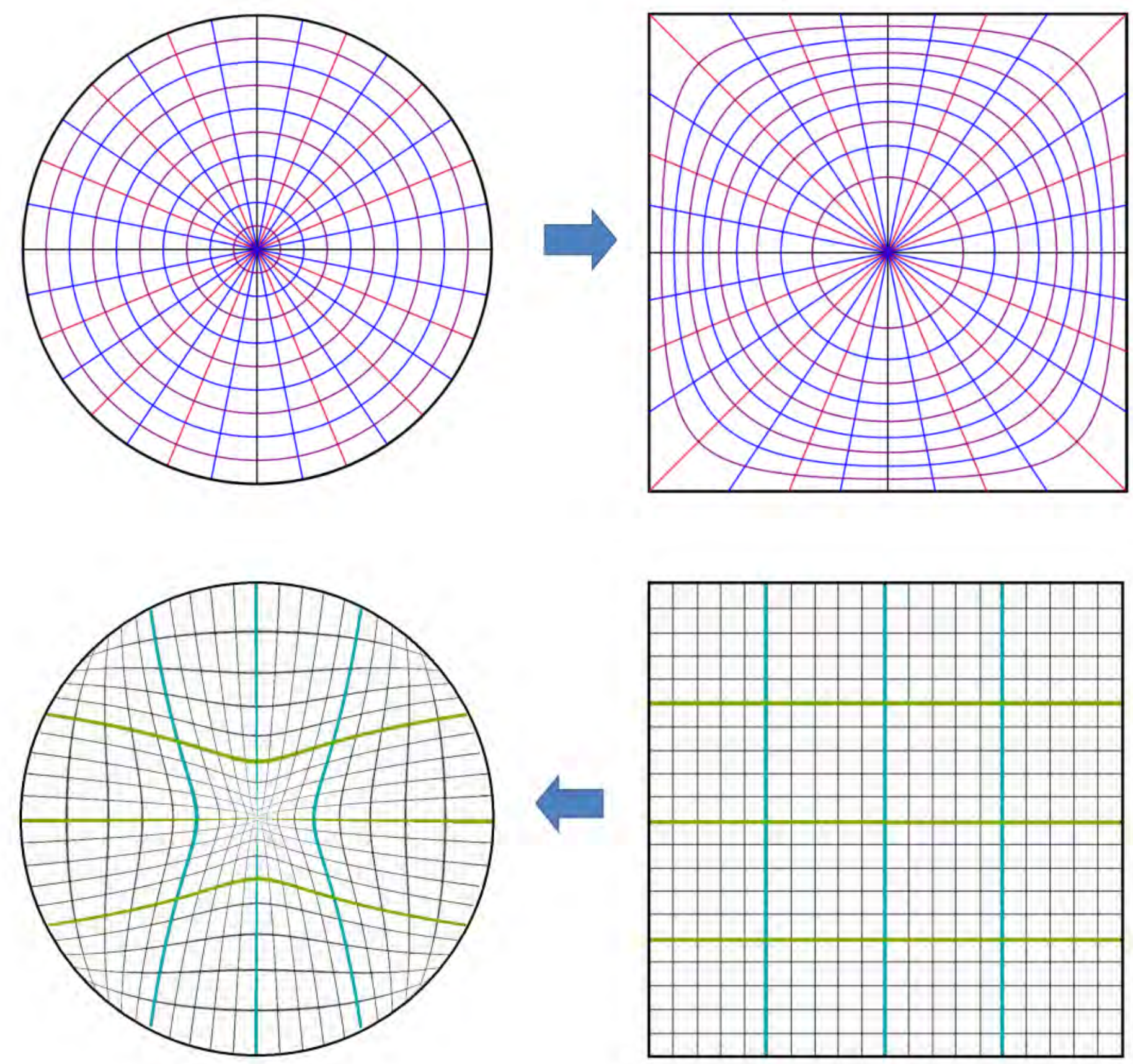

Disc to square mapping:

$$x = \begin{cases} \dfrac{sgn(uv)}{v\ \sqrt[4]{2}} \left( u^2 + v^2 - 2u^2v^2 - \sqrt{(u^2 + v^2 - 4u^2v^2)(u^2 + v^2)} \right)^{\frac{1}{4}} & when\ v \neq 0 \\ sgn(u)\sqrt{|u|} & otherwise \end{cases}$$

$$y = \begin{cases} \dfrac{sgn(uv)}{u\ \sqrt[4]{2}} \left( u^2 + v^2 - 2u^2v^2 - \sqrt{(u^2 + v^2 - 4u^2v^2)(u^2 + v^2)} \right)^{\frac{1}{4}} & when\ u \neq 0 \\ sgn(v)\sqrt{|v|} & otherwise \end{cases}$$

Square to disc mapping:

$$u = \frac{x\sqrt{x^2 + y^2}}{1 + x^2y^2} \qquad\qquad v = \frac{y\sqrt{x^2 + y^2}}{1 + x^2y^2}$$

# Non-axial ½-Punch Mapping

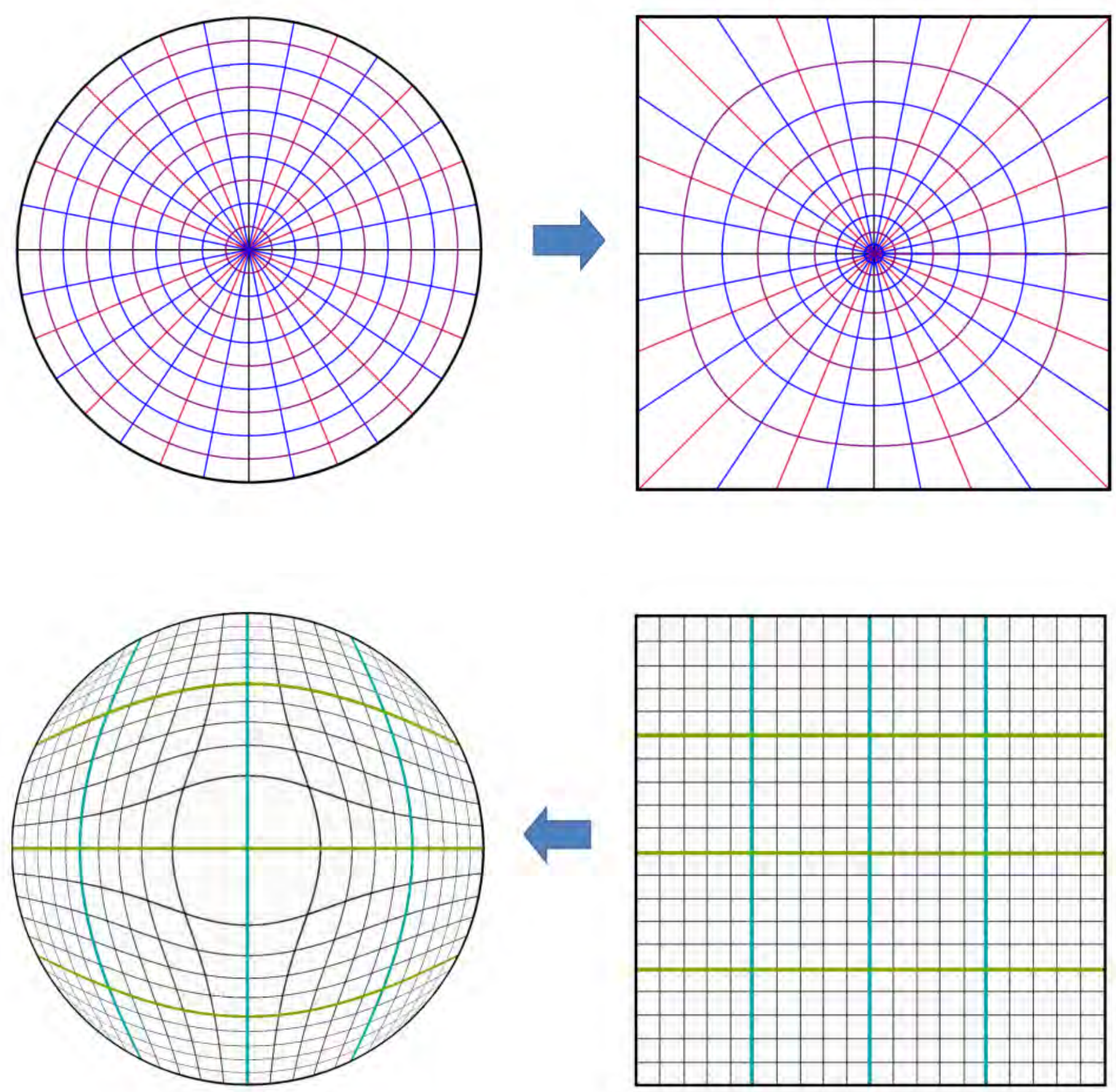

Disc to square mapping:

$$x = \begin{cases} \dfrac{sgn(uv)}{v}\sqrt{\dfrac{1-\sqrt{1-4u^2v^2(u^2+v^2)^2}}{2(u^2+v^2)}} & when \;\; v \neq 0 \\ \\ sgn(u)\, u^2 & otherwise \end{cases}$$

$$y = \begin{cases} \dfrac{sgn(uv)}{u}\sqrt{\dfrac{1-\sqrt{1-4u^2v^2(u^2+v^2)^2}}{2(u^2+v^2)}} & when \;\; u \neq 0 \\ \\ sgn(v)\, v^2 & otherwise \end{cases}$$

Square to disc mapping:

$$u = \frac{x}{(x^2+y^2)^{1/4}\;(1+x^2y^2)^{1/4}} \qquad\qquad v = \frac{y}{(x^2+y^2)^{1/4}\;(1+x^2y^2)^{1/4}}$$

# Sham Quartic Mapping

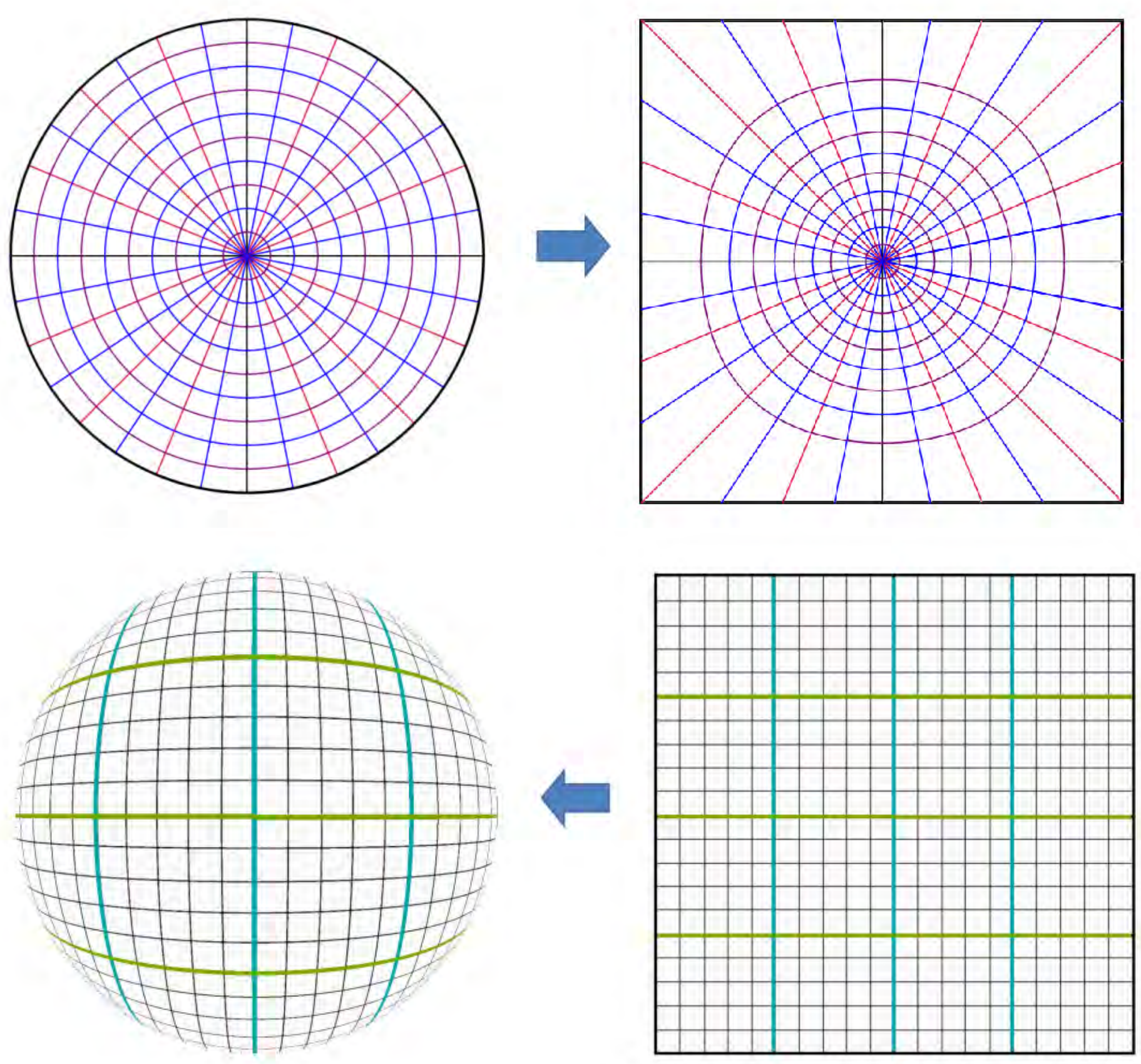

Disc to square mapping (requires quartic formula):

Let $q_0$ be the smallest non-negative real root of the quartic equation:

$$v^4 \boldsymbol{q}^4 - 2v^2 \boldsymbol{q}^3 + (u^2 + v^2 + 2u^2 v^2)\boldsymbol{q}^2 - 2u^2 \boldsymbol{q} + u^4 = 0$$

$$x = sgn(u)\sqrt{q_0} \qquad\qquad y = sgn(v)\left|\frac{v}{u}\right|\sqrt{q_0}$$

Square to disc mapping:

$$u = \frac{x\sqrt{2 + 2x^2y^2 - x^2 - y^2}}{1 + x^2y^2} \qquad\qquad v = \frac{y\sqrt{2 + 2x^2y^2 - x^2 - y^2}}{1 + x^2y^2}$$

## 1.7 Rectangle to Ellipse Mapping

Given the various square-to-disc mappings, we can extend these to handle rectangles and ellipses. First, let us cover some notation. Let the input rectangular coordinates be $(x_o,y_o)$ and the output elliptical coordinates be $(u_o,v_o)$. Here are the steps to get equations relating $(x_o,y_o)$ with $(u_o,v_o)$.

1) Input: rectangular coordinates $(x_o,y_o)$

2) Given the rectangular coordinates, remove the eccentricity and squeeze the coordinates into square coordinates (x,y)

$$x = \frac{x_o}{a} \qquad\qquad y = \frac{y_o}{b}$$

3) Choose an appropriate square-to-disc mapping from the list previously provided. Denote this mapping as two functions $g_o$ and $h_o$ such that

$$u = g_o(x,y) \qquad \mathrm{v} = h_o(x,y)$$

4) Convert circular coordinates into elliptical coordinates by reintroducing the eccentricity back.

$$u_o = au = a\, g_o\left(\frac{x_o}{a},\frac{y_o}{b}\right)$$

$$v_o = bv = b\, h_o(\frac{x_o}{a},\frac{y_o}{b})$$

5) Output: elliptical coordinates $(u_o,v_o)$ computed from the equations given above.

These steps are summarized in the diagram provided below.

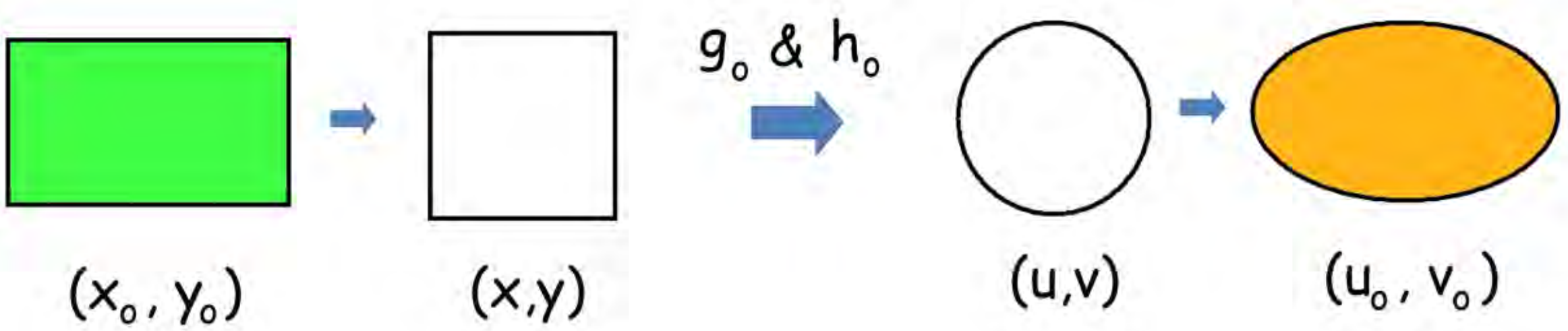

Figure 4: Rectangle to ellipse pipeline

## 1.8 Ellipse to Rectangle Mapping

We can now reverse the mapping using a similar approach and notation. Let the input elliptical coordinates be $(u_1,v_1)$ and the output rectangular coordinates be $(x_1,y_1)$. Here are the steps to get equations relating $(u_1,v_1)$ with $(x_1,y_1)$.

1) Input: elliptical coordinates $(u_1,v_1)$

2) Given the elliptical coordinates, remove the eccentricity and squeeze the coordinates into circular coordinates (u,v)

$$u = \frac{u_1}{a} \qquad v = \frac{v_1}{b}$$

3) Choose an appropriate disc-to-square mapping from the list previously provided. Denote this mapping as two functions $g_1$ and $h_1$ such that

$$x = g_1(u,v) \qquad y = h_1(u,v)$$

4) Convert square coordinates into rectangular coordinates by reintroducing the eccentricity back.

$$x_1 = ax = a\, g_1\left(\frac{u_1}{a},\frac{v_1}{b}\right)$$

$$y_1 = by = b\, h_1\left(\frac{u_1}{a},\frac{v_1}{b}\right)$$

5) Output: rectangular coordinates $(x_1, y_1)$ computed from the equations given above.

These steps are summarized in the diagram provided below.

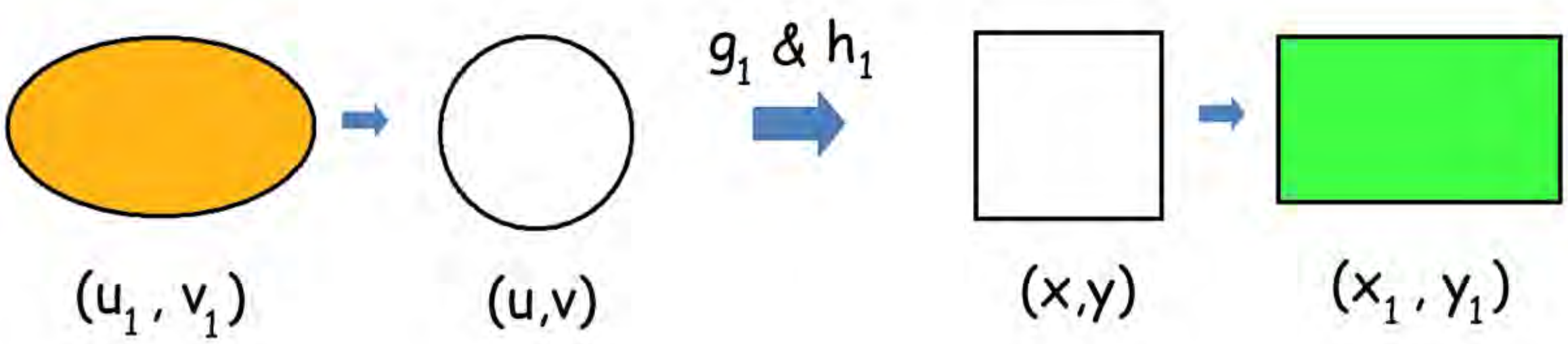

Figure 5: Ellipse to rectangle pipeline

# 2 Results

We provide elliptification results for several images in Figures 6 & 7. Of course, the simplest way to convert a rectangular image to an elliptical one is by cropping out the corner regions. For many images, this method can produce decent results. In fact, this method has been used for centuries on many paintings and photographs. However, there are several disadvantages to this approach. First, it loses significant detail near the corners. As a matter of fact, for imagery with plenty of features near the corners, this might not be acceptable. Second, this approach is not reversible. It is impossible to invert the process to get back a rectangular image because regions are entirely cropped out.

We have used each of the extended mappings mentioned in the previous section to convert rectangular imagery into elliptical regions. The mappings produce very similar results near the central regions of the images. It is only towards the corner regions that the differences are more pronounced. Even though this might not be evident for some of the resulting images, each of the mappings provided with this paper are unique. They all have distinct forward and inverse equations.

We would like to mention that applying our eccentric extension to the Schwarz-Christoffel mapping is technically not equivalent to the mapping developed by Schwarz and Christoffel in the 1860s. Their work only covers mapping the upper half-plane or unit disc to a polygonal region in the complex plane. Their work does not cover the ellipse at all. Therefore, we shall call this eccentric extension of their mapping as the "Stretched Schwarz-Christoffel"

## 2.1 United States flag

The elliptification of flags is a popular form of artistic stylization. In fact, doing an internet search for national flags depicted as ellipses produces many results for different nations. Most of these results just resort to cropping the flag. The U.S. flag is not quite amenable to cropping because it contains much detail near the top left corner. As Figure 6 shows, only 41 of 50 stars in the flag are discernable after cropping. On the other hand, the other mappings in the figure show all 50 stars in the flag. Of course, this comes with a compromise. There are all sorts of size and shape distortions in these mappings. In our opinion, the Tapered2 and the Tapered4 mappings produce the best results for this case.

## 2.2 A Sunday Afternoon on the Island of La Grande Jatte

We believe that the elliptification of paintings is a legitimate and meaningful form of artistic expression. In figure 7, we show different mappings applied to a world-famous painting. Georges Seurat's "A Sunday Afternoon on the Island of La Grande Jatte" is a well-known painting that is considered to be the origin of the Pointillism movement in Neo-Impressionist art. It is a beautiful painting that took Seurat over 2 years to complete.

The painting depicts a busy park along the banks of the river Seine. It is rich with corner to corner details that will simply be lost if cropped. Consequently, this painting as a prime example of a painting that requires more sophisticated methods if converted into an ellipse. The top right image in figure 7 shows much of the detailed richness of painting gone if simple cropped. These include interesting features such as the lady's umbrella, the sail boat, and most of the pipe-smoking man's legs. On the other hand, the other mappings preserve most of these details, albeit with some noticeable distortions.

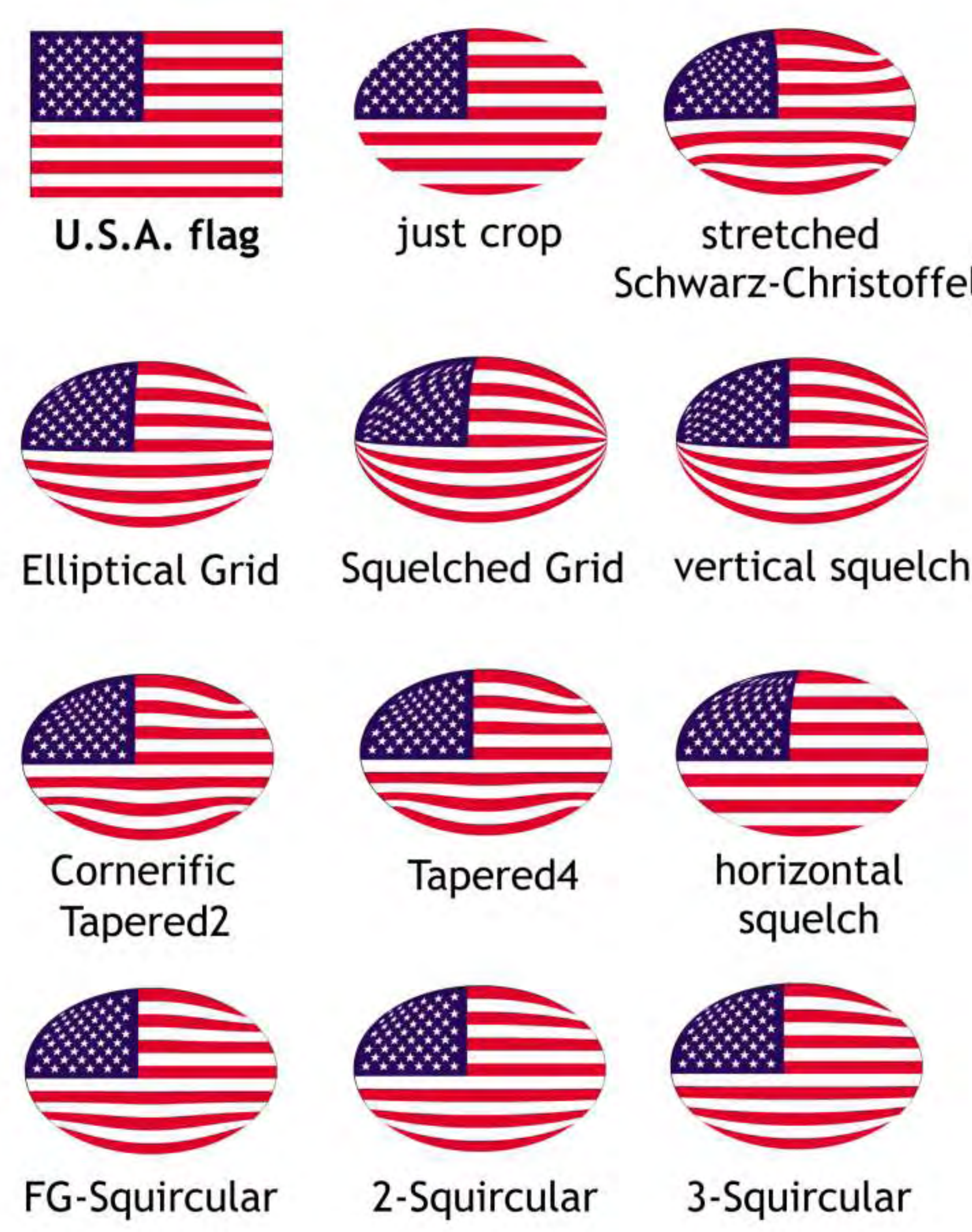

Figure 6: Elliptification of the U.S. flag

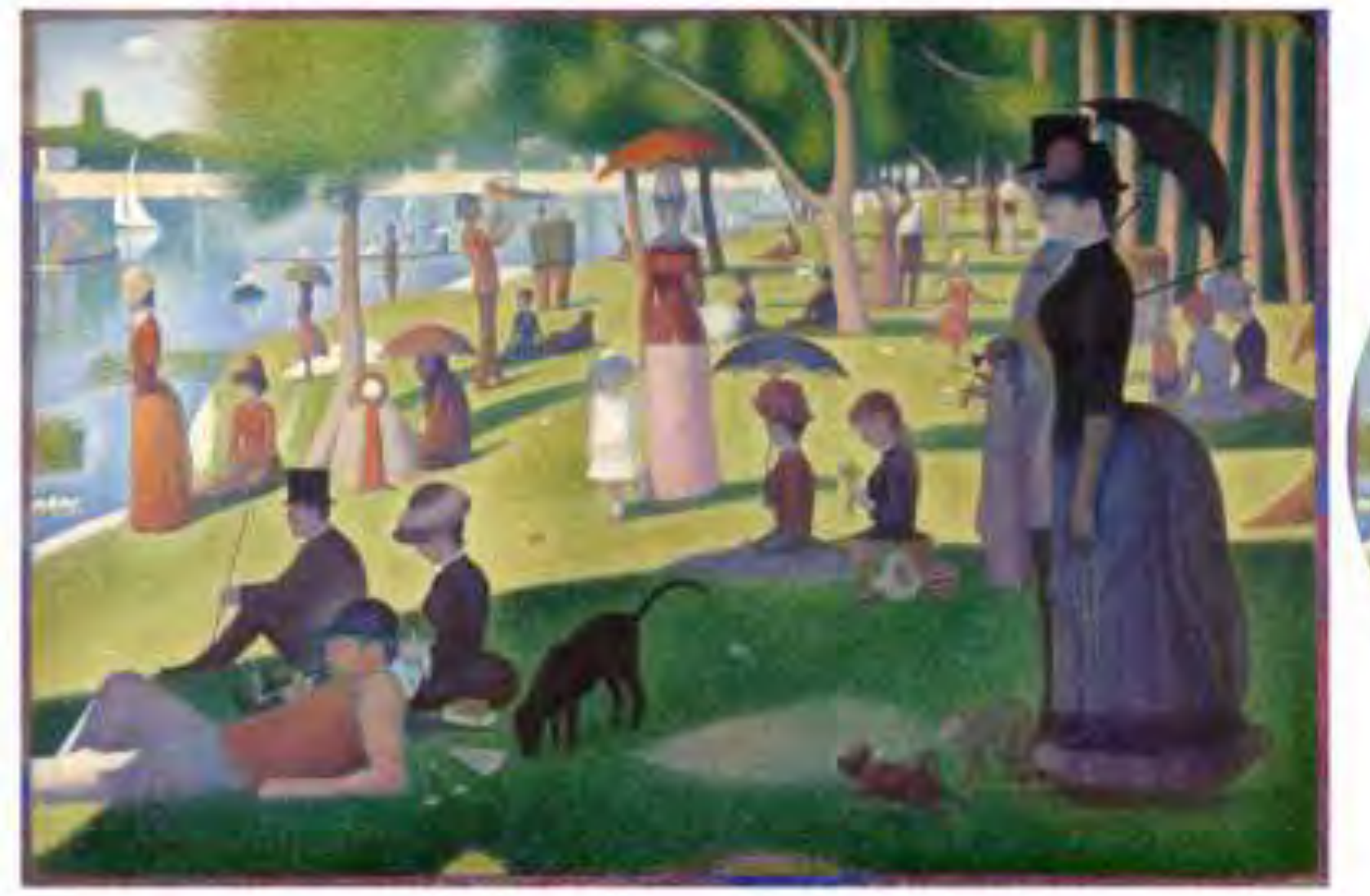

original

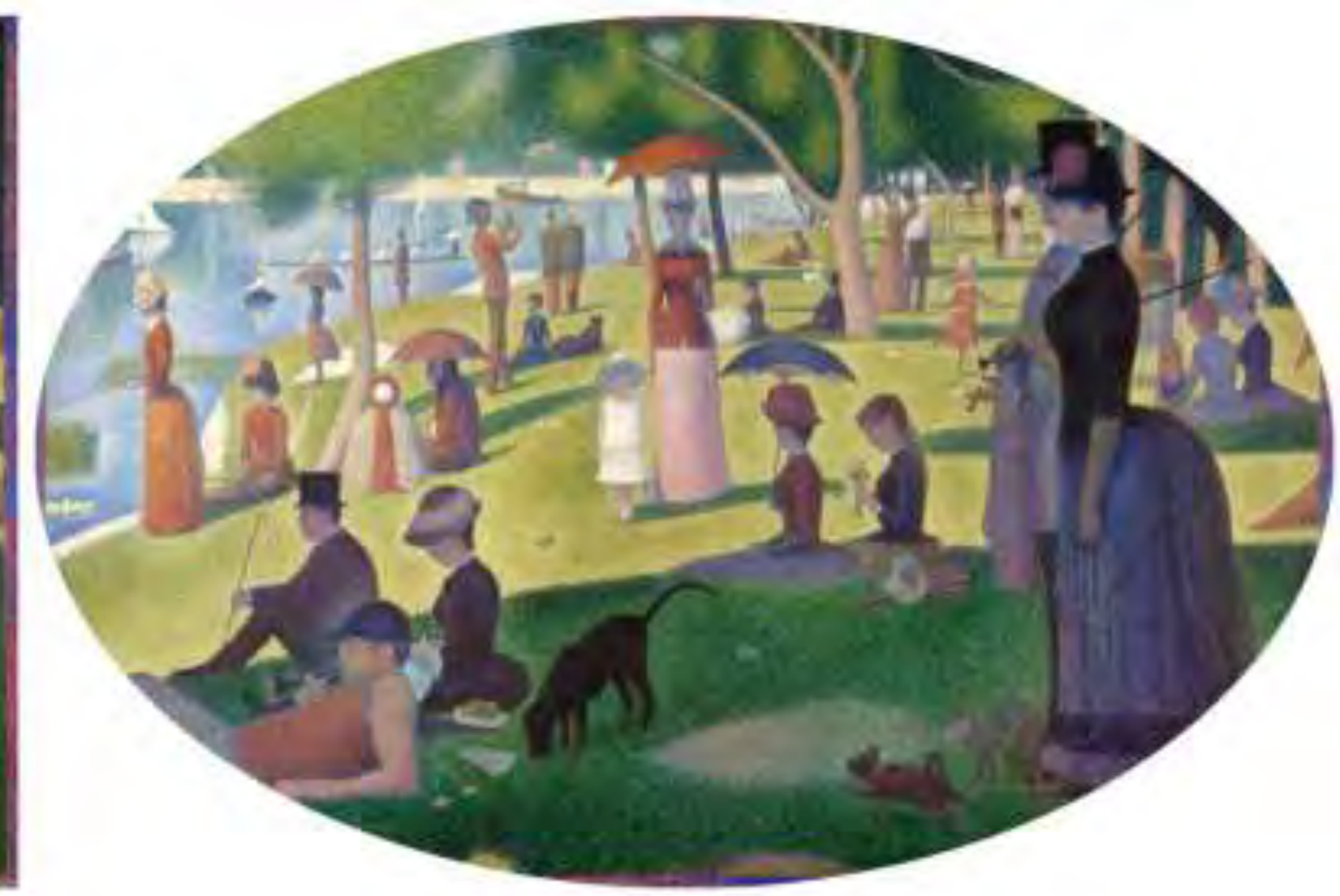

just crop

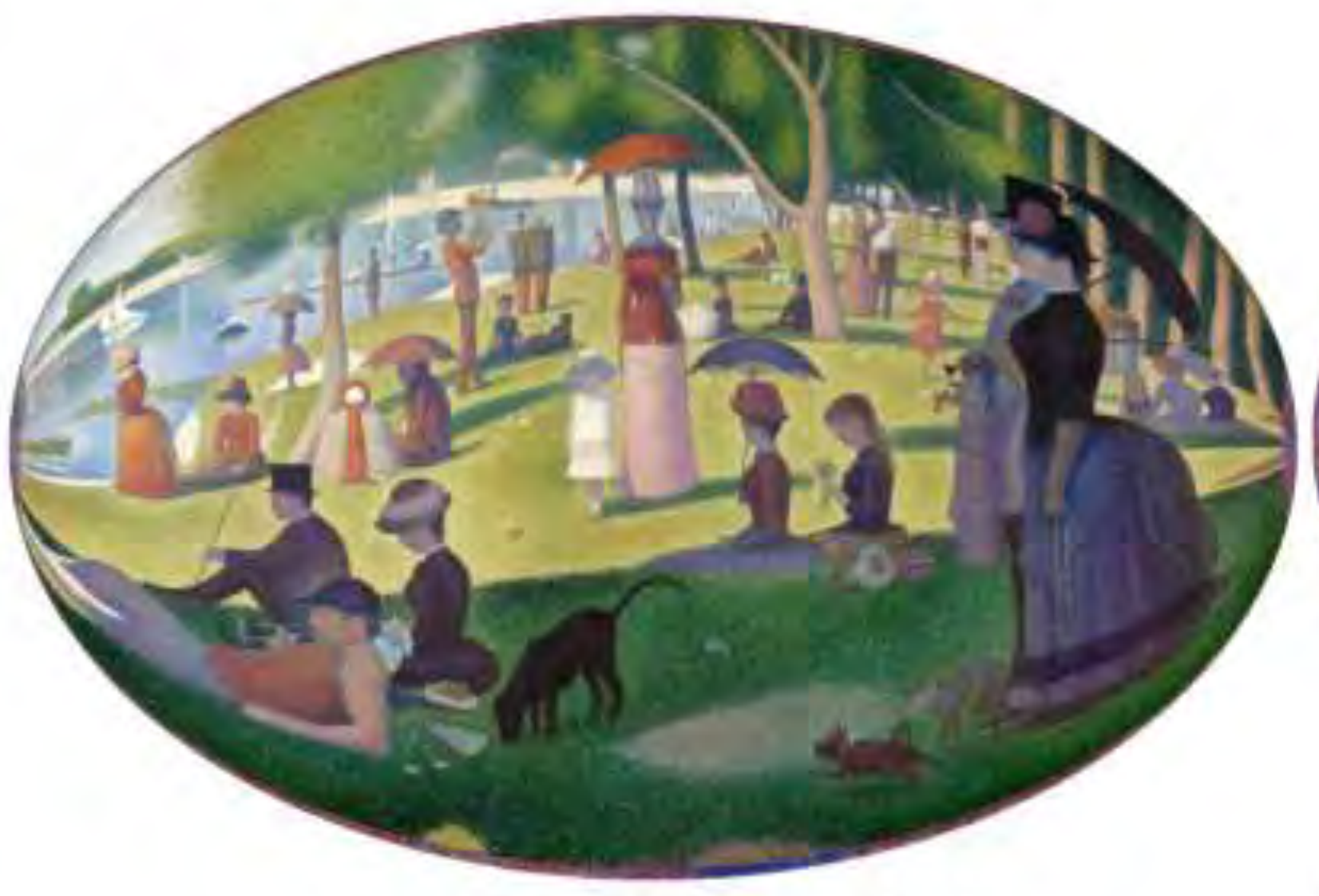

vertical squelch

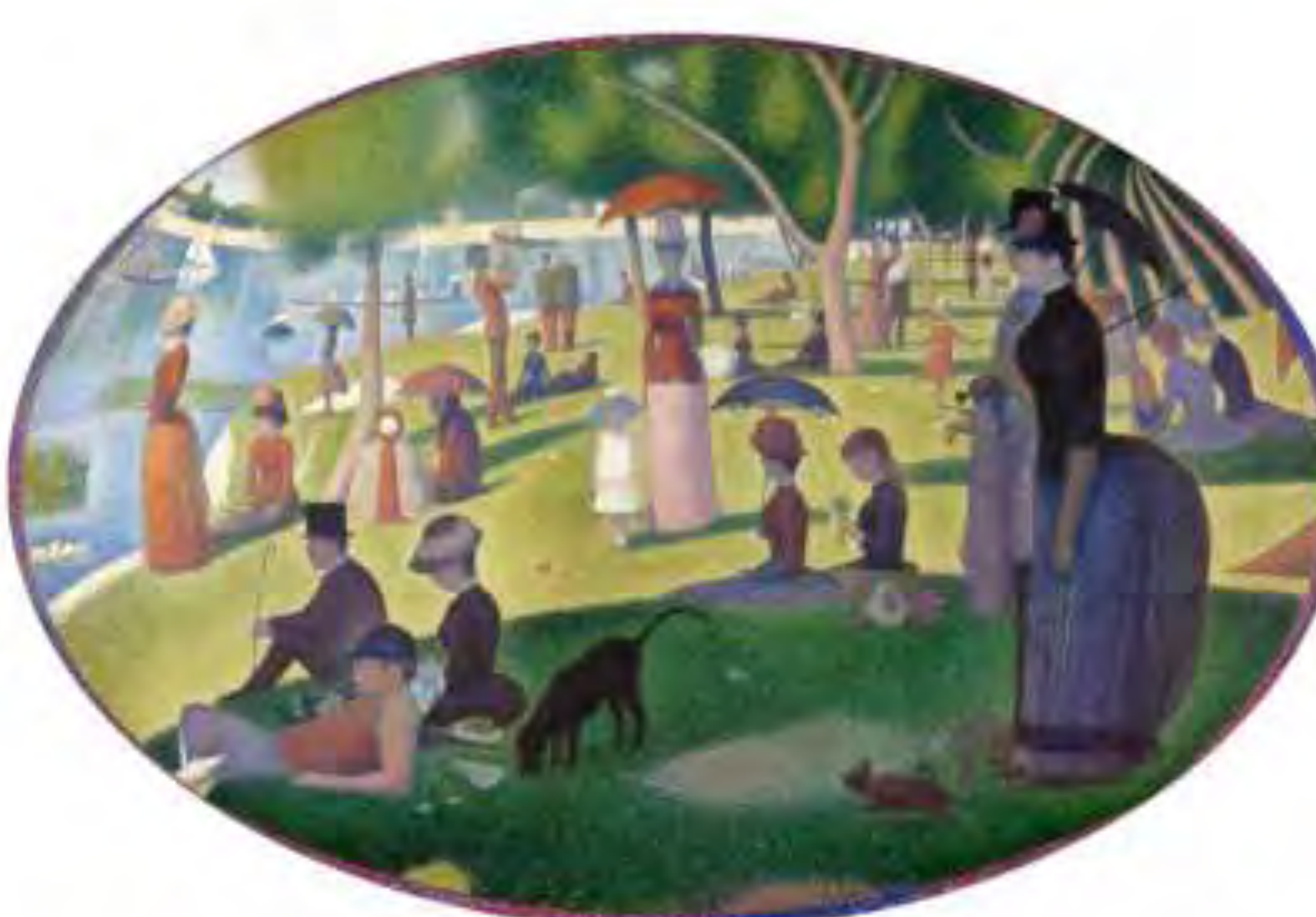

stretched Schwarz-Christoffel

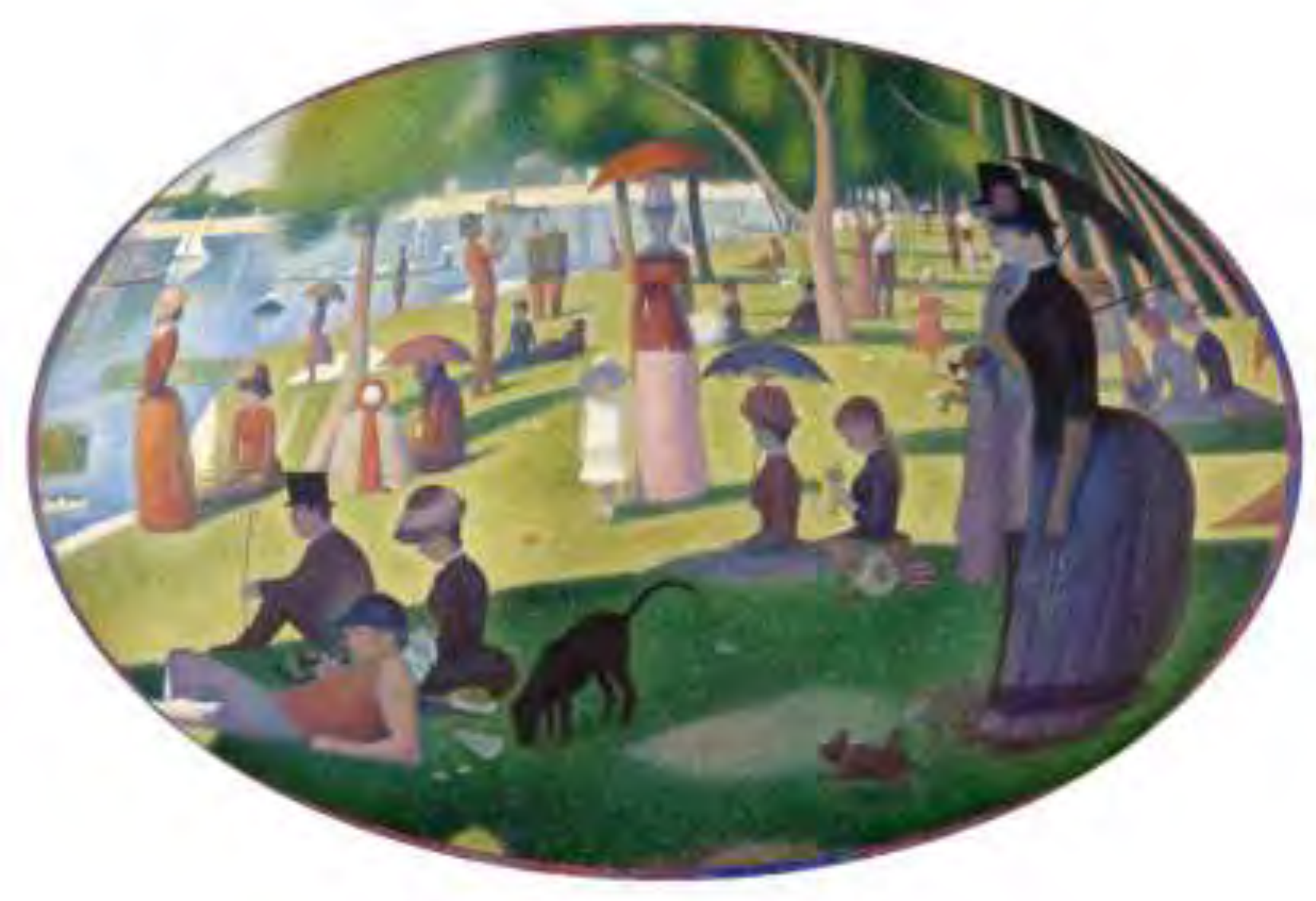

FG-Squircular

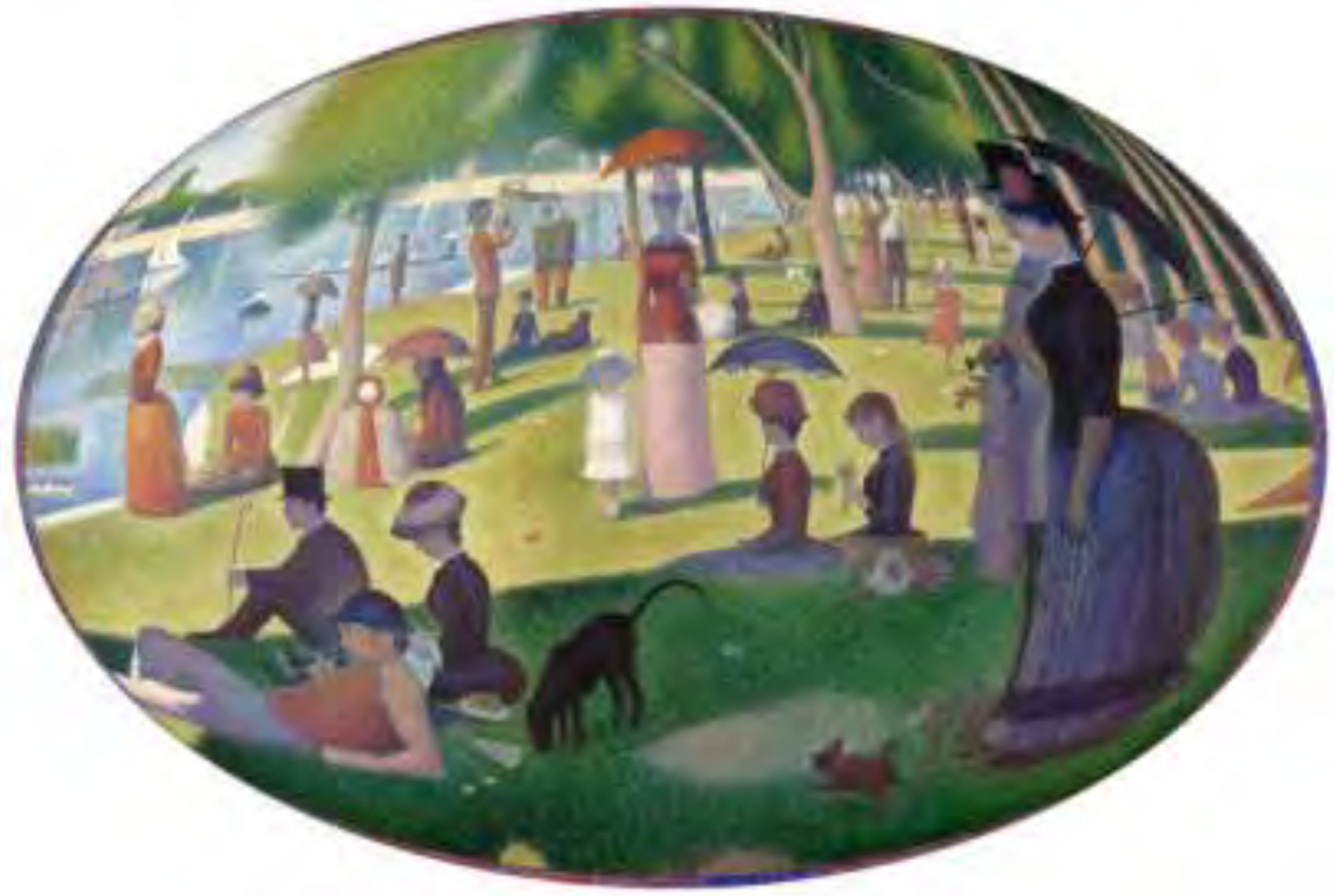

Elliptical Grid

Figure 7: Elliptified painting

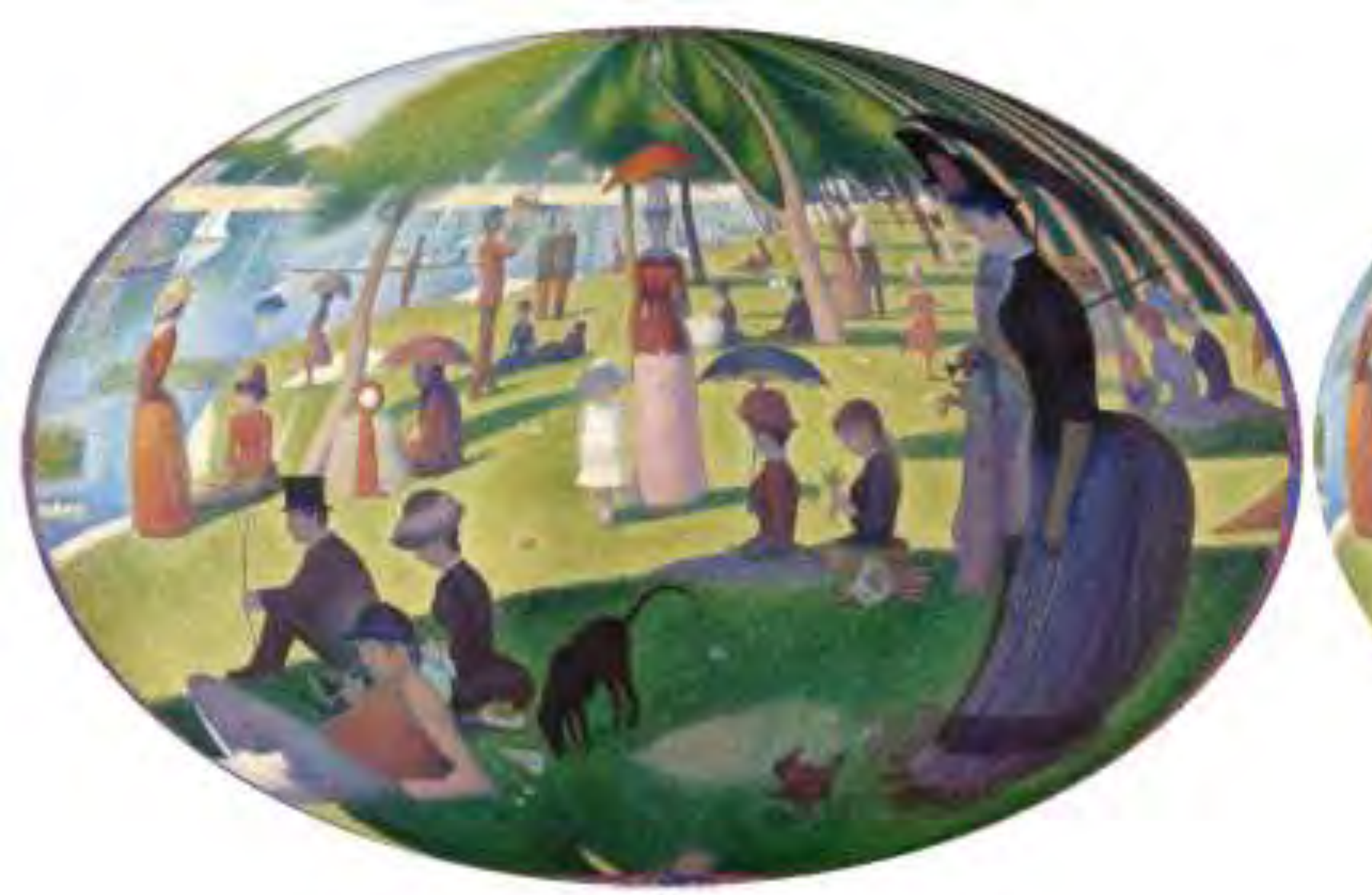

horizontal squelch

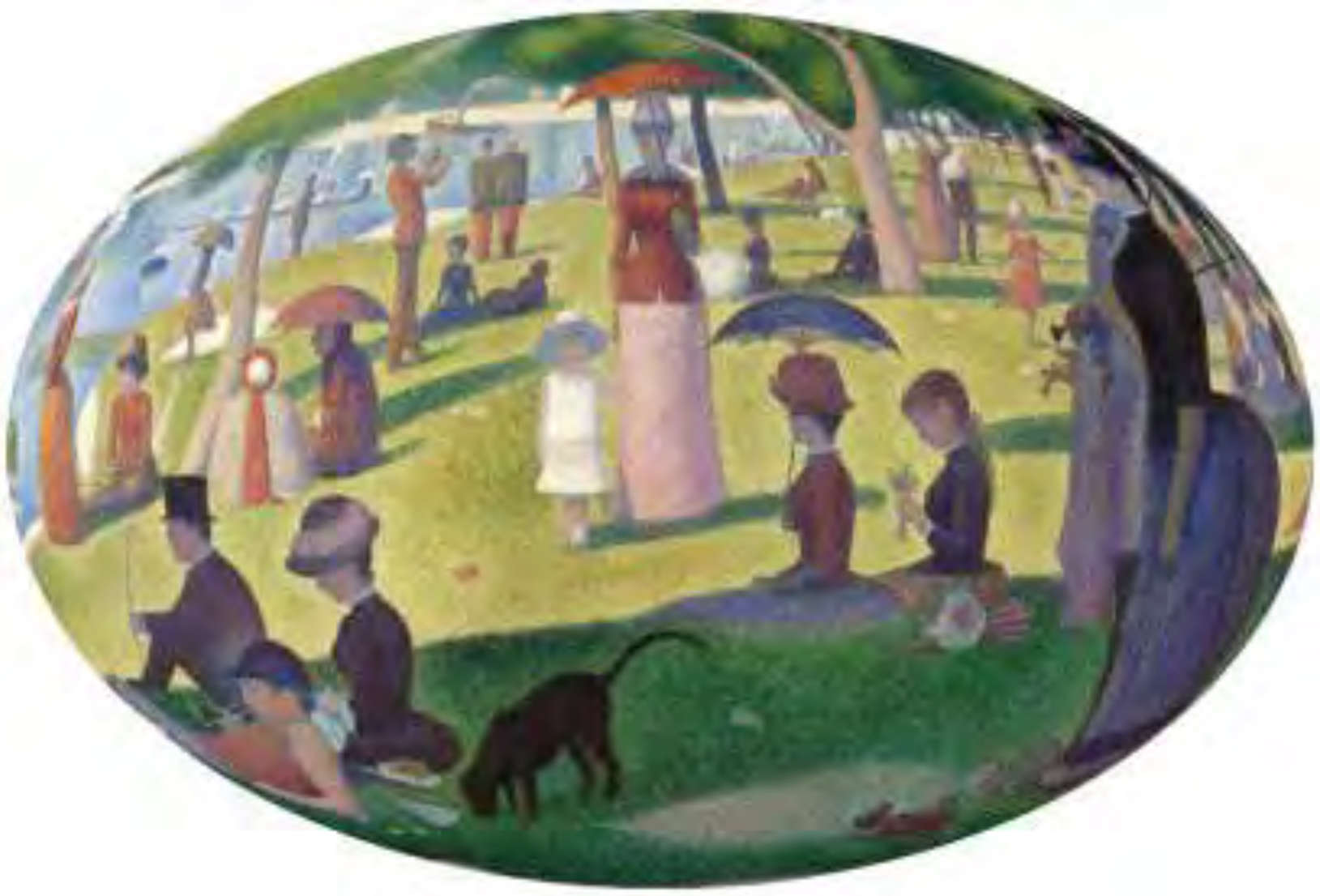

Sham Quartic

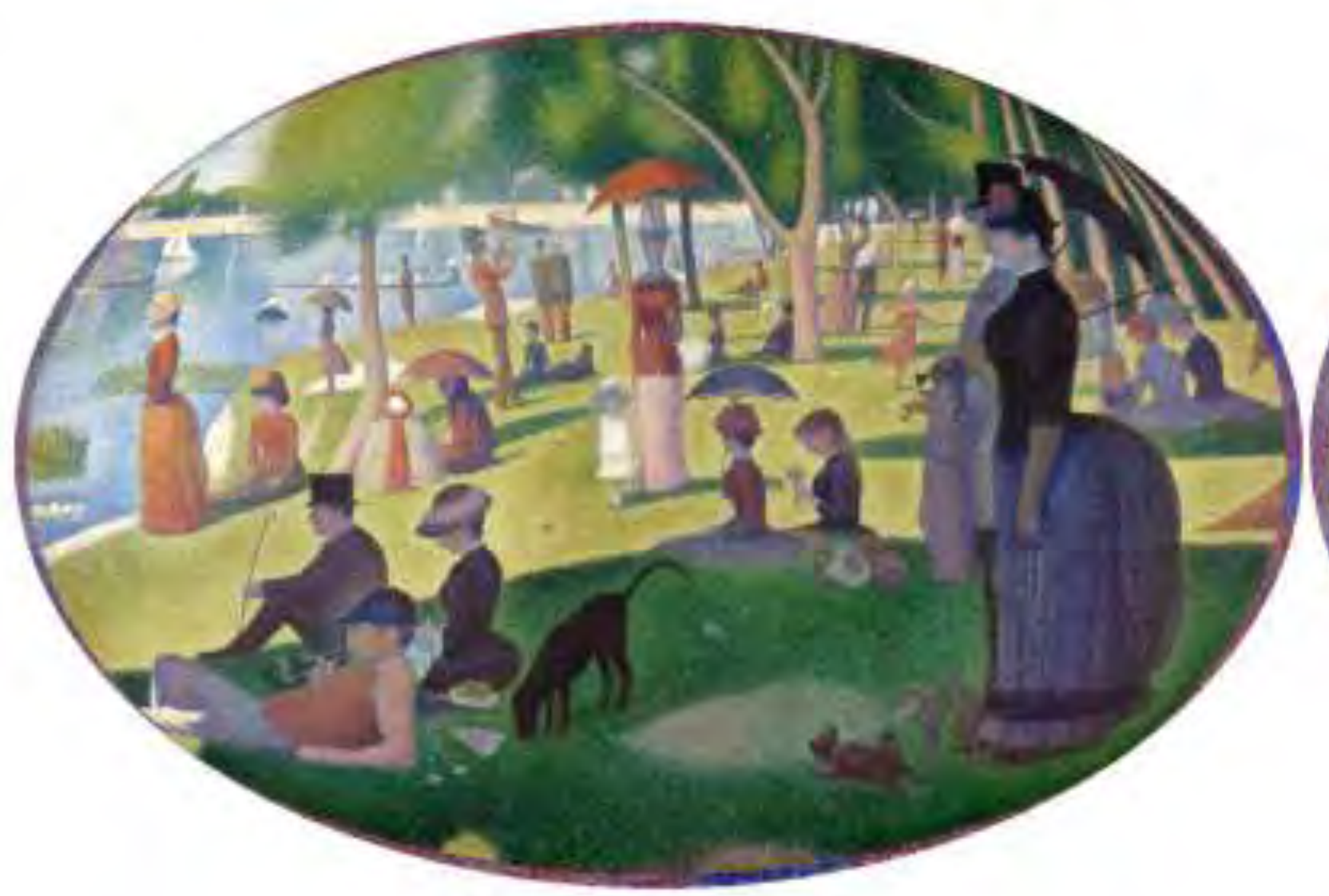

2-Squircular

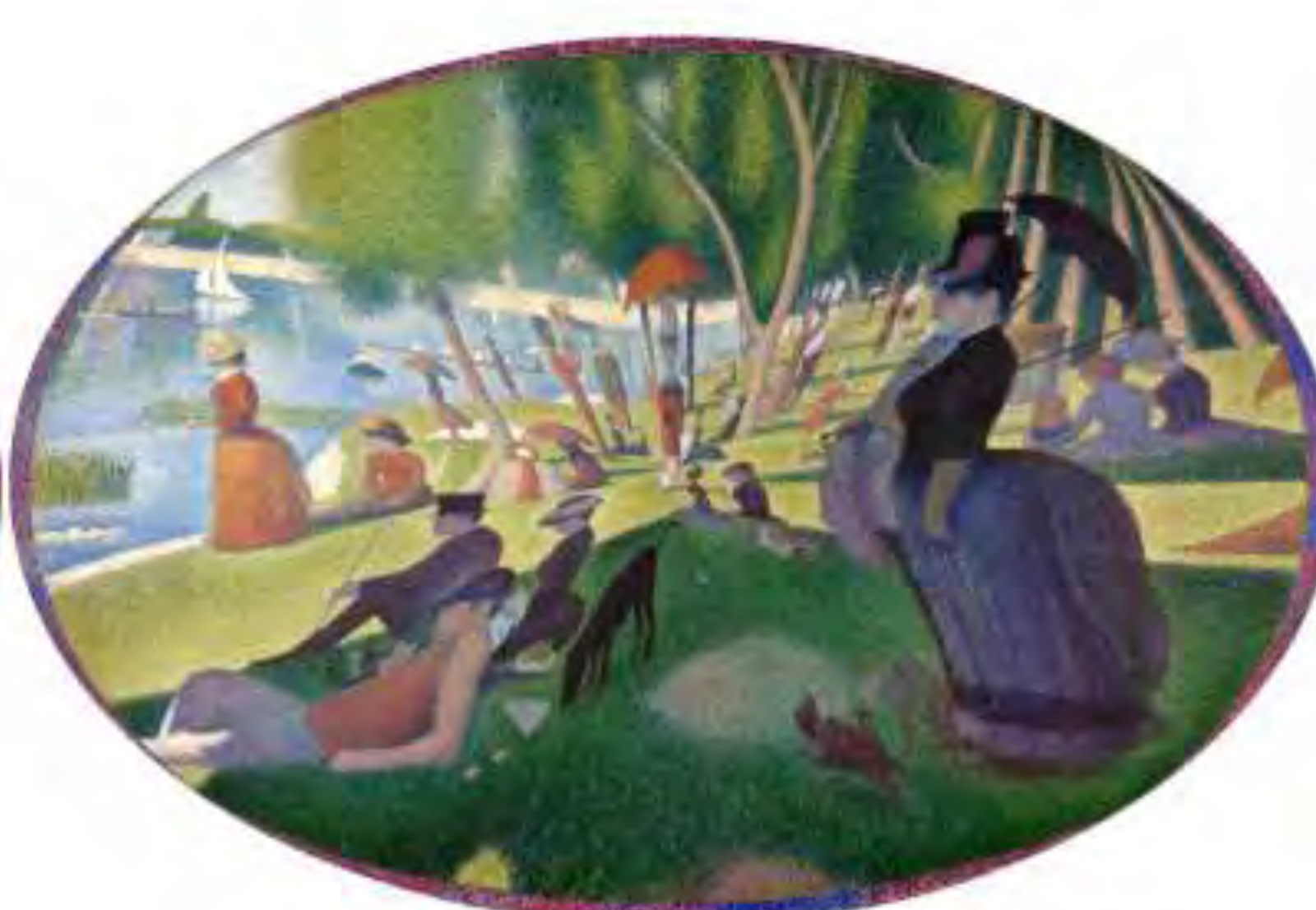

Non-axial 2-Pinch

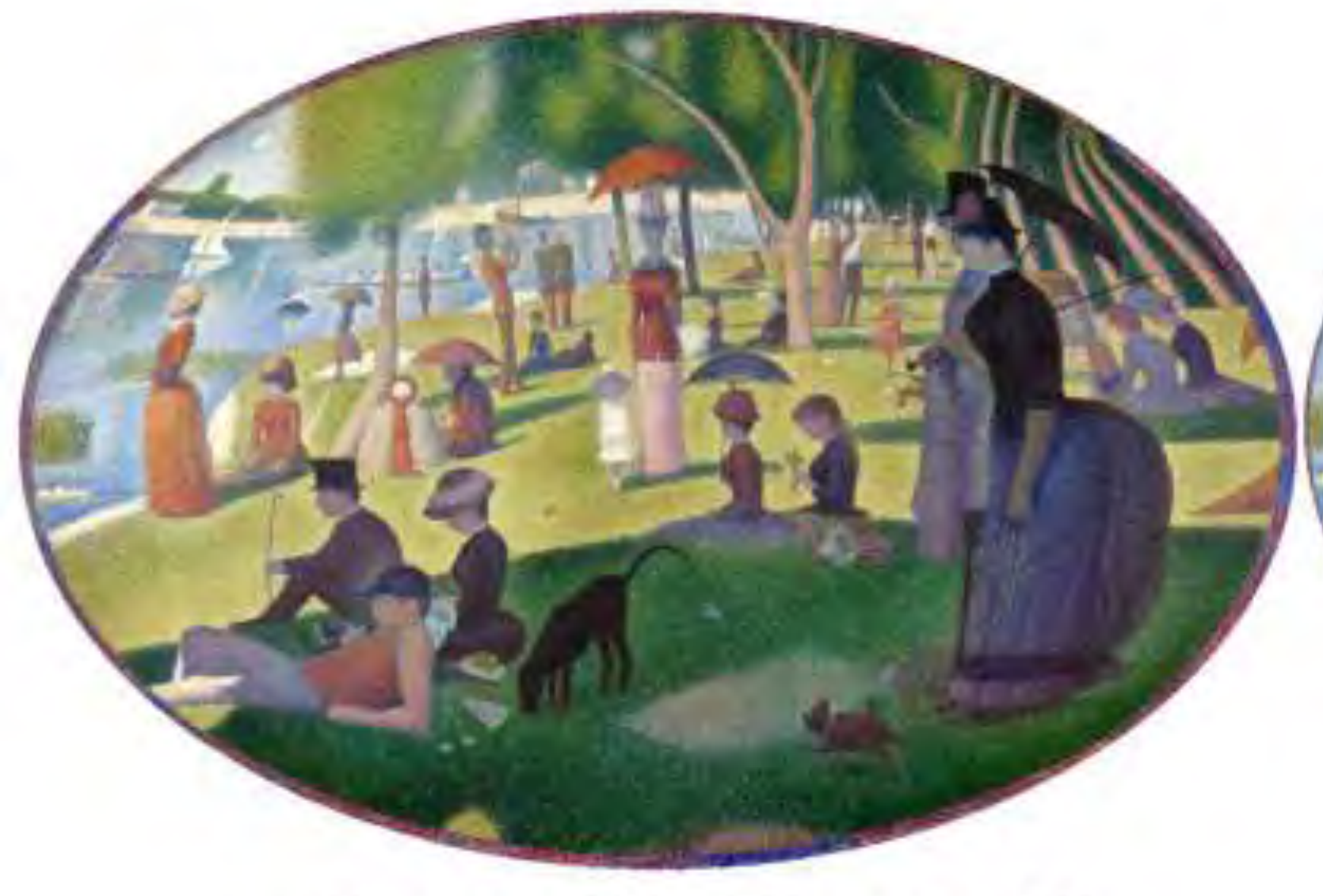

Cornerific Tapered2

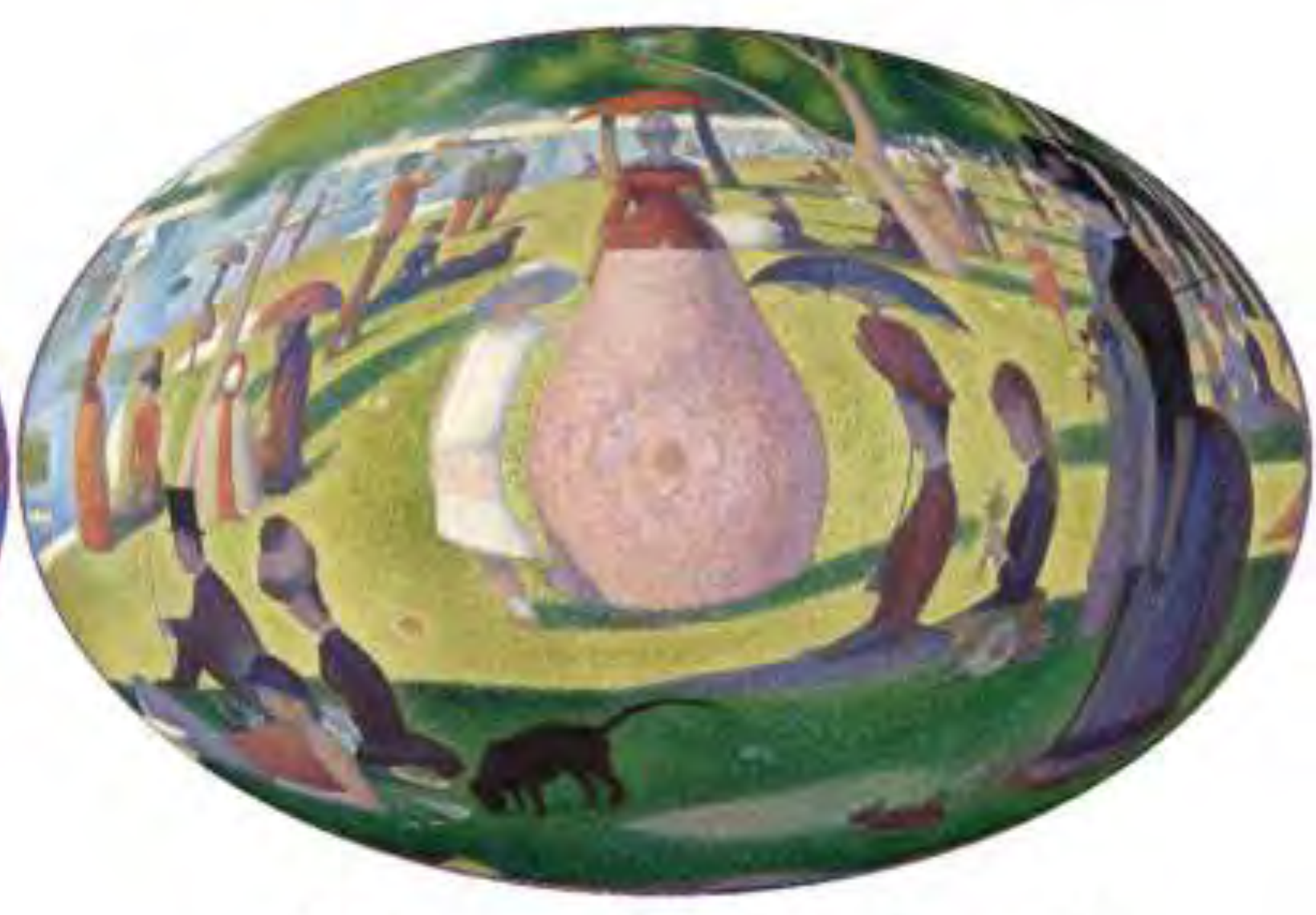

Non-axial ½- Punch

Figure 7 ( continued )

# Part II. Mathematical Details

## 3.1 Review of Concepts and Basic Equations

This paper is the spiritual successor of the paper "*Analytical Methods for Squaring the Disc*" [Fong 2014]. In this section, we will review some important concepts and equations from that paper.

The *Fernandez-Guasti squircle* [Fernandez-Guasti 1992] is an intermediate shape between the circle and the square. It is a quartic curve with the equation

$$x^2 + y^2 - \frac{s^2}{t^2} x^2 y^2 = t^2$$

*Linear squircular continuum of the square.* This follows from setting $s = t$ in the squircle equation with

$$x^2 + y^2 - x^2 y^2 = t^2 \qquad for \ \ t \in [0,1]$$

In other words, the square and its interior can be represented by this equation

$$x^2 + y^2 - x^2 y^2 \leq 1$$

*Radial constraint of mappings.* All radial disc-to-square mappings satisfy the constraint $y = \frac{v}{u} x$

*Radial mapping linear parametric equation.* This is the general form of equation for all linear radial square-to-disc mappings.

$$u = \frac{x}{\sqrt{x^2+y^2}} \ t \qquad\qquad v = \frac{y}{\sqrt{x^2+y^2}} \ t$$

The parameter *t* is some suitable function of x and y with $0 \leq t \leq 1$.

## 3.2 Rampant Functions

There is a family of functions that we consider important in the study of disc-to-square mappings. In fact, they are so important that we decided to give this type of functions a name.

*Definition*: We define a *rampant* function as a monotonic and continuous function *f* that is valid in the interval [0, 1] and having these properties

$$f(0) = 0 \qquad\qquad and \qquad\qquad f(1) = 1$$

*Example*: The function $f(t) = t^n$ for n>0 is a rampant function. It is easy to check that $0^n = 0 \ \ and \ 1^n = 1$ for all *n>0*. When *n≤0*, these conditions will no longer hold.

*Example*: The function $f(t) = t\sqrt{2 - t^2}$ is a rampant function. It can be verified that $f'(t) = -\frac{2(x^2-1)}{\sqrt{2-x^2}} > 0$ in the interval [0,1] so *f* is a monotonic function in that interval.

*Example*: The function $f(t) = t\sqrt{\frac{3}{2} - \frac{1}{2}t^4}$ is a rampant function. It can be verified that $f'(t) = \frac{3-3x^4}{\sqrt{6-2x^2}} > 0$ in the interval [0,1] so *f* is a monotonic function in that interval

## 3.3 Exploiting Nonlinear Variations of the FG-Squircular Mapping

There are several avenues for exploiting nonlinearities in the derivation of the FG-Squircular mapping [Fong 2014] to come-up with new disc-to-square mappings. We will study two types of nonlinearities in this paper.

The 1st type of nonlinearity to exploit is the relationship between the radius of the squircle and its squareness. This boils down to the relationship between $s$ and $t$ in the squircle equation.

$$x^2 + y^2 - \frac{s^2}{t^2} x^2 y^2 = t^2$$

For the FG-Squircular mapping [Fong 2014], a simple linear relationship of $s=t$ was imposed. But, it is possible to replace this with a nonlinear relationship $s=f(t)$ where $f$ is a rampant function. We shall call these nonlinear relationships between $s$ and $t$ as *squircular nonlinearities.* For this paper, we will specifically explore several squircular nonlinearities to produce new mappings between the circular disc and the square

- $s = t^2$
- $s = t^3$
- $s = t\sqrt{2 - t^2}$
- $s = t\sqrt{\frac{3}{2} - \frac{1}{2}\, t^4}$

The 2nd type of nonlinearity to exploit is within the radial mapping parametric equation. The FG-Squircular mapping uses linear parametric equations in the form of

$$u = \frac{x}{\sqrt{x^2 + y^2}}\; t$$

$$v = \frac{y}{\sqrt{x^2 + y^2}}\; t$$

These equations can be made nonlinear by introducing a nonlinear function $m$ on the parameter $t$ to get

$$u = \frac{x}{\sqrt{x^2 + y^2}}\; m(t)$$

$$v = \frac{y}{\sqrt{x^2 + y^2}}\; m(t)$$

where $m$ is a rampant function

We will call such nonlinearities as *axial nonlinearities*. In a latter section of this paper, we shall derive three new mappings that are based on this type of nonlinearity. Specifically, we shall use the following functions

- $m(t) = t^2$
- $m(t) = \sqrt{t}$
- $m(t) = t\sqrt{2 - t^2}$

## 4.1 A Simpler Variation of the FG-Squircular Mapping

The FG-Squircular mapping [Fong 2014] is by no means the only radial mapping between the circular disc and the square that uses the Fernandez-Guasti squircle. In this section, we shall introduce another radial mapping based on the Fernandez-Guasti squircle which has much simpler equations.

1[st] recall the *radial mapping linear parametric equations* for mappings between the circular disc and the square. All radial mappings between the disc and the square have an equation of this form, where the parameter *t* can be expressed as a suitable function of *x* and *y*.

$$u = t\,\frac{x}{\sqrt{x^2+y^2}} \qquad\qquad v = t\,\frac{y}{\sqrt{x^2+y^2}}$$

Next, recall our condition for the *linear squircular continuum* of the square. We start from the squircle equation $x^2 + y^2 - \frac{s^2}{t^2}x^2y^2 = t^2$ and set *s=t.* This gives the square as $\mathcal{S} = \{(x,y) \in \mathbb{R}^2 \mid x^2 + y^2 - x^2y^2 = t^2\ ,\ \ 0 \le t \le 1\}$ In other words, we have this basic equation for the FG-squircular mapping:

$$x^2 + y^2 - x^2y^2 = t^2$$

We can actually use any rampant function *f* for the squircular continuum of the square where *s=f(t).* For example, consider what happens if we set $s = t^2$ in the squircle equation instead, we get this condition

$$x^2 + y^2 - t^2x^2y^2 = t^2$$

We collect the $t^2$ terms into one side of the equation to isolate *t*

$$t^2(1 + x^2y^2) = x^2 + y^2 \quad\Rightarrow\quad t = \sqrt{\frac{x^2+y^2}{1+x^2y^2}}$$

and plug back this value of *t* to the *radial mapping linear parametric equations*.

$$u = \sqrt{\frac{x^2+y^2}{1+x^2y^2}}\,\frac{x}{\sqrt{x^2+y^2}} \qquad\qquad v = \sqrt{\frac{x^2+y^2}{1+x^2y^2}}\,\frac{y}{\sqrt{x^2+y^2}}$$

This simplifies to

$$\boldsymbol{u = \frac{x}{\sqrt{1 + x^2y^2}}} \qquad\qquad \boldsymbol{v = \frac{y}{\sqrt{1 + x^2y^2}}}$$

We now solve for the inverse of these mapping equations.

$$u = \frac{x}{\sqrt{1 + x^2y^2}} \quad\Rightarrow\quad u\sqrt{1 + x^2y^2} = x \quad\Rightarrow\quad u^2(1 + x^2y^2) = x^2 \quad\Rightarrow\quad 1 + x^2y^2 = \frac{x^2}{u^2}$$

$$\Longrightarrow\quad x^2y^2 = \frac{x^2}{u^2} - 1 \quad\Rightarrow\quad y^2 = \frac{1}{u^2} - \frac{1}{x^2} \quad\Rightarrow\quad y = \sqrt{\frac{1}{u^2} - \frac{1}{x^2}}$$

Substitute y into the other equation

$$v = \frac{y}{\sqrt{1 + x^2y^2}} = \frac{\sqrt{\frac{1}{u^2} - \frac{1}{x^2}}}{\sqrt{1 + x^2\left(\frac{1}{u^2} - \frac{1}{x^2}\right)}} = \frac{\sqrt{\frac{1}{u^2} - \frac{1}{x^2}}}{\sqrt{\frac{x^2}{u^2}}} = \frac{u}{x}\sqrt{\frac{1}{u^2} - \frac{1}{x^2}} = \frac{u}{x}\sqrt{\frac{x^2 - u^2}{u^2x^2}} = \frac{\sqrt{x^2 - u^2}}{x^2}$$

We now find a polynomial equation for x in terms of u and v.

$$v^2 = \frac{x^2 - u^2}{x^4} \quad\Rightarrow\quad x^4 = \frac{x^2 - u^2}{v^2} \quad\Rightarrow\quad x^4v^2 - x^2 + u^2 = 0$$

This is a biquadratic equation in which we could solve for $x^2$ using the standard quadratic equation with coefficients: $a = v^2$ $b = -1$ and $c = u^2$

$$x^2 = \frac{1 \pm \sqrt{1 - 4u^2v^2}}{2v^2}$$

We can then get a quadrant-aware expression for x by taking the square root.

$$\boldsymbol{x} = \frac{\boldsymbol{sgn(uv)}}{\boldsymbol{v\sqrt{2}}} \sqrt{\boldsymbol{1 - \sqrt{1 - 4u^2v^2}}} = \frac{sgn(uv)}{v} \sqrt{\frac{1}{2} - \sqrt{\frac{1}{4} - u^2v^2}}$$

Similarly, we get an expression for y as

$$\boldsymbol{y} = \frac{\boldsymbol{sgn(uv)}}{\boldsymbol{u\sqrt{2}}} \sqrt{\boldsymbol{1 - \sqrt{1 - 4u^2v^2}}} = \frac{sgn(uv)}{u} \sqrt{\frac{1}{2} - \sqrt{\frac{1}{4} - u^2v^2}}$$

In summary, we were able to derive another radial mapping based on the Fernandez-Guasti squircle by using simple algebraic manipulations on the squircular continuum of the square. We were also able to find simple inverse equations for the mapping. We shall name this mapping as the *2-Squircular mapping***.**

Note that this case produces mapping equations that are much simpler than the case $s=t$. The qualitative results of both mappings are very similar. In fact, it is quite difficult to tell the two mappings apart visually. See the figures from part 1 of this paper to compare and contrast this mapping with the FG-Squircular mapping.

These mapping equations can be made compact by using vector notation. Here are the mappings rewritten in vector format.

Disc to square mapping:

$$\begin{bmatrix} \boldsymbol{x} \\ \boldsymbol{y} \end{bmatrix} = \frac{\boldsymbol{sgn(uv)}}{\boldsymbol{\sqrt{2}}} \sqrt{\boldsymbol{1 - \sqrt{1 - 4u^2v^2}}} \begin{bmatrix} \dfrac{\boldsymbol{1}}{\boldsymbol{v}} \\ \\ \dfrac{\boldsymbol{1}}{\boldsymbol{u}} \end{bmatrix}$$

Square to disc mapping:

$$\begin{bmatrix} \boldsymbol{u} \\ \boldsymbol{v} \end{bmatrix} = \frac{\boldsymbol{1}}{\sqrt{\boldsymbol{1 + x^2y^2}}} \begin{bmatrix} \boldsymbol{x} \\ \boldsymbol{y} \end{bmatrix}$$

## 4.2 A Quasi-Symmetric Variation of the FG-Squircular Mapping

In the previous section, we derived a mapping by using the assumption $s=t^2$. We can take this step further by considering the case when $s=t^3$. The squircle equation then reduces to $x^2+y^2-t^4x^2y^2=t^2$ or equivalently, we have this biquadratic polynomial equation in $t$:

$$x^2y^2\,\boldsymbol{t}^4+\boldsymbol{t}^2-x^2-y^2=0$$

We can solve for $t^2$ using the quadratic equation to get

$$t^2=\frac{\pm\sqrt{4x^4y^2+4x^2y^4+1}-1}{2x^2y^2}$$

or

$$t=\pm\sqrt{\frac{\pm\sqrt{4x^4y^2+4x^2y^4+1}-1}{2x^2y^2}}=\frac{\pm\sqrt{\pm\sqrt{1+4x^2y^2(x^2+y^2)}-1}}{xy\sqrt{2}}$$

Plugging back into the *radial mapping linear parametric equations* and accounting for the 4 different quadrants, we get these square-to-disc mapping equations

$$\boldsymbol{u}=\frac{\boldsymbol{sgn(xy)}}{\boldsymbol{y}}\sqrt{\frac{\mathbf{-1}+\sqrt{\mathbf{1}+\mathbf{4}\boldsymbol{x}^4\boldsymbol{y}^2+\mathbf{4}\boldsymbol{x}^2\boldsymbol{y}^4}}{\mathbf{2}\,(\boldsymbol{x}^2+\boldsymbol{y}^2)}}$$

$$\boldsymbol{v}=\frac{\boldsymbol{sgn(xy)}}{\boldsymbol{x}}\sqrt{\frac{\mathbf{-1}+\sqrt{\mathbf{1}+\mathbf{4}\boldsymbol{x}^4\boldsymbol{y}^2+\mathbf{4}\boldsymbol{x}^2\boldsymbol{y}^4}}{\mathbf{2}\,(\boldsymbol{x}^2+\boldsymbol{y}^2)}}$$

For the inverse of these mapping equations, we can start by manipulating the **u** equation

$$uy\sqrt{2(x^2+y^2)}=\sqrt{-1+\sqrt{1+4x^2y^2(x^2+y^2)}}\qquad\Rightarrow$$

$$2u^2y^2(x^2+y^2)=-1+\sqrt{1+4x^2y^2(x^2+y^2)}\qquad\Rightarrow$$

$$(2u^2y^2(x^2+y^2)+1)^2=1+4x^2y^2(x^2+y^2)\qquad\Rightarrow$$

$$4u^4y^4(x^2+y^2)^2+4u^2y^2(x^2+y^2)+1=4x^4y^2+4x^2y^4+1\qquad\Rightarrow$$

$$4u^4y^4(x^4+2x^2y^2+y^4)+4u^4x^2y^2+4u^4y^4=4x^4y^2+4x^2y^4$$

We can divide the equation by $4y^2$ to get

$$u^4x^4y^2+2u^4x^2y^4+u^4y^6+u^4x^2+u^4y^2=x^4+x^2y^2$$

Substitute $y=\frac{v}{u}x$, to get

$$u^2v^2x^6+2v^4x^6+\frac{v^6}{u^2}x^6+u^2x^2+v^2x^2=x^4+\frac{v^2}{u^2}x^4$$

Multiply the equation by $\frac{u^2}{x^2}$ to get

$$u^4v^2x^4+2u^2v^4x^4+v^6x^4+u^4+u^2v^2=u^2x^2+v^2x^2$$

Collect the terms to get this biquadratic equation in x

$$(u^4v^2 + 2u^2v^4 + v^6)\boldsymbol{x}^4 - (u^2+v^2)\boldsymbol{x}^2 + u^4 + u^2v^2 = 0 \qquad \Rightarrow$$
$$v^2(u^2+v^2)^2\boldsymbol{x}^4 - (u^2+v^2)\boldsymbol{x}^2 + u^2(u^2+v^2) = 0$$

We can then divide the equation by $(u^2+v^2)$ to get this simple biquadratic equation in *x*

$$v^2(u^2+v^2)\boldsymbol{x}^4 - \boldsymbol{x}^2 + u^2 = 0$$

Thus, we can solve for $x^2$ using the quadratic equation.

$$x^2 = \frac{1 \pm \sqrt{1 - 4u^2v^2(u^2+v^2)}}{2v^2(u^2+v^2)} = \frac{1 \pm \sqrt{1 - 4u^4v^2 - 4u^2v^4}}{2v^2(u^2+v^2)}$$

Taking the square root of this and accounting for the 4 quadrants, we get this equation in x.

$$\boldsymbol{x = \frac{sgn(uv)}{v}\sqrt{\frac{1 - \sqrt{1 - 4u^4v^2 - 4u^2v^4}}{2(u^2+v^2)}}}$$

Substitute $y = \frac{v}{u}x$, we get the equation for y as

$$\boldsymbol{y = \frac{sgn(uv)}{u}\sqrt{\frac{1 - \sqrt{1 - 4u^4v^2 - 4u^2v^4}}{2(u^2+v^2)}}}$$

If we compare these inverse equations with the forward equations derived earlier, we can see that the two sets of equations look very similar. In fact, they are almost symmetrical if not for a few changes in sign involved in the addition/subtraction of terms in the numerator. This quasi-symmetrical property of the forward and inverse equations is noteworthy and surprising.

Since this mapping comes from the nonlinear relationship $s=t^3$ in the Fernandez-Guasti squircle, we shall denote this mapping as the 3-Squircular mapping. This mapping is also radial.

These mapping equations can be made compact by using vector notation. Here are the mappings rewritten in vector format.

Disc to square mapping:

$$\boldsymbol{\begin{bmatrix} x \\ y \end{bmatrix} = sgn(uv)\sqrt{\frac{1 - \sqrt{1 - 4u^4v^2 - 4u^2v^4}}{2(u^2+v^2)}}\begin{bmatrix} \frac{1}{v} \\ \frac{1}{u} \end{bmatrix}}$$

Square to disc mapping:

$$\boldsymbol{\begin{bmatrix} u \\ v \end{bmatrix} = sgn(xy)\sqrt{\frac{-1 + \sqrt{1 + 4x^4y^2 + 4x^2y^4}}{2(x^2+y^2)}}\begin{bmatrix} \frac{1}{y} \\ \frac{1}{x} \end{bmatrix}}$$

## 5 More Variations of the Radial Squircular Mapping

We can further generalize the result in Section 3.1 by setting $s=t^n$ for different exponents $n$. Not all exponents are amenable to simplification so we have to carefully select those that produce soluble polynomials. Also, we have to restrict ourselves to n>0 for the exponents.

Note that we will only solve for square-to-disc mapping equations in this section. We have deliberately left out the inverse equations because these are particularly gnarly, if at all possible to be written in the form of an explicit equation. For now, we shall relegate the disc-to-square mapping equations as future work.

### 5.1 The $\frac{3}{2}$-Squircular Mapping (case $s=t^{3/2}$)

In the case $s=t^{3/2}$, the squircle equation reduces to $x^2 + y^2 - tx^2y^2 = t^2$ or equivalently, we have this quadratic polynomial equation in $t$:

$$\boldsymbol{t}^2 + x^2y^2\,\boldsymbol{t} - x^2 - y^2 = 0$$

We can solve for $t$ using the quadratic equation to get

$$t = \pm\frac{1}{2}\sqrt{x^4y^4+4x^2+4y^2} - \frac{1}{2}x^2y^2$$

Thus, the square-to-disc mapping equations are

$$\boldsymbol{u} = \frac{\boldsymbol{x}}{2\sqrt{\boldsymbol{x}^2+\boldsymbol{y}^2}}\left(\sqrt{\boldsymbol{x}^4\boldsymbol{y}^4+4\boldsymbol{x}^2+4\boldsymbol{y}^2} - \boldsymbol{x}^2\boldsymbol{y}^2\right) \qquad \boldsymbol{v} = \frac{\boldsymbol{y}}{2\sqrt{\boldsymbol{x}^2+\boldsymbol{y}^2}}\left(\sqrt{\boldsymbol{x}^4\boldsymbol{y}^4+4\boldsymbol{x}^2+4\boldsymbol{y}^2} - \boldsymbol{x}^2\boldsymbol{y}^2\right)$$

### 5.2 The ½-Squircular Mapping (case $s=t^{1/2}$)

In the case $s=t^{1/2}$, the squircle equation reduces to $x^2 + y^2 - \frac{x^2y^2}{t} = t^2$ or equivalently, we have this depressed cubic polynomial equation in t:

$$\boldsymbol{t}^3 - (x^2+y^2)\,\boldsymbol{t} + x^2y^2 = 0$$

We can solve for $t$ using the cubic formula to get

$$t = \frac{1}{3}\left(\frac{1}{2}\sqrt{729x^4y^4-4(3x^2+3y^2)^3} - \frac{27}{2}x^2y^2\right)^{\frac{1}{3}} + \frac{x^2+y^2}{\left(\frac{1}{2}\sqrt{729x^4y^4-4(3x^2+3y^2)^3} - \frac{27}{2}x^2y^2\right)^{\frac{1}{3}}}$$

Thus, the square-to-disc mapping equations are

$$u = \frac{\boldsymbol{x}}{\sqrt{\boldsymbol{x}^2+\boldsymbol{y}^2}}\left(\frac{1}{3}\left(\frac{1}{2}\sqrt{729x^4y^4-4(3x^2+3y^2)^3} - \frac{27}{2}x^2y^2\right)^{\frac{1}{3}} + \frac{x^2+y^2}{\left(\frac{1}{2}\sqrt{729x^4y^4-4(3x^2+3y^2)^3} - \frac{27}{2}x^2y^2\right)^{\frac{1}{3}}}\right)$$

$$v = \frac{\boldsymbol{y}}{\sqrt{\boldsymbol{x}^2+\boldsymbol{y}^2}}\left(\frac{1}{3}\left(\frac{1}{2}\sqrt{729x^4y^4-4(3x^2+3y^2)^3} - \frac{27}{2}x^2y^2\right)^{\frac{1}{3}} + \frac{x^2+y^2}{\left(\frac{1}{2}\sqrt{729x^4y^4-4(3x^2+3y^2)^3} - \frac{27}{2}x^2y^2\right)^{\frac{1}{3}}}\right)$$

### 5.3 The 4-Squircular Mapping (case s=$t^4$)

In the case $s=t^4$, the squircle equation reduces to $x^2 + y^2 - t^6x^2y^2 = t^2$ or equivalently, we have this soluble sextic polynomial equation in t:

$$x^2y^2\,\boldsymbol{t}^6 + \boldsymbol{t^2} - x^2 - y^2 = 0$$

We can solve for $t^2$ using the cubic formula to get

$$t^2 = \frac{\left(\frac{27}{2}x^6y^4 + \frac{27}{2}x^4y^6 + \frac{3}{2}\sqrt{12x^6y^6 + 81(x^6y^4 + x^4y^6)^2}\right)^{\frac{1}{3}}}{3x^2y^2} - \frac{1}{\left(\frac{27}{2}x^6y^4 + \frac{27}{2}x^4y^6 + \frac{3}{2}\sqrt{12x^6y^6 + 81(x^6y^4 + x^4y^6)^2}\right)^{\frac{1}{3}}}$$

Thus, the square-to-disc mapping equations are

$$u = \frac{\boldsymbol{x}}{\sqrt{\boldsymbol{x^2} + \boldsymbol{y^2}}}\sqrt{\frac{\left(\frac{27}{2}x^6y^4 + \frac{27}{2}x^4y^6 + \frac{3}{2}\sqrt{12x^6y^6 + 81(x^6y^4 + x^4y^6)^2}\right)^{\frac{1}{3}}}{3x^2y^2} - \frac{1}{\left(\frac{27}{2}x^6y^4 + \frac{27}{2}x^4y^6 + \frac{3}{2}\sqrt{12x^6y^6 + 81(x^6y^4 + x^4y^6)^2}\right)^{\frac{1}{3}}}}$$

$$v = \frac{\boldsymbol{y}}{\sqrt{\boldsymbol{x^2} + \boldsymbol{y^2}}}\sqrt{\frac{\left(\frac{27}{2}x^6y^4 + \frac{27}{2}x^4y^6 + \frac{3}{2}\sqrt{12x^6y^6 + 81(x^6y^4 + x^4y^6)^2}\right)^{\frac{1}{3}}}{3x^2y^2} - \frac{1}{\left(\frac{27}{2}x^6y^4 + \frac{27}{2}x^4y^6 + \frac{3}{2}\sqrt{12x^6y^6 + 81(x^6y^4 + x^4y^6)^2}\right)^{\frac{1}{3}}}}$$

### 5.4 Other Soluble Exponents

One can easily see from previous section that the mapping equations get very complicated quickly when we solve polynomial equations with higher exponents. Moreover, the Abel-Ruffini theorem states that we cannot get explicit solutions for general polynomials with degree greater than or equal to 5 (i.e. the quintic). In other words, we can only choose exponents for $s=t^n$ where the effective polynomial equation derived from the squircular continuum is linear, quadratic, cubic, or quartic.

Furthermore, all these complicated mappings probably produce qualitative results not much different from the FG-Squircular mapping or the 2-Squircular mapping. This means that we are getting diminishing returns and hardly any improvement for more complicated formulas.

Therefore, we shall not continue examining other exponents and will only list down the different possible exponents that produce soluble polynomials equations in *t*. The table on the next page shows a list of these soluble exponents along with their polynomial equations. All of these exponents except the last entry can be used to produce explicit square-to-disc mapping equations.

| **function $s=h(t)$** | **squircular continuum equation** | **effective type** | **equivalent polynomial equation** |
|---|---|---|---|
| $s=t$ | $x^2+y^2-x^2y^2=t^2$ | simple quadratic | $\boldsymbol{t}^2 = x^2+y^2-x^2y^2$ |
| $s=t^2$ | $x^2+y^2-t^2x^2y^2=t^2$ | simple quadratic | $\boldsymbol{t}^2(1+x^2y^2) = x^2+y^2$ |
| $s=t^{3/2}$ | $x^2+y^2-tx^2y^2=t^2$ | general quadratic | $\boldsymbol{t}^2+x^2y^2\,\boldsymbol{t}-x^2-y^2=0$ |
| $s=t^{1/2}$ | $x^2+y^2-\frac{x^2y^2}{t}=t^2$ | depressed cubic | $\boldsymbol{t}^3-(x^2+y^2)\,\boldsymbol{t}+x^2y^2=0$ |
| $s=t^{5/2}$ | $x^2+y^2-t^3x^2y^2=t^2$ | cubic | $x^2y^2\,\boldsymbol{t}^3+\boldsymbol{t}^2-x^2-y^2=0$ |
| $s=t^3$ | $x^2+y^2-t^4x^2y^2=t^2$ | biquadratic | $x^2y^2\,\boldsymbol{\tau}^2+\boldsymbol{\tau}-x^2-y^2=0$<br>where $\tau=t^2$ |
| $s=t^4$ | $x^2+y^2-t^6x^2y^2=t^2$ | depressed bi-cubic | $x^2y^2\,\boldsymbol{\tau}^3+\boldsymbol{\tau}-x^2-y^2=0$<br>where $\tau=t^2$ |
| $s=t^5$ | $x^2+y^2-t^8x^2y^2=t^2$ | depressed bi-quartic | $x^2y^2\,\boldsymbol{\tau}^4+\boldsymbol{\tau}-x^2-y^2=0$<br>where $\tau=t^2$ |
| $s=t^{5/4}$ | $x^2+y^2-x^2y^2\sqrt{t}=t^2$ | depressed quartic | $\boldsymbol{\tau}^4+x^2y^2\,\boldsymbol{\tau}-x^2-y^2=0$<br>where $\tau=\sqrt{t}$ |
| $s=t^{4/3}$ | $x^2+y^2-t^{\frac{2}{3}}x^2y^2=t^2$ | depressed cubic | $\boldsymbol{\tau}^3+x^2y^2\,\boldsymbol{\tau}-x^2-y^2=0$<br>where $\tau^3=t^2$ |
| $s=t^n$ | $x^2+y^2-t^{2n-2}\,x^2y^2=t^2$ | polynomial | $x^2y^2\,\boldsymbol{t}^{2n-2}+\boldsymbol{t}^2-x^2-y^2=0$ |

## 6.1 The Cornerific Tapered2 Mapping

Recall the squircle equation:

$$x^2 + y^2 - \frac{s^2}{t^2} x^2 y^2 = t^2$$

Consider the rampant function $s = t\sqrt{2 - t^2}$. We can use this relationship for s and t. Substituting, we get

$$x^2 + y^2 - \frac{\left(t\sqrt{2 - t^2}\right)^2}{t^2} x^2 y^2 = t^2 \qquad \Rightarrow \qquad x^2 + y^2 - 2x^2y^2 + x^2y^2t^2 = t^2$$

We can then solve for t in terms of x and y

$$t = \sqrt{\frac{x^2 + y^2 - 2x^2y^2}{1 - x^2y^2}}$$

We can substitute this *t* value to the *radial mapping linear parametric equations* to get these mapping equations

$$\boldsymbol{u = x\sqrt{\frac{x^2+y^2-2x^2y^2}{(x^2+y^2)(1-x^2y^2)}}} \qquad \boldsymbol{v = y\sqrt{\frac{x^2+y^2-2x^2y^2}{(x^2+y^2)(1-x^2y^2)}}}$$

To invert this mapping, we start by squaring the equation for u

$$u^2 = \frac{(x^2 + y^2 - 2x^2y^2)x^2}{(1 - x^2y^2)(x^2 + y^2)} \qquad \Rightarrow \qquad u^2(1 - x^2y^2)(x^2 + y^2) = (x^2 + y^2 - 2x^2y^2)x^2 \qquad \Rightarrow$$

$$u^2(x^2 + y^2 - x^4y^2 - x^2y^4) = x^4 + x^2y^2 - 2x^4y^2 \qquad \Rightarrow$$

$$u^2x^2 + u^2y^2 - u^2x^4y^2 - u^2x^2y^4 = x^4 + x^2y^2 - 2x^4y^2$$

Substitute $y = \frac{v}{u}x$ to get

$$u^2x^2 + v^2x^2 - v^2x^6 - \frac{v^4}{u^2}x^6 = x^4 + \frac{v^2}{u^2}x^4 - 2\frac{v^2}{u^2}x^6$$

Multiply the equation by $\frac{u^2}{x^2}$

$$u^4 + u^2v^2 - u^2v^2x^4 - v^4x^4 = u^2x^2 + v^2x^2 - 2v^2x^4$$

Collect all terms as a polynomial in x

$$0 = (v^4 + u^2v^2 - 2v^2)x^4 + (u^2 + v^2)x^2 - u^2(u^2 + v^2)$$

This is a biquadratic equation in x with coefficients

$$a = v^4 + u^2v^2 - 2v^2 \qquad b = u^2 + v^2 \qquad c = -u^2(u^2 + v^2)$$

We can then solve for $x^2$

$$x^2 = \frac{-u^2 - v^2 \pm \sqrt{(u^2 + v^2)^2 + 4u^2(u^2 + v^2)(v^4 + u^2v^2 - 2v^2)}}{2\,(v^4 + u^2v^2 - 2v^2)}$$

$$\Rightarrow \qquad x^2 = \frac{-u^2 - v^2 \pm \sqrt{(u^2 + v^2)(u^2 + v^2 - 4u^2v^2(u^2 + v^2 - 2))}}{2v^2(u^2 + v^2 - 2)}$$

We can then simplify to get a quadrant-aware equation for x as

$$x = \frac{sgn(uv)}{v}\sqrt{\frac{-u^2 - v^2 \pm \sqrt{(u^2+v^2)(u^2+v^2-4u^2v^2(u^2+v^2-2))}}{2(u^2+v^2-2)}}$$

Using $y = \frac{v}{u}x$, we can solve for y as

$$y = \frac{sgn(uv)}{u}\sqrt{\frac{-u^2 - v^2 \pm \sqrt{(u^2+v^2)(u^2+v^2-4u^2v^2(u^2+v^2-2))}}{2(u^2+v^2-2)}}$$

These mapping equations can be made compact in vector notation. Here are the mappings equations rewritten in vector format with degenerate cases at the corners and the center of square included.

Disc to square mapping:

$$\begin{bmatrix} x \\ y \end{bmatrix} = \begin{cases} sgn(uv)\sqrt{\dfrac{-u^2 - v^2 + \sqrt{(u^2+v^2)[u^2+v^2+4u^2v^2(u^2+v^2-2)]}}{2(u^2+v^2-2)}}\begin{bmatrix} \dfrac{1}{v} \\ \dfrac{1}{u} \end{bmatrix} & if\ u \neq 0,\quad v \neq 0 \\ \begin{bmatrix} u \\ v \end{bmatrix} & otherwise \end{cases}$$

Square to disc mapping:

$$\begin{bmatrix} u \\ v \end{bmatrix} = \begin{cases} \sqrt{\dfrac{x^2+y^2-2x^2y^2}{(x^2+y^2)(1-x^2y^2)}}\begin{bmatrix} x \\ y \end{bmatrix} & if\ (x,y) \neq (0,0),\quad (x,y) \neq (\pm1,\pm1) \\ \begin{bmatrix} sgn(x)\dfrac{\sqrt{2}}{2} \\ sgn(y)\dfrac{\sqrt{2}}{2} \end{bmatrix} & if\ (x,y) = (\pm1,\pm1) \\ \begin{bmatrix} 0 \\ 0 \end{bmatrix} & if\ (x,y) = (0,0) \end{cases}$$

We know that this mapping converts circular contours inside the circular disc to squircles inside the square. However, it is important to see what sort of equation comes out of circles after the mapping given the relationship $s = t\sqrt{2-t^2}$ for the mapping.

$$u^2 + v^2 = x^2\frac{x^2+y^2-2x^2y^2}{(x^2+y^2)(1-x^2y^2)} + y^2\frac{x^2+y^2-2x^2y^2}{(x^2+y^2)(1-x^2y^2)} = (x^2+y^2)\frac{x^2+y^2-2x^2y^2}{(x^2+y^2)(1-x^2y^2)}$$

$$\Rightarrow \qquad u^2 + v^2 = \frac{x^2+y^2-2x^2y^2}{1-x^2y^2}$$

Thus, from the nonlinear relationship of $s = t\sqrt{2-t^2}$, we get an equation relating *(u,v)* with *(x,y)*. We will encounter this equation again later, so it is handy to give it a name. We shall call this as the *cornerific squircular continuum.*

## 6.2 The Tapered4 Squircular Mapping

Recall the squircle equation:

$$x^2 + y^2 - \frac{s^2}{t^2} x^2 y^2 = t^2$$

Consider the rampant function $s = t\sqrt{\frac{3}{2} - \frac{1}{2}\,t^4}$ . We can use this relationship for s and t. Substituting, we get

$$x^2 + y^2 - \frac{\left(t\sqrt{\frac{3}{2} - \frac{1}{2}t^4}\right)^2}{t^2} x^2 y^2 = t^2$$

This will simplify into a biquadratic equation in t

$$\tfrac{1}{2}x^2y^2t^4 - t^2 + x^2 + y^2 - \tfrac{3}{2}x^2y^2 = 0$$

We can then solve for $t^2$ using the quadratic equation

$$t^2 = \frac{1 \pm \sqrt{1 - 2x^2y^2(x^2 + y^2 - \frac{3}{2}x^2y^2)}}{x^2y^2} = \frac{1 \pm \sqrt{1 - 2x^4y^2 - 2x^2y^4 + 3x^4y^4}}{x^2y^2}$$

$$\Rightarrow \qquad t = \sqrt{\frac{1 - \sqrt{1 - 2x^4y^2 - 2x^2y^4 + 3x^4y^4}}{x^2y^2}}$$

We can then substitute this *t* value to the *radial mapping linear parametric equations* to get these mapping equations

$$\boldsymbol{u = \frac{sgn(xy)}{y\sqrt{x^2 + y^2}}\sqrt{1 - \sqrt{1 - 2x^4y^2 - 2x^2y^4 + 3x^4y^4}}}$$

$$\boldsymbol{v = \frac{sgn(xy)}{x\sqrt{x^2 + y^2}}\sqrt{1 - \sqrt{1 - 2x^4y^2 - 2x^2y^4 + 3x^4y^4}}}$$

To invert this mapping, we start by squaring the equation for u

$$u^2 = \frac{1 - \sqrt{1 - 2x^4y^2 - 2x^2y^4 + 3x^4y^4}}{y^2(x^2 + y^2)}$$

$$u^2y^2(x^2 + y^2) = 1 - \sqrt{1 - 2x^4y^2 - 2x^2y^4 + 3x^4y^4}$$

$$u^2y^2(x^2 + y^2) - 1 = -\sqrt{1 - 2x^4y^2 - 2x^2y^4 + 3x^4y^4}$$

$$(u^2y^2(x^2 + y^2) - 1)^2 = 1 - 2x^4y^2 - 2x^2y^4 + 3x^4y^4$$

$$u^4y^4(x^2 + y^2)^2 - 2u^2y^2(x^2 + y^2) + 1 = 1 - 2x^4y^2 - 2x^2y^4 + 3x^4y^4$$

$$u^4y^4(x^4 + 2x^2y^2 + y^4) - 2u^2y^2(x^2 + y^2) = -2x^4y^2 - 2x^2y^4 + 3x^4y^4$$

$$u^4x^4y^4 + 2u^4x^2y^6 + u^4y^8 - 2u^2x^2y^2 - 2u^2y^4 = -2x^4y^2 - 2x^2y^4 + 3x^4y^4$$

Substitute $y = \frac{v}{u}x$

$$u^4x^4\frac{v^4}{u^4}x^4 + 2u^4x^2\frac{v^6}{u^6}x^6 + u^4\frac{v^8}{u^8}x^8 - 2u^2x^2\frac{v^2}{u^2}x^2 - 2u^2\frac{v^4}{u^4}x^4 = -2x^4\frac{v^2}{u^2}x^2 - 2x^2\frac{v^4}{u^4}x^4 + 3x^4\frac{v^4}{u^4}x^4$$

Multiply the equation by $\frac{u^4}{v^2x^4}$

$$u^4v^2x^4 + 2u^2v^4x^4 + v^6x^4 - 2u^4 - 2u^2v^2 = -2u^2x^2 - 2v^2x^2 + 3v^2x^4$$

Collect terms as a polynomial in x

$$0 = v^2(u^4 + 2u^2v^2 + v^4 - 3)x^4 + 2(u^2 + v^2)x^2 - 2u^2(u^2 + v^2)$$

This is a biquadratic equation in x with coefficients

$$a = v^2(u^4 + 2u^2v^2 + v^4 - 3) \qquad b = 2(u^2 + v^2\ ) \qquad c = -2u^2(u^2 + v^2)$$

We can then solve for $x^2$

$$x^2 = \frac{-2(u^2 + v^2) \pm \sqrt{4(u^2 + v^2)^2 + 8u^2v^2(u^2 + v^2)(u^4 + 2u^2v^2 + v^4 - 3)]}}{2v^2(u^4 + 2u^2v^2 + v^4 - 3)}$$

$$\Rightarrow \qquad x^2 = \frac{-u^2 - v^2 \pm \sqrt{(u^2 + v^2)[u^2 + v^2 + 2u^2v^2(u^4 + 2u^2v^2 + v^4 - 3)]}}{v^2(u^4 + 2u^2v^2 + v^4 - 3)}$$

We can then simplify to get a quadrant-aware equation for x as

$$\boldsymbol{x = \frac{sgn(uv)}{v}\sqrt{\frac{u^2 + v^2 - \sqrt{(u^2 + v^2)[u^2 + v^2 - 2u^2v^2(3 - u^4 - 2u^2v^2 - v^4)]}}{3 - u^4 - 2u^2v^2 - v^4}}}$$

Using $y = \frac{v}{u}x$, we can solve for y as

$$\boldsymbol{y = \frac{sgn(uv)}{u}\sqrt{\frac{u^2 + v^2 - \sqrt{(u^2 + v^2)[u^2 + v^2 - 2u^2v^2(3 - u^4 - 2u^2v^2 - v^4)]}}{3 - u^4 - 2u^2v^2 - v^4}}}$$

These mapping equations can be made compact in vector notation. Here are the mappings rewritten in vector format.

Disc to square mapping:

$$\boldsymbol{\begin{bmatrix} x \\ y \end{bmatrix} = sgn(uv)\sqrt{\frac{u^2 + v^2 - \sqrt{(u^2 + v^2)[u^2 + v^2 - 2u^2v^2(3 - u^4 - 2u^2v^2 - v^4)]}}{3 - u^4 - 2u^2v^2 - v^4}}\begin{bmatrix} \frac{1}{v} \\ \frac{1}{u} \end{bmatrix}}$$

Square to disc mapping:

$$\boldsymbol{\begin{bmatrix} u \\ v \end{bmatrix} = \frac{sgn(xy)}{\sqrt{x^2 + y^2}}\sqrt{1 - \sqrt{1 - 2x^4y^2 - 2x^2y^4 + 3x^4y^4}}\begin{bmatrix} \frac{1}{y} \\ \frac{1}{x} \end{bmatrix}}$$

## 7.1 A Cursory Review of the Elliptical Grid Mapping

In this section, we shall come up with variations of the Elliptical Grid mapping [Nowell 2005] [Fong 2014]. Before we do this, let us do a quick cursory review of the Elliptical Grid mapping. There are two main ideas in the derivation of the mapping.

The 1st idea is to map vertical lines segments inside the square to vertically-oriented elliptical arcs inside the circular disc. This is illustrated in Figure 7 where a vertical line and its corresponding elliptical arc are shown in red.

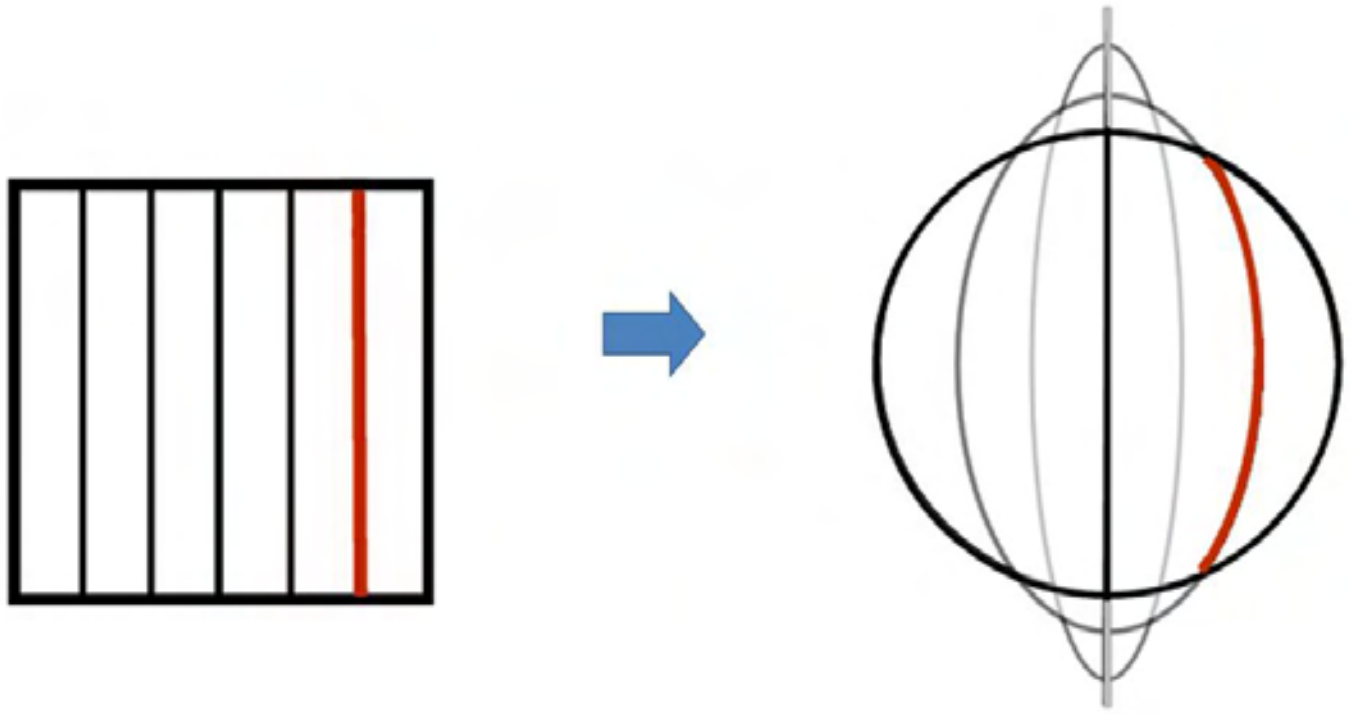

Figure 7: Vertical constraint of the Elliptical Grid mapping

This mapping relationship can be summarized by an equation which we shall call the *vertical constraint* of the mapping.

$$1 = \frac{u^2}{x^2} + \frac{v^2}{2 - x^2}$$

Essentially, for each vertical line segment of constant *x* inside the square, we have a corresponding equation of an ellipse inside the circle. The semi-major and semi-minor axial lengths of the ellipse will vary with *x*. The left and right vertex tips of the ellipse are set equal to *x*. The top and bottom vertex tips of the ellipse vary as $\sqrt{2 - x^2}$.

The 2nd main idea of the mapping is to map horizontal lines inside the square to sideways-oriented elliptical arcs inside the circular disc. This is illustrated in Figure 8 where a horizontal line and its corresponding elliptical arc are shown in red.

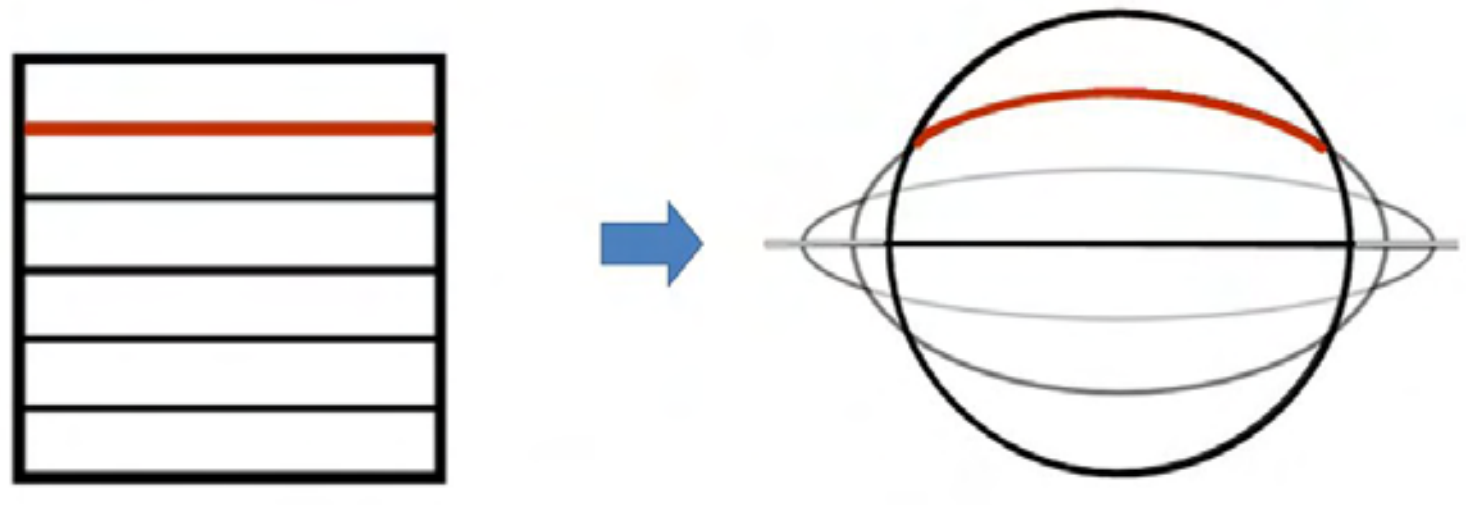

Figure 8: Horizontal constraint of the Elliptical Grid mapping

This mapping relationship can also be summarized by an equation which we shall call the *horizontal constraint* of the mapping.

$$1 = \frac{u^2}{2 - y^2} + \frac{v^2}{y^2}$$

Essentially, for each horizontal line segment of constant $y$ in the square, we have a corresponding equation of an ellipse inside the circle. The semi-major and semi-minor axis lengths of the ellipse vary with $y$. The top and bottom vertex tips of the ellipse are set equal to $y$. The left and right vertex tips of the ellipse vary as $\sqrt{2-y^2}$ .

Basically, we are mapping the grid of perpendicular vertical and horizontal lines in the square to a grid of elliptical arcs inside the circular disc. This is illustrated in Figure 9. Note that the ellipses get more and more eccentric as x or y approach zero. Also, the ellipses get more circular as x or y approaches ±1.

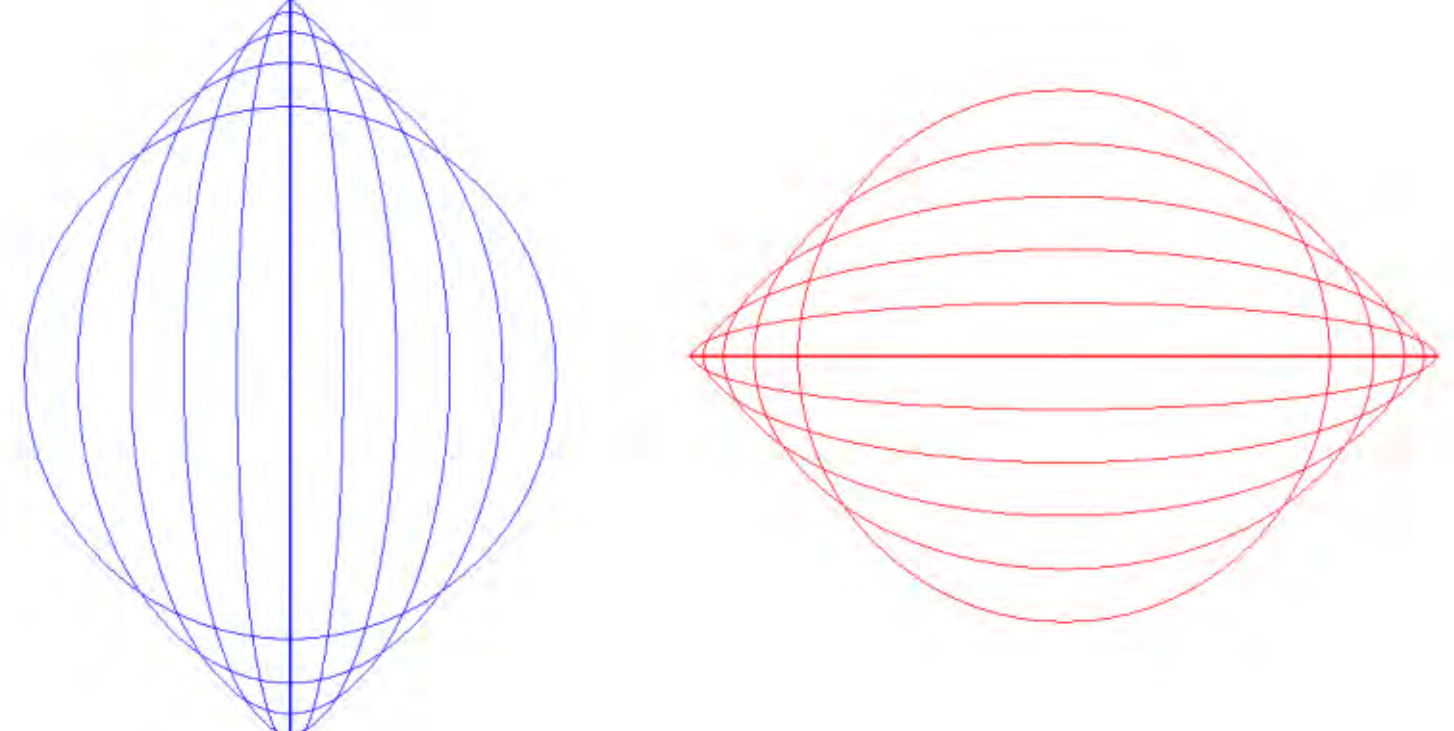

Figure 9: Vertical and horizontal grids from the constraints of the mapping

We can then superimpose the vertically-oriented and horizontally-oriented elliptical arcs inside the circular disc together to create a curvilinear grid of elliptical arcs. This is shown in Figure 10. This figure is a visual overview of the how the Elliptical Grid mapping works

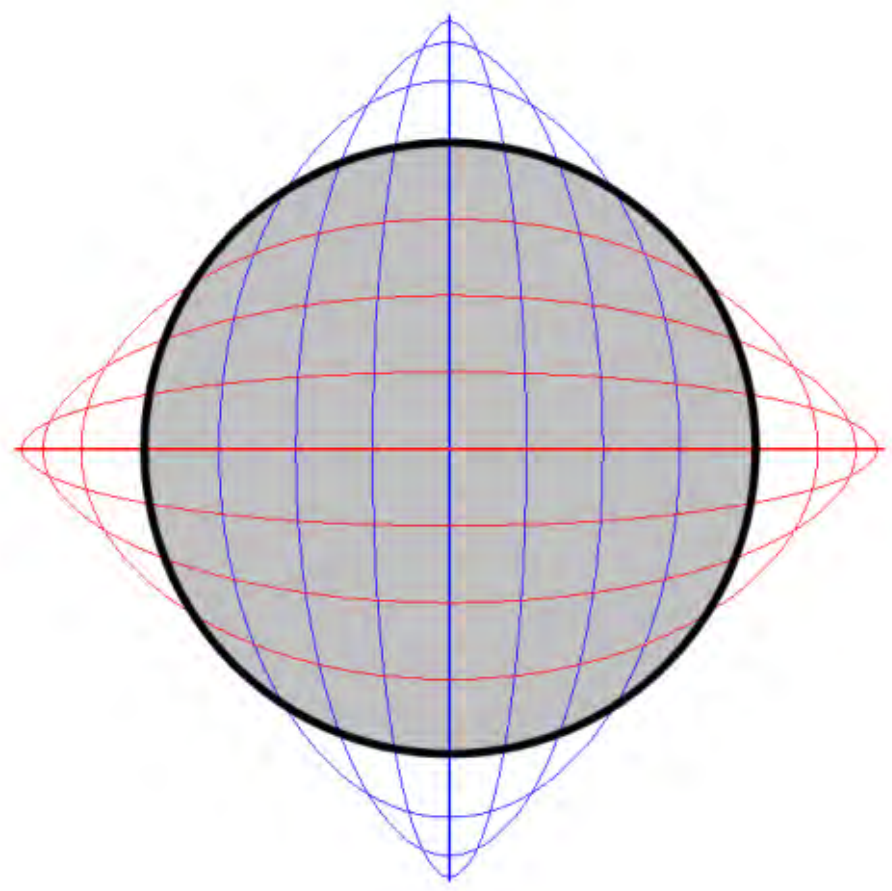

Figure 10: Superimposed vertical and horizontal grids

Mathematically, we want to mix the vertical constraint equation with the horizontal constraint equation to get algebraic expressions for u and v in terms of x and y.

We can do this by starting with the vertical constraint equation and do some manipulations to isolate $u^2$.

$$1 = \frac{u^2}{x^2} + \frac{v^2}{2-x^2} \qquad \Rightarrow \qquad 1 - \frac{v^2}{2-x^2} = \frac{u^2}{x^2} \qquad \Rightarrow \qquad u^2 = x^2\left(1 - \frac{v^2}{2-x^2}\right)$$

We can then plug this $u^2$ value into the horizontal constraint equation

$$1 = \frac{u^2}{2-y^2} + \frac{v^2}{y^2} \qquad \Rightarrow \qquad 1 = \frac{x^2\left(1 - \frac{v^2}{2-x^2}\right)}{2-y^2} + \frac{v^2}{y^2}$$

Multiply both sides of the equation by $(2-x^2)(2-y^2)y^2$ to remove the fractions and get

$$(2-x^2)(2-y^2)y^2 = x^2y^2(2-x^2-v^2) + v^2(2-x^2)(2-y^2)$$

After which, we can isolate $v^2$ into one side of the equation

$$(2-x^2)(2-y^2)y^2 - x^2y^2(2-x^2) = -x^2y^2v^2 + (2-x^2)(2-y^2)v^2 \qquad \Rightarrow$$

$$v^2\big((2-x^2)(2-y^2) - x^2y^2\big) = (2-x^2)y^2(2-y^2-x^2) \qquad \Rightarrow$$

$$v^2 = y^2\frac{(2-x^2)(2-y^2-x^2)}{(2-x^2)(2-y^2)-x^2y^2} \qquad \Rightarrow \qquad v^2 = y^2\frac{(2-x^2)(2-y^2-x^2)}{4-2x^2-2y^2} = y^2\frac{2-x^2}{2} \qquad \Rightarrow$$

$$v = y\sqrt{\frac{2-x^2}{2}} \qquad \Rightarrow \qquad \boldsymbol{v = y\sqrt{1-\frac{x^2}{2}}}$$

Similarly in a symmetric fashion, we can solve for u as

$$\boldsymbol{u = x\sqrt{1-\frac{y^2}{2}}}$$

This completes our derivation of the Elliptical Grid mapping. We will not derive the inverse equations here since this has already been done in the paper "*Analytical Methods for Squaring the Disc*" [Fong 2014]

## 7.2 A Simple Squelched Version of the Elliptical Grid mapping

Using the same idea of mapping vertical and horizontal lines to elliptical arcs, we can come up with a simpler version of the Elliptical grid mapping. We shall name this variant as the *Squelched Grid open mapping*.

Recall the vertical constraint equation given in the previous section. We can actually make it simpler by fixing the top and bottom vertex tip locations of the ellipse at coordinates (0, ±1) to get this constraint equation.

$$1 = \frac{u^2}{x^2} + \frac{v^2}{1}$$

The effect of this simplified vertical constraint is shown in Figure 11. All the ellipses inside the circular disc effectively have identical top and bottom vertex tips.

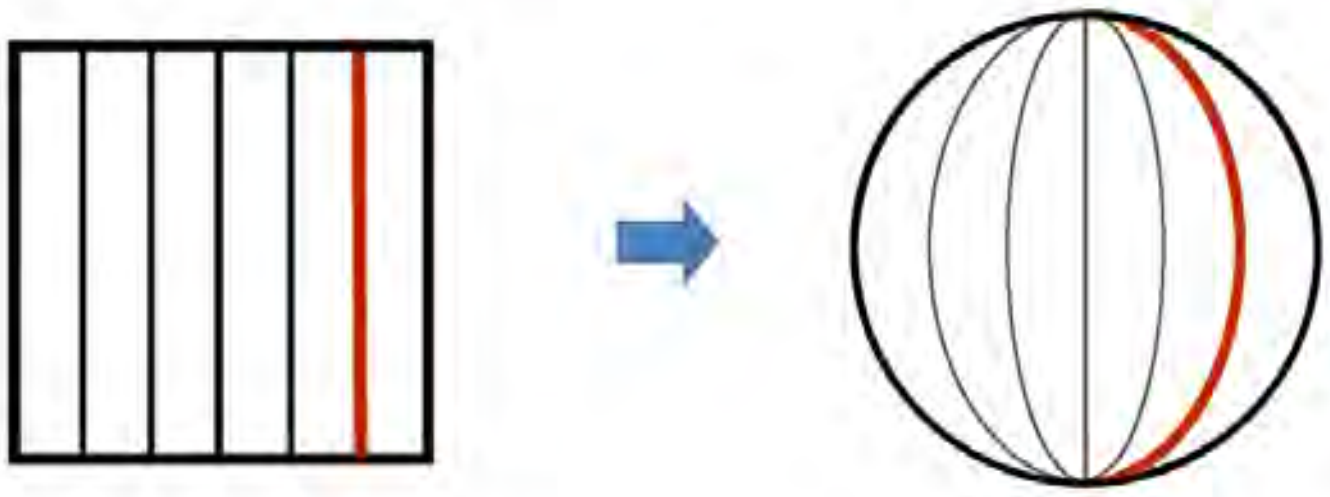

Figure 11: Vertical constraint of the Squelched Grid open mapping

Similarly, we can do the same with the horizontal constraint equation from the previous section. We can fix the left and right vertex tip locations of the ellipse at coordinates (±1,0) to get this constraint equation.

$$1 = \frac{u^2}{1} + \frac{v^2}{y^2}$$

The effect of this simplified horizontal constraint is shown in Figure 12. All the ellipses inside the circular disc effectively have identical left and right vertex tips.

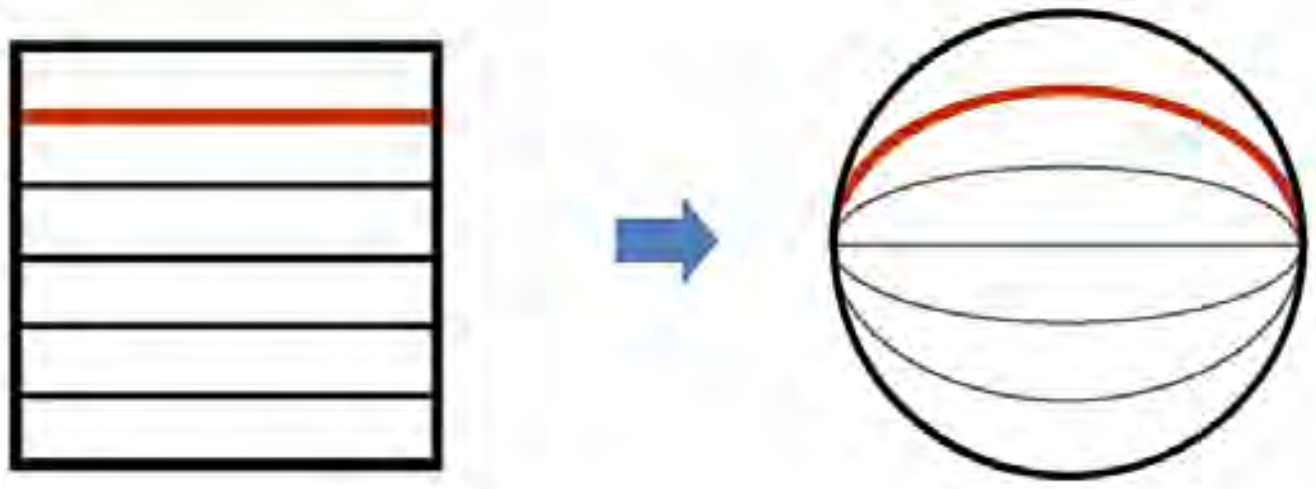

Figure 12: Horizontal constraint of the Squelched Grid open mapping

From these simpler constraint equations we can easily derive the equations for a disc-to-square mapping

$$1 = \frac{u^2}{x^2} + \frac{v^2}{1} \quad \Rightarrow \quad \frac{u^2}{x^2} = 1 - v^2 \quad \Rightarrow \quad \frac{x^2}{u^2} = \frac{1}{1 - v^2} \quad \Rightarrow x^2 = \frac{u^2}{1 - v^2} \quad \Rightarrow \qquad \boldsymbol{x} = \frac{\boldsymbol{u}}{\sqrt{\boldsymbol{1} - \boldsymbol{v}^2}}$$

Likewise,

$$1 = \frac{u^2}{1} + \frac{v^2}{y^2} \quad \Rightarrow \quad \frac{v^2}{y^2} = 1 - u^2 \quad \Rightarrow \quad \frac{y^2}{v^2} = \frac{1}{1 - u^2} \quad \Rightarrow y^2 = \frac{v^2}{1 - u^2} \quad \Rightarrow \qquad \boldsymbol{y} = \frac{\boldsymbol{v}}{\sqrt{\boldsymbol{1} - \boldsymbol{u}^2}}$$

To invert this mapping, we start with the vertical constraint equation and isolate $u^2$

$$1 = \frac{u^2}{x^2} + \frac{v^2}{1} \quad \Rightarrow \quad \frac{u^2}{x^2} = 1 - v^2 \quad \Rightarrow \quad u^2 = x^2(1 - v^2)$$

And plug it into the horizontal constraint equation

$$1 = x^2(1 - v^2) + \frac{v^2}{y^2}$$

We can then isolate for v to get

$$1 = x^2 + \left(-x^2 + \frac{1}{y^2}\right) v^2 \quad \Rightarrow \quad 1 - x^2 = \left(\frac{1}{y^2} - x^2\right) v^2 \quad \Rightarrow \quad v^2 = \frac{y^2(1 - x^2)}{(1 - x^2y^2)} \quad \Rightarrow$$

$$\boldsymbol{v = y\sqrt{\frac{1 - x^2}{1 - x^2y^2}}}$$

Similarly in a symmetric fashion, we can solve for u as

$$\boldsymbol{u = x\sqrt{\frac{1 - y^2}{1 - x^2y^2}}}$$

## 7.3 Singular points in the Squelched Grid Open Mapping

By looking at the forward and inverse equations of the Squelched Grid mapping, we can see that it is possibly the simplest disc-square mapping there is when compared to other mappings. Unfortunately, the Squelched Grid open mapping has a serious drawback. Strictly speaking, it is not a bijective mapping. There exist 4 singular points on the circular disc and 4 singular points on the square for this mapping. These singular points fall on the coordinates (±1,0) and (0, ±1) for the circular disc, and they fall on the corners of the square at (±1, ±1). These singular points will produce a division by zero in the mapping equations.

In essence, the disc-to-square equations map a perimeter arc segment in a quadrant of the circular disc to a single corner point on the square. Likewise, the square-to-disc equations map a whole line segment along the perimeter of the square to vertical or horizontal extremum of the circular disc. To remedy these unwanted singularities, we restrict the mapping be an open-type mapping. In doing so, we exclude the bounding rims of the circular disc and the square from the mapping.

## 7.4 Squircularity of the Squelched Grid Open Mapping

We already know from the paper "*Analytical Methods for Squaring the Disc*" [Fong 2014] that the Elliptical Grid mapping converts circular contours inside the disc to Fernandez-Guasti squircles inside the square. We now pose the same question for the Squelched Grid open mapping.

It turns out that the Squelched Grid open mapping also does the same. We shall show this here. Start with the expression $u^2+v^2$ and substitute using the square-to-disc mapping equations for the mapping.

$$u^2 + v^2 \;=\; x^2\frac{1 - y^2}{1 - x^2y^2} + y^2\frac{1 - x^2}{1 - x^2y^2} = \frac{x^2(1 - y^2) + y^2(1 - x^2)}{1 - x^2y^2} \;=\; \frac{x^2 + y^2 - 2x^2y^2}{1 - x^2y^2}$$

We have already encountered this equation previously in our derivation of the Cornerific Tapered2 mapping in Section 6.1. This equation is the *cornerific squircular continuum*. This means that just like the Cornerific Tapered2 mapping, the Squelched Grid open mapping converts circular contours inside the disc to Fernandez-Guasti squircles inside the square.

## 7.5 Vertical Squelch Open Mapping

The simplified vertical constraint and horizontal constraint equations from the Squelched Grid open mapping are actually independent of each other and hence separable. It is quite possible to come up with mappings using the constraint equations independently.

For example, the simplified horizontal constraint equation from the Squelched Grid open mapping

$$1 = \frac{u^2}{1} + \frac{v^2}{y^2}$$

can be used in conjunction with the simple relation $\boldsymbol{u = x}$ to produce another square-to-disc mapping, which we shall call the *Vertical Squelch open mapping*.

$$1 = \frac{u^2}{1} + \frac{v^2}{y^2} \quad \Rightarrow \quad \boldsymbol{y = \frac{v}{\sqrt{1-u^2}}}$$

It is quite easy to get the inverse equation.

$$y = \frac{v}{\sqrt{1-u^2}} \quad \Rightarrow \quad v = y\sqrt{1-u^2} \quad \Rightarrow \quad \boldsymbol{v = y\sqrt{1-x^2}}$$

Furthermore, we can show that this mapping also maps circular contours inside the disc to Fernandez-Guasti squircles inside the square. Using the equations for $u$ and $v$ given above, we show that

$$u^2 + v^2 \;=\; x^2 + y^2(1-x^2) \;=\; x^2 + y^2 - x^2y^2$$

This is just the equation for the *linear squircular continuum*.

## 7.6 Horizontal Squelch Open Mapping

Similarly, the simplified vertical constraint equation

$$1 = \frac{u^2}{x^2} + \frac{v^2}{1}$$

can be used in conjunction with the simple relation $\boldsymbol{v = y}$ to produce another square-to-disc mapping, which we shall call the *Horizontal Squelch open mapping*.

$$1 = \frac{u^2}{x^2} + \frac{v^2}{1} \quad \Rightarrow \quad \boldsymbol{x = \frac{u}{\sqrt{1-v^2}}}$$

It is also easy to get the inverse equation.

$$x = \frac{u}{\sqrt{1-v^2}} \quad \Rightarrow \quad u = x\sqrt{1-v^2} \quad \Rightarrow \quad \boldsymbol{u = x\sqrt{1-y^2}}$$

Likewise, it is easy to show that this mapping produces the *linear squircular continuum*.

$$u^2 + v^2 \;=\; x^2(1-y^2) + y^2 \;=\; x^2 + y^2 - x^2y^2$$

## 7.7 Generalized Elliptical Grid mapping

We can further generalize the Elliptical Grid mapping by introducing suitable functions *Q(x)* and *P(y)* into the vertical and horizontal constraint equations discussed earlier in Section 7.1 of this paper. Specifically, we have this generalized vertical constraint equation

$$1 = \frac{u^2}{x^2} + \frac{v^2}{Q^2}$$

and this generalized horizontal constraint equation.

$$1 = \frac{u^2}{P^2} + \frac{v^2}{y^2}$$

We proceed to do some manipulations on the generalized vertical constraint equation and isolate $u^2$,

$$1 = \frac{u^2}{x^2} + \frac{v^2}{Q^2} \quad \Rightarrow \quad 1 - \frac{v^2}{Q^2} = \frac{u^2}{x^2} \quad \Rightarrow \quad u^2 = x^2\left(1 - \frac{v^2}{Q^2}\right)$$

We can then plug this $u^2$ value into the generalized horizontal constraint equation

$$1 = \frac{u^2}{P^2} + \frac{v^2}{y^2} \quad \Rightarrow \quad 1 = \frac{x^2\left(1 - \frac{v^2}{Q^2}\right)}{P^2} + \frac{v^2}{y^2}$$

Multiply both sides of the equation by $P^2Q^2y^2$ to remove the fractions and get

$$P^2Q^2y^2 = x^2y^2(Q^2 - v^2) + v^2P^2Q^2$$

After which, we can isolate $v^2$ into one side of the equation

$$P^2Q^2y^2 = x^2y^2(Q^2 - v^2) + v^2P^2Q^2 = x^2y^2Q^2 - x^2y^2v^2 + v^2P^2Q^2 \quad \Rightarrow$$

$$v^2P^2Q^2 - x^2y^2v^2 = P^2Q^2y^2 - x^2y^2Q^2 \quad \Rightarrow \quad v^2 = y^2\frac{P^2Q^2 - x^2Q^2}{P^2Q^2 - x^2y^2} = y^2\frac{x^2Q^2 - P^2Q^2}{x^2y^2 - P^2Q^2} \quad \Rightarrow$$

$$\boldsymbol{v = yQ\sqrt{\frac{x^2 - P^2}{x^2y^2 - P^2Q^2}}}$$

Similarly in a symmetric fashion, we can derive an equation for *u* in terms *x, P,* and Q

$$\boldsymbol{u = xP\sqrt{\frac{y^2 - Q^2}{x^2y^2 - P^2Q^2}}}$$

## 7.8 Blended E-Grid mapping

In and of itself, the Squelched Grid open mapping is not interesting, but it is possible to generalize it to produce an interesting family of mappings. Using the equations for *u* and *v* in the previous section, we can blend the Squelched Grid open mapping with the Elliptical Grid mapping to come up with a mapping that generalizes both. Consider setting functions *P* and *Q* as follows

$$Q(x) = \sqrt{\beta + 1 - \beta x^2} \qquad\qquad P(y) = \sqrt{\beta + 1 - \beta y^2}$$

This gives us these vertical and horizontal constraint equations

$$1 = \frac{u^2}{x^2} + \frac{v^2}{\beta + 1 - \beta x^2} \qquad and \qquad 1 = \frac{u^2}{\beta + 1 - \beta y^2} + \frac{v^2}{y^2}$$

Here $\beta$ is a blend parameter that has a value between 0 and 1. When $\beta$ is zero, the mapping becomes the Squelched Grid open mapping. When $\beta$ is one, the mapping becomes the Elliptical Grid mapping. In between, it is a blended average of the two mappings.

It is very easy to derive the square-to-disc mapping equations by using the equations provided in the last part of the previous section and substituting $\sqrt{\beta + 1 - \beta y^2}$ and $\sqrt{\beta + 1 - \beta x^2}$ for *P* and *Q*, respectively to get

$$\boldsymbol{u = x\sqrt{\frac{y^2(\beta + 1 - \beta y^2) - (\beta + 1 - \beta x^2)(\beta + 1 - \beta y^2)}{x^2y^2 - (\beta + 1 - \beta x^2)(\beta + 1 - \beta y^2)}}}$$

$$\boldsymbol{v = y\sqrt{\frac{x^2(\beta + 1 - \beta x^2) - (\beta + 1 - \beta x^2)(\beta + 1 - \beta y^2)}{x^2y^2 - (\beta + 1 - \beta x^2)(\beta + 1 - \beta y^2)}}}$$

To invert this mapping, we start with the vertical constraint equation and isolate u

$$1 = \frac{u^2}{x^2} + \frac{v^2}{\beta + 1 - \beta x^2} \quad \Rightarrow \quad \frac{u^2}{x^2} = 1 - \frac{v^2}{\beta + 1 - \beta x^2} \quad \Rightarrow \quad u^2 = x^2 - \frac{x^2v^2}{\beta + 1 - \beta x^2} \quad \Rightarrow$$

$$u^2 = \frac{(\beta + 1)x^2 - \beta x^4 - x^2v^2}{\beta + 1 - \beta x^2} \quad \Rightarrow \quad (\beta + 1)u^2 - \beta u^2x^2 = (\beta + 1)x^2 - \beta x^4 - x^2v^2 \quad \Rightarrow$$

$$\beta x^4 - (\beta + 1 + \beta u^2 - v^2)x^2 + u^2(\beta + 1) = 0$$

This is a biquadratic equation in *x* with coefficients

$$a = \beta \qquad\qquad b = -(\beta + 1 + \beta u^2 - v^2) \qquad\qquad c = u^2(\beta + 1)$$

We can then solve for *x* as

$$\boldsymbol{x = \frac{sgn(u)}{\sqrt{2\beta}}\sqrt{\beta + 1 + \beta u^2 - v^2 - \sqrt{(\beta + 1 + \beta u^2 - v^2)^2 - 4\beta(\beta + 1)u^2}}}$$

Similarly in a symmetric fashion, we can solve for *y* as

$$\boldsymbol{y = \frac{sgn(v)}{\sqrt{2\beta}}\sqrt{\beta + 1 - u^2 + \beta v^2 - \sqrt{(\beta + 1 - u^2 + \beta v^2)^2 - 4\beta(\beta + 1)v^2}}}$$

Figure 13 shows the Blended E-Grid mapping at various values for the blend parameter. Note that the mapping equations are restricted to having the parameter $\neq 0$ , since there is an unwanted division by zero for that case. One can simply revert to using the much simpler Squelched Grid open mapping equations for that case.

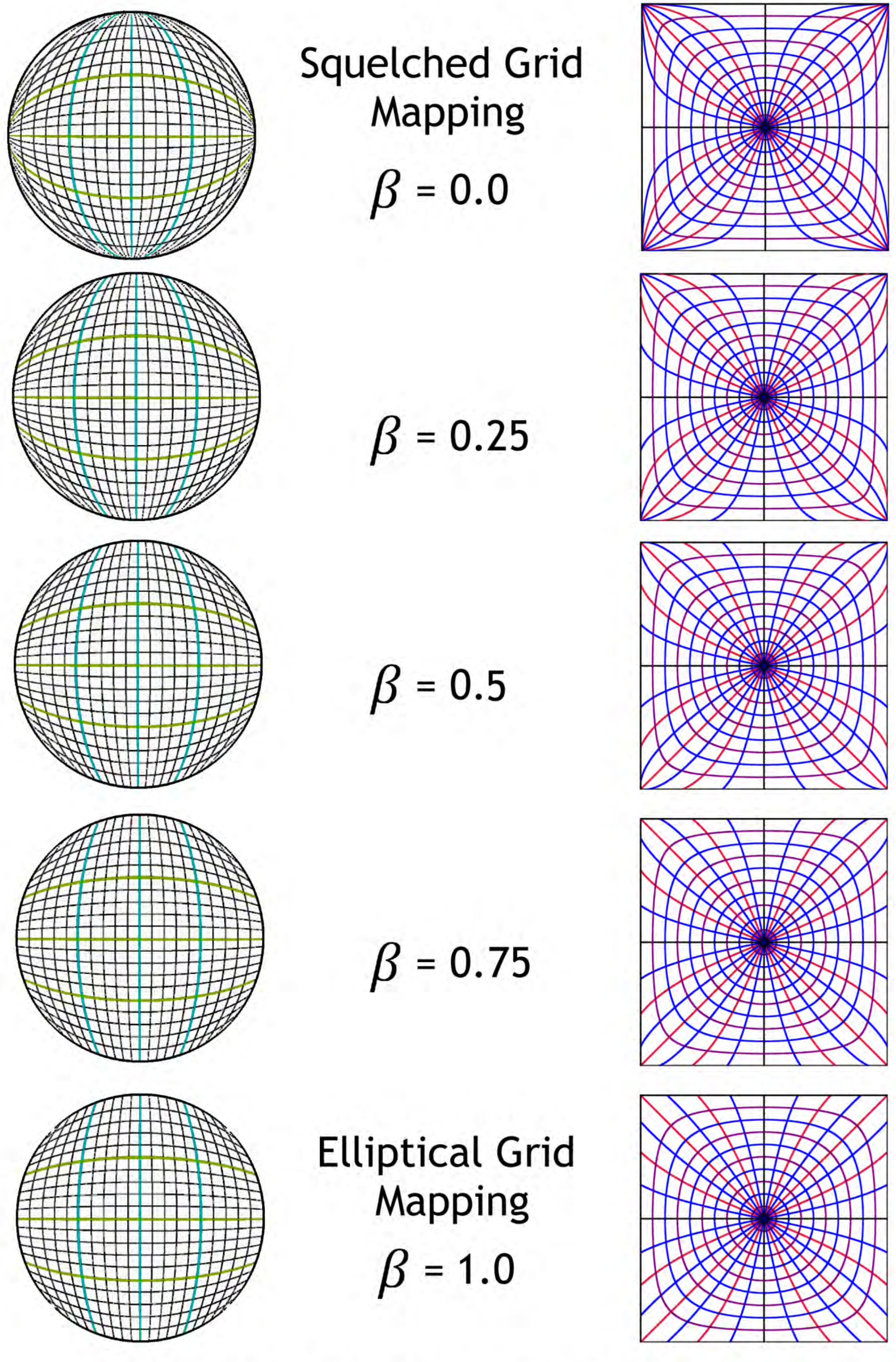

Figure 13: Blended E-Grid Mapping

## 7.9 Blending the Vertical Squelch with the Squelched Grid Open Mapping

It is also possible to blend the Squelched Grid open mapping with the Vertical Squelch open mapping to get a parameterized family of mappings in between the two. In a way, this is akin to introducing vertical bias to the Squelched Grid open mapping. This can be done by introducing a blend parameter $\beta$ to the vertical constraint equation. The blending is summarized in the table below.

| Mapping | blend value | vertical constraint equation | equation for $x$ |
|---|---|---|---|
| Vertical Squelch | 0 | $u = x$ | $x = u$ |
| Squelched Grid | 1 | $1 = \frac{u^2}{x^2} + \frac{v^2}{1}$ | $x = \frac{u}{\sqrt{1 - v^2}}$ |
| blended | $\beta$ | $1 = \frac{u^2}{x^2} + \beta\frac{v^2}{1}$ | $x = \frac{u}{\sqrt{1 - \beta v^2}}$ |

When $\beta$=0, the blended mapping reduces to the Vertical Squelch open mapping. When $\beta$=1, the blended mapping reduces to the Squelched Grid open mapping. In between, we have a blended mapping which shares the same horizontal constraint equation:

$$1 = \frac{u^2}{1} + \frac{v^2}{y^2}$$

The derivation of the square-to-disc mapping equations follows from substituting $P(y) = 1$ and $Q(x) = {}^{1}/_{\beta}$ into the Generalized Elliptical Grid mapping equations:

$$u = xP\sqrt{\frac{y^2 - Q^2}{x^2y^2 - P^2Q^2}} \qquad v = yQ\sqrt{\frac{x^2 - P^2}{x^2y^2 - P^2Q^2}}$$

Here are the mapping equations for this blended open mapping

Disc to square mapping:

$$\boldsymbol{x = \frac{u}{\sqrt{1 - \beta v^2}}} \qquad \boldsymbol{y = \frac{v}{\sqrt{1 - u^2}}}$$

Square to disc mapping:

$$\boldsymbol{u = x\sqrt{\frac{1 - \beta y^2}{1 - \beta x^2 y^2}}} \qquad \boldsymbol{v = y\sqrt{\frac{1 - x^2}{1 - \beta x^2 y^2}}}$$

## 7.10 Blending the Horizontal Squelch with the Squelched Grid Open Mapping

It is also possible to blend the Squelched Grid open mapping with the Horizontal Squelch open mapping to get a parameterized family of mappings in between the two. In a way, this is akin to introducing horizontal bias to the Squelched Grid open mapping. This can be done by introducing a blend parameter $\beta$ to the horizontal constraint equation. The blending is summarized in the table below.

| Mapping | blend value | horizontal constraint equation | equation for $y$ |
|---|---|---|---|
| Horizontal Squelch | 0 | $v = y$ | $y = v$ |
| Squelched Grid | 1 | $1 = \frac{u^2}{1} + \frac{v^2}{y^2}$ | $y = \frac{v}{\sqrt{1-u^2}}$ |
| blended | $\beta$ | $1 = \beta\frac{u^2}{1} + \frac{v^2}{y^2}$ | $y = \frac{v}{\sqrt{1-\beta u^2}}$ |

When $\beta$=0, the blended mapping reduces to the Horizontal Squelch open mapping. When $\beta$=1, the blended mapping reduces to the Squelched Grid open mapping. In between, we have a blended mapping which shares the same vertical constraint equation:

$$1 = \frac{u^2}{x^2} + \frac{v^2}{1}$$

The derivation of the square-to-disc mapping equations follows from substituting $P(y) = {}^{1}/_{\beta}$ and $Q(x) = 1$ into the Generalized Elliptical Grid mapping equations:

$$u = xP\sqrt{\frac{y^2 - Q^2}{x^2y^2 - P^2Q^2}} \qquad v = yQ\sqrt{\frac{x^2 - P^2}{x^2y^2 - P^2Q^2}}$$

Here are the mapping equations for this blended open mapping

Disc to square mapping:

$$\boldsymbol{x = \frac{u}{\sqrt{1-v^2}}} \qquad \boldsymbol{y = \frac{v}{\sqrt{1-\beta u^2}}}$$

Square to disc mapping:

$$\boldsymbol{u = x\sqrt{\frac{1-y^2}{1-\beta x^2y^2}}} \qquad \boldsymbol{v = y\sqrt{\frac{1-\beta x^2}{1-\beta x^2y^2}}}$$

# 8 Mappings with Axial Nonlinearities

All of the mappings derived in this paper so far are *axial mappings*. We define *axial mappings* mathematically as mappings that satisfy these conditions

$u = x$ whenever y is 0
$v = y$ whenever x is 0

In other words, when $y$ is zero, the mapping matches $u$ with $x$ exactly on the horizontal axis of the Cartesian plane. Likewise, when $x$ is zero, the mapping matches $v$ with $y$ exactly on the vertical axis of the Cartesian plane. Axial mappings between the circular disc and the square match points exactly on the horizontal and vertical axes of the Cartesian plane. Moreover, these inverse conditions also hold for axial mappings

$x = u$ whenever v is 0
$y = v$ whenever u is 0

Radial axial mappings arise from the *radial mapping linear parametric equation* mentioned in Section 3.1

$$u = \frac{x}{\sqrt{x^2 + y^2}} \; t$$

$$v = \frac{y}{\sqrt{x^2 + y^2}} \; t$$

These are parametric equations for $u$ & $v$ coordinates in terms of linear parameter $t$. We can relax these parametric equations and allow for nonlinearities by substituting the linear parameter $t$ with a modulator function $m(t)$ where $m$ is a rampant function of $t$.

$$u = \frac{x}{\sqrt{x^2 + y^2}} \; m(t)$$
$$v = \frac{y}{\sqrt{x^2 + y^2}} \; m(t)$$

Introducing this nonlinearity will allow us to come up with more mappings between the circular disc and the square. These new mappings are non-axial mappings.

Non-axial mappings have a non-linear relationship between $u$ and $x$ on the horizontal axis of the Cartesian plane. They also have a non-linear relationship between $v$ and $y$ on the vertical axis of the Cartesian plane. There are infinitely many non-axial mappings between the circular disc and the square. For example, the Schwarz-Christoffel conformal mapping [Fong 2014] is a non-axial mapping. In the following sections, we will derive and discuss three non-axial mappings that have explicit and invertible mapping equations. These non-axial mappings are derived using these nonlinearities

- $m(t) = t^2$
- $m(t) = \sqrt{t}$
- $m(t) = t\sqrt{2 - t^2}$

## 8.1 The Non-Axial 2-Pinch Mapping

Consider the modulator function $m(t) = t^2$, the nonlinear radial mapping parametric equation for this is

$$u = \frac{x}{\sqrt{x^2+y^2}}t^2 \qquad\qquad v = \frac{y}{\sqrt{x^2+y^2}}t^2$$

Furthermore, if we use a nonlinear squircular continuum relationship between $s$ and $t$ in the squircle equation, i.e. $s=t^2$, just as in section 4.1, we get

$$t = \sqrt{\frac{x^2+y^2}{1+x^2y^2}}$$

We can then plug this into the modulator function to get these square-to-disc mapping equations

$$u = \frac{x}{\sqrt{x^2+y^2}}\frac{(x^2+y^2)}{(1+x^2y^2)} \qquad\qquad v = \frac{y}{\sqrt{x^2+y^2}}\frac{(x^2+y^2)}{(1+x^2y^2)}$$

This simplifies to

$$\boldsymbol{u = \frac{x\sqrt{x^2+y^2}}{1+x^2y^2}} \qquad\qquad \boldsymbol{v = \frac{y\sqrt{x^2+y^2}}{1+x^2y^2}}$$

To invert this mapping, start with the equation for $u$

$$u = \frac{x\sqrt{x^2+y^2}}{1+x^2y^2} \quad \Rightarrow \quad u\,(1+x^2y^2) = x\sqrt{x^2+y^2}$$

$$\Rightarrow \quad u^2(1+x^2y^2)^2 = x^2(x^2+y^2) \quad \Rightarrow \quad u^2(1+2x^2y^2+x^4y^4) = x^4+x^2y^2$$

Use the radial constraint of the mappings $y = \frac{v}{u}x$ and substitute

$$u^2\left(1+2x^2\frac{v^2}{u^2}x^2+x^4\frac{v^4}{u^4}x^4\right) = x^4+\frac{v^2}{u^2}x^2 \quad \Rightarrow \quad u^4+2u^2v^2x^4+v^4x^8 = u^2x^4+v^2x^4$$

Collecting the polynomial in x to one side of the equation, we get

$$v^4\boldsymbol{x^8}+(2u^2v^2-u^2-v^2)\boldsymbol{x^4}+u^4 = 0$$

We can then use the quadratic formula to solve for $x^4$

$$x^4 = \frac{-2(u^2v^2-u^2-v^2)\pm\sqrt{(2u^2v^2-u^2-v^2)^2-4u^4v^4}}{2v^4} = \frac{u^2+v^2-2u^2v^2\pm\sqrt{(u^2+v^2-2u^2v^2)^2-4u^4v^4}}{2v^4}$$

which gives us an for expression for x as

$$\boldsymbol{x = \frac{sgn(uv)}{v\sqrt[4]{2}}\left(u^2+v^2-2u^2v^2-\sqrt{(u^2+v^2-4u^2v^2)(u^2+v^2)}\right)^{\frac{1}{4}}}$$

Furthermore, the expression for y can be calculated using the radial constraint $y = \frac{v}{u}x$ to get

$$\boldsymbol{y = \frac{sgn(uv)}{u\sqrt[4]{2}}\left(u^2+v^2-2u^2v^2-\sqrt{(u^2+v^2-4u^2v^2)(u^2+v^2)}\right)^{\frac{1}{4}}}$$

## 8.2 The Non-Axial ½-Punch Mapping

Consider the modulator function $m(t) = t^{1/2}$, the nonlinear radial mapping parametric equation for this is

$$u = \frac{x}{\sqrt{x^2+y^2}}\sqrt{t} \qquad\qquad v = \frac{y}{\sqrt{x^2+y^2}}\sqrt{t}$$

Furthermore, if we use a nonlinear squircular continuum relationship between *s* and *t* in the squircle equation, i.e. $s=t^2$, we get

$$t = \sqrt{\frac{x^2+y^2}{1+x^2y^2}}$$

We can then plug this into the modulator function to get these square-to-disc mapping equations

$$u = \frac{x}{\sqrt{x^2+y^2}}\sqrt{\sqrt{\frac{x^2+y^2}{1+x^2y^2}}} \qquad\qquad v = \frac{y}{\sqrt{x^2+y^2}}\sqrt{\sqrt{\frac{x^2+y^2}{1+x^2y^2}}}$$

This simplifies to

$$\boldsymbol{u = \frac{x}{(x^2+y^2)^{1/4}\ (1+x^2y^2)^{1/4}}} \qquad\qquad \boldsymbol{v = \frac{y}{(x^2+y^2)^{1/4}\ (1+x^2y^2)^{1/4}}}$$

To invert this mapping, start with the equation for *u*

$$u = \frac{x}{(x^2+y^2)^{1/4}\ (1+x^2y^2)^{1/4}} \;\Rightarrow\; u^4(x^2+y^2)(1+x^2y^2) = x^4 \;\Rightarrow\; u^4(x^2+y^2+x^4y^2+x^2y^4) = x^4$$

Use the radial constraint of the mappings $y = \frac{v}{u}x$ and substitute

$$u^4\left(x^2+\frac{v^2}{u^2}x^2+x^4\frac{v^2}{u^2}x^2+x^2\frac{v^4}{u^4}x^4\right) = x^4 \qquad \Rightarrow \qquad x^2(u^2+v^2)+x^6(u^2v^2+v^4) = x^4$$

Divide by $x^2$ and collect the terms as a polynomial in x

$$(u^2v^2+v^4)\boldsymbol{x^4} - \boldsymbol{x^2} + u^2v^2 + u^4 = 0$$

This is a biquadratic equation for which we can solve for $x^2$

$$x^2 = \frac{1 \pm \sqrt{1-4(u^2v^2+v^4)(u^2v^2+u^4)}}{2(u^2v^2+v^4)}$$

We can then get a quadrant-aware expression for x

$$\boldsymbol{x = \frac{sgn(uv)}{v}\sqrt{\frac{1-\sqrt{1-4u^2v^2(u^2+v^2)^2}}{2(u^2+v^2)}}}$$

Furthermore, the expression for y can be calculated using the radial constraint $y = \frac{v}{u}x$ to get

$$\boldsymbol{y = \frac{sgn(uv)}{u}\sqrt{\frac{1-\sqrt{1-4u^2v^2(u^2+v^2)^2}}{2(u^2+v^2)}}}$$

## 8.3 The Sham Quartic Mapping

Consider the modulator function $m(t) = t\sqrt{2-t^2}$ , the nonlinear radial mapping parametric equation for this is

$$u = \frac{x}{\sqrt{x^2+y^2}}\, t\sqrt{2-t^2} \qquad\qquad v = \frac{y}{\sqrt{x^2+y^2}}\, t\sqrt{2-t^2}$$

Furthermore, if we use a nonlinear squircular continuum relationship between *s* and *t* in the squircle equation, i.e. $s=t^2$, we get

$$t = \sqrt{\frac{x^2+y^2}{1+x^2y^2}}$$

We can then plug this into the modulator function to get these square-to-disc mapping equations

$$u = \frac{x}{\sqrt{x^2+y^2}}\sqrt{\frac{x^2+y^2}{1+x^2y^2}}\sqrt{2-\frac{x^2+y^2}{1+x^2y^2}} \qquad\qquad v = \frac{y}{\sqrt{x^2+y^2}}\sqrt{\frac{x^2+y^2}{1+x^2y^2}}\sqrt{2-\frac{x^2+y^2}{1+x^2y^2}}$$

This simplifies to

$$\boldsymbol{u = \frac{x\sqrt{2+2x^2y^2-x^2-y^2}}{1+x^2y^2}} \qquad\qquad \boldsymbol{v = \frac{y\sqrt{2+2x^2y^2-x^2-y^2}}{1+x^2y^2}}$$

Unfortunately, the inverse of this mapping is not as simple. Start with the expression for *u*

$$u = \frac{x\sqrt{2+2x^2y^2-x^2-y^2}}{1+x^2y^2} \quad\Rightarrow\quad u(1+x^2y^2) = x\sqrt{2+2x^2y^2-x^2-y^2}$$

$$\Rightarrow \qquad u^2(1+x^2y^2)^2 \;=\; x^2(2+2x^2y^2-x^2-y^2)$$

$$\Rightarrow \qquad u^2(1+2x^2y^2+x^4y^4) \;=\; 2x^2+2x^4y^2-x^4-x^2y^2$$

Use the radial constraint of the mappings $y = \frac{v}{u}x$ and substitute

$$u^2\left(1+2x^2\frac{v^2}{u^2}x^2+x^4\frac{v^4}{u^4}x^4\right) = 2x^2+2x^4\frac{v^2}{u^2}x^2-x^4-x^2\frac{v^2}{u^2}x^2$$

Multiply the equation by $u^2$ to get

$$u^4+2u^2v^2x^4+v^4x^8 = 2u^2x^2+2v^2x^6-u^2x^4-v^2x^4$$

Collect the terms as a polynomial in x

$$v^4\boldsymbol{x^8}-2v^2\boldsymbol{x^6}+(u^2+v^2+2u^2v^2)\boldsymbol{x^4}-2u^2\boldsymbol{x^2}+u^4 = 0$$

This is an octic polynomial in *x* with no terms of odd degree. This polynomial equation is equivalent to

$$v^4\boldsymbol{q^4}-2v^2\boldsymbol{q^3}+(u^2+v^2+2u^2v^2)\boldsymbol{q^2}-2u^2\boldsymbol{q}+u^4 = 0$$

where $q=x^2$. The roots of this polynomial can be calculated explicitly using the quartic formula. After which, we can get formulas for *x* and *y* as the inverse equations of the mapping. However, the full quartic formula is quite complicated [Shmakov 2011] [Blinn 2002], so we will not write down the equations here. Instead, we will assume that there is already an existing subroutine than can provide the roots of a quartic equation.

Denote $a_n$ as the coefficient that corresponds to the $q^n$ power term in the quartic polynomial equation. The polynomial coefficients to be used an input to the quartic formula are

$$a_4 = v^4$$

$$a_3 = -2v^2$$

$$a_2 = u^2 + v^2 + 2u^2v^2$$

$$a_1 = -2u^2$$

$$a_0 = u^4$$

Let $q_0$ be the smallest non-negative root of the given quartic equation. We can then solve for $x$ using the formula

$$\boldsymbol{x = sgn(u)\sqrt{q_0}}$$

Likewise, using the radial constraint of the mappings $y = \frac{v}{u}x$, we get this equation

$$y = sgn(u)\frac{v}{u}\sqrt{q_0}$$

or equivalently

$$\boldsymbol{y = sgn(v)\left|\frac{v}{u}\right|\sqrt{q_0}}$$

As already shown, the disc-to-square equations for this mapping require the quartic formula. This is partly the reason why the mapping is called the Sham Quartic mapping. Unfortunately, the equations for the full quartic formula are too unwieldy to include here.

Incidentally, when a square grid is mapped to a circular disc using this mapping, the circular disc appears like a perspective view of a sphere! In actuality, this grid on the circular disc is flat and only appears like a fake (or sham) sphere. Nevertheless, the Sham Quartic mapping produces excellent results when used as a spherize filter. An example result of this spherize application is shown Figure 14.

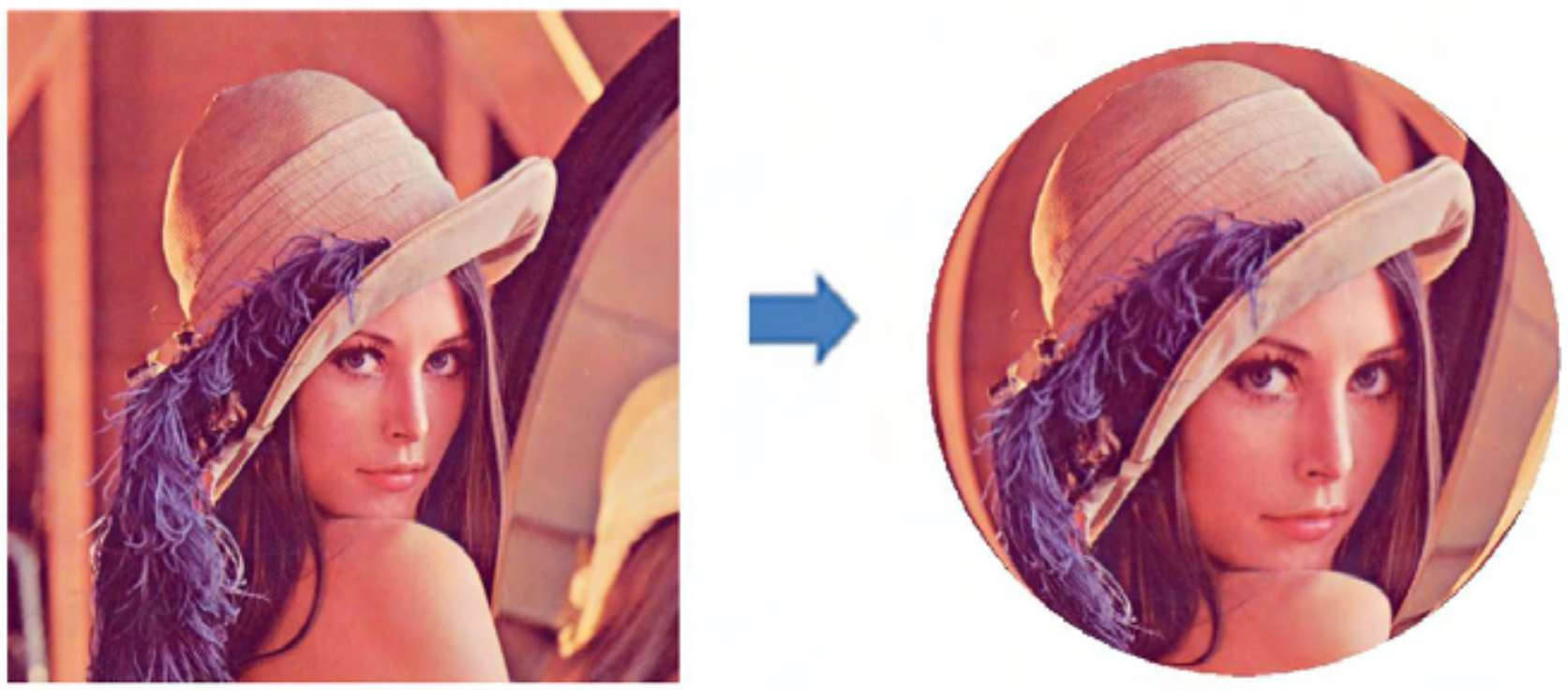

Figure 14: The Sham Quartic mapping as a spherize filter

## 9.1 detJ Surfaces

In this section, we will introduce a 3D surface arising from a specific square-to-disc mapping. Using any of the square-to-disc mappings mentioned in this paper along with the Jacobian determinant of the mapping, it is possible to create an interesting 3D surface useful for visualization.

Suppose we have a given square-to-disc mapping with equations

$$u = g(x, y)$$

$$v = h(x, y)$$

The Jacobian determinant of the mapping is defined as

$$\det J = \begin{vmatrix} \frac{\partial g}{\partial x} & \frac{\partial g}{\partial y} \\ \frac{\partial h}{\partial x} & \frac{\partial h}{\partial y} \end{vmatrix} = \frac{\partial g}{\partial x}\frac{\partial h}{\partial y} - \frac{\partial g}{\partial y}\frac{\partial h}{\partial x}$$

After simplification, the Jacobian determinant of the mapping is a mathematical expression written in terms of variables x and y.

We define the ***detJ surface*** of a mapping as the 3D surface with parametric equations

$$u = g(x, y)$$

$$v = h(x, y)$$

$$w = \det J = \frac{\partial g}{\partial x}\frac{\partial h}{\partial y} - \frac{\partial g}{\partial y}\frac{\partial h}{\partial x}$$

For example, consider the 2-Squircular mapping where

$$u = g(x, y) = \frac{x}{\sqrt{1 + x^2 y^2}}$$

$$v = h(x, y) = \frac{y}{\sqrt{1 + x^2 y^2}}$$

To solve for the Jacobian determinant, we need to compute the partial derivatives

$$\frac{\partial g}{\partial x} = \frac{1}{\sqrt{1 + x^2 y^2}} + x\frac{-xy^2}{(1 + x^2 y^2)^{\frac{3}{2}}} = \frac{1 + x^2 y^2}{(1 + x^2 y^2)^{\frac{3}{2}}} - \frac{x^2 y^2}{(1 + x^2 y^2)^{\frac{3}{2}}} = \frac{1}{(1 + x^2 y^2)^{\frac{3}{2}}}$$

$$\frac{\partial g}{\partial y} = \frac{-x}{2(1 + x^2 y^2)^{\frac{3}{2}}} \quad 2x^2 y = \frac{-x^3 y}{(1 + x^2 y^2)^{\frac{3}{2}}}$$

$$\frac{\partial h}{\partial x} = \frac{-y}{2(1 + x^2 y^2)^{\frac{3}{2}}} \quad 2xy^2 = \frac{-xy^3}{(1 + x^2 y^2)^{\frac{3}{2}}}$$

$$\frac{\partial h}{\partial y} = \frac{1}{\sqrt{1+x^2y^2}} + y\frac{-x^2y}{(1+x^2y^2)^{\frac{3}{2}}} = \frac{1+x^2y^2}{(1+x^2y^2)^{\frac{3}{2}}} - \frac{x^2y^2}{(1+x^2y^2)^{\frac{3}{2}}} = \frac{1}{(1+x^2y^2)^{\frac{3}{2}}}$$

Thus, the Jacobian matrix for the mapping is

$$J = \begin{bmatrix} \frac{\partial g}{\partial x} & \frac{\partial g}{\partial y} \\ \frac{\partial h}{\partial x} & \frac{\partial h}{\partial y} \end{bmatrix} = \frac{1}{(1+x^2y^2)^{\frac{3}{2}}} \begin{bmatrix} 1 & -x^3y \\ -xy^3 & 1 \end{bmatrix}$$

and the Jacobian determinant is

$$\det J = \frac{1}{(1+x^2y^2)^3}(1-x^4y^4) = \frac{1}{(1+x^2y^2)^3}(1-x^2y^2)(1+x^2y^2) = \frac{1-x^2y^2}{(1+x^2y^2)^2}$$

So the detJ surface for the 2-Squircular mapping are given by these parametric equations

$$u = \frac{x}{\sqrt{1+x^2y^2}} \qquad v = \frac{y}{\sqrt{1+x^2y^2}} \qquad w = \frac{1-x^2y^2}{(1+x^2y^2)^2}$$

As a matter of convention in order to avoid confusion, we are going to rename our *u,v,w* and *x,y* variables to different counterparts. The reason for this is because many software that handle parametric surfaces use a notation where the surface is represented with *x,y,z* independent variables in $\mathbb{R}^3$ that are parameterized by variables *u* and *v*. Unfortunately, this notation is the exact opposite of the notation used in this paper. In order to avoid confusion from ambiguity, it is necessary to rename our variables. For this section on detJ surfaces, we shall adopt this renaming scheme

$\boldsymbol{u} \longrightarrow \hat{X}$ (capital X hat)

$\boldsymbol{v} \longrightarrow \hat{Y}$ (capital Y hat)

$\boldsymbol{w} \longrightarrow \hat{Z}$ (capital Z hat)

$\boldsymbol{x} \longrightarrow \boldsymbol{\lambda}$ (Greek letter lambda)

$\boldsymbol{y} \longrightarrow \boldsymbol{\gamma}$ (Greek letter gamma)

Here are the parametric equations for the detJ surface of the 2-Squircular mapping after renaming the variables to their Greek counterparts.

$$\hat{X} = \frac{\lambda}{\sqrt{1+\lambda^2\gamma^2}} \qquad \hat{Y} = \frac{\gamma}{\sqrt{1+\lambda^2\gamma^2}} \qquad \hat{Z} = \frac{1-\lambda^2\gamma^2}{(1+\lambda^2\gamma^2)^2}$$

These parametric equations can then be plotted using mathematical visualization software such as Geogebra, CalcPlot3D, or Mathematica. We recommend using the built-in ParametricPlot3D command in Mathematica [Wolfram 1999] [Torrence 2009] for visualizing this surface.

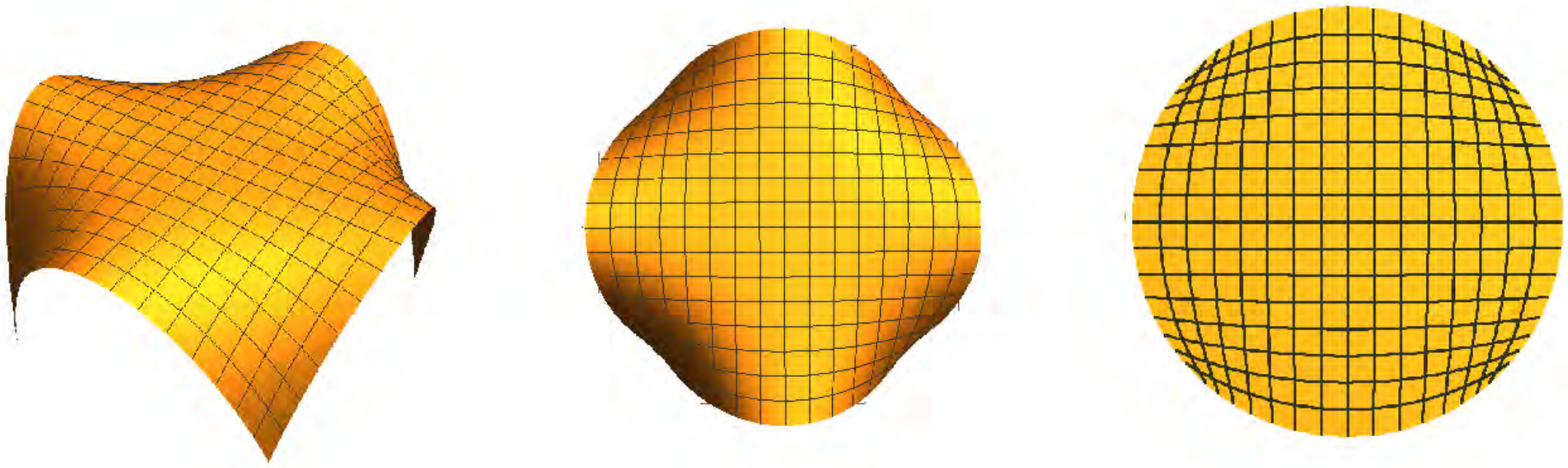

Figure 15: Different perspectives of the detJ surface for the 2-Squircular mapping

The diagrams in Figure 15 show the detJ surface at different viewing angles. On the left side of Figure 15 we have perspective view of the 3D surface that is slightly askew from the top. On the center, we have an almost top view perspective. On the right side of Figure 15, we have a full orthographic top view of the 3D surface.

The 3D perspective plots can be viewed in Mathematica using the following command:

```
ParametricPlot3D[{ λ / Sqrt[1 + λ* λ* γ * γ],
                   γ / Sqrt[1 + λ* λ* γ * γ],
                  (1- λ* λ* γ * γ) / ( 1 + λ* λ* γ * γ)^2},
                  { λ, -1, 1}, { γ, -1,  1}]
```

The orthographic top view can be generated in Mathematica using the following command:

```
ParametricPlot3D[{ λ / Sqrt[1 + λ* λ* γ * γ],
                   γ / Sqrt[1 + λ* λ* γ * γ],
                   0},{ λ, -1, 1}, { γ, -1,  1}]
```

The diagrams in Figure 15 show something noteworthy about detJ surfaces. The orthographic top view of detJ surfaces appear like the 2D plot of a square grid mapped into a circular disc. This phenomenon is best explained by Grant Sanderson (3blue1brown) in his Khan Academy lecture series on multivariable calculus [Sanderson 2017]. The Jacobian determinant is just a measure of the distortion of area after performing a mapping.

One can think of a detJ surface as the deformation of a flat 2D square grid in three dimensional space. The square grid is deformed in such a way that it appears circular from the top view. The Jacobian determinant height comes from the stretching factor of area within the mapping.

## 9.2 More detJ Surfaces

Without going through the lengthy mathematical manipulation required for computing Jacobian determinant of each of the mappings in this paper, we summarize in Figure 16 the other detJ surfaces arising from the mappings. The derivation of the Jacobian determinant for each mapping is very similar to how we computed it for the 2-Squircular mapping. This includes computing partial derivatives and a 2x2 matrix determinant followed by detailed algebraic simplification. Each set of parametric equations for the detJ surfaces in Figure 16 comes with a 3D plot of the arising surface. Note that all of these detJ surfaces will have a circular base when viewed from an orthographic top view.

The detJ surfaces appearing in Figure 16 have different interesting features and idiosyncrasies. We will briefly discuss some of these. Most of the detJ surfaces have zero Jacobian determinant at $(\lambda, \gamma) = (\pm 1, \pm 1)$. These points correspond to the corners of the square. The 3-Squircular detJ surface looks very similar to the 2-Squircular detJ surface. The vertical and horizontal squelch detJ surfaces are just rotated versions of each other. The Non-Axial ½-Punch mapping has a singularity at the top of the origin where the Jacobian determinant is infinite.

## FG-Squircular

$$\widehat{X} = \frac{\lambda\sqrt{\lambda^2+\gamma^2-\lambda^2\gamma^2}}{\sqrt{\lambda^2+\gamma^2}}$$

$$\hat{Y} = \frac{\gamma\sqrt{\lambda^2+\gamma^2-\lambda^2\gamma^2}}{\sqrt{\lambda^2+\gamma^2}}$$

$$\hat{Z} = 1 - \frac{2\lambda^2\gamma^2}{\lambda^2+\gamma^2}$$

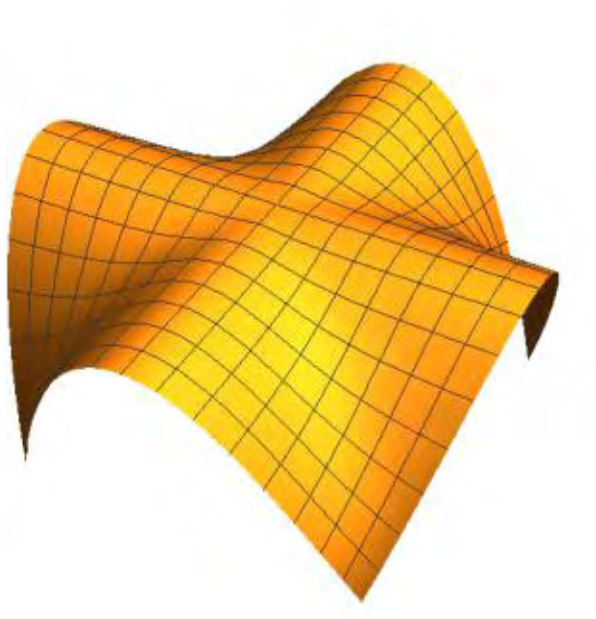

## Elliptical Grid

$$\widehat{X} = \lambda\sqrt{1-\frac{\gamma^2}{2}}$$

$$\hat{Y} = \gamma\sqrt{1-\frac{\lambda^2}{2}}$$

$$\hat{Z} = \frac{2-\lambda^2-\gamma^2}{\sqrt{2-\lambda^2}\sqrt{2-\gamma^2}}$$

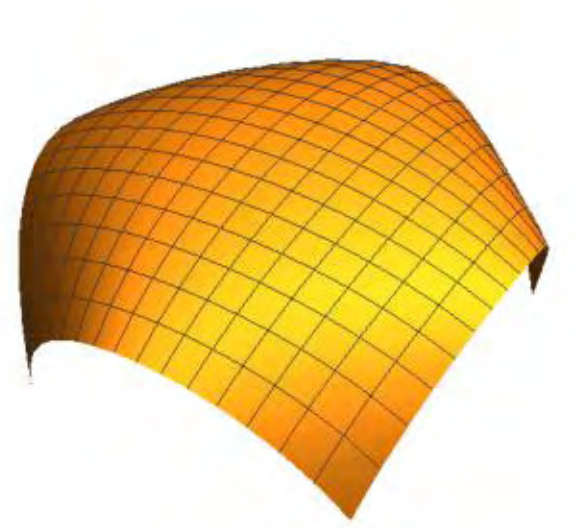

## Non-axial 2-Pinch

$$\widehat{X} = \frac{\lambda\sqrt{\lambda^2+\gamma^2}}{1+\lambda^2\gamma^2}$$

$$\hat{Y} = \frac{\gamma\sqrt{\lambda^2+\gamma^2}}{1+\lambda^2\gamma^2}$$

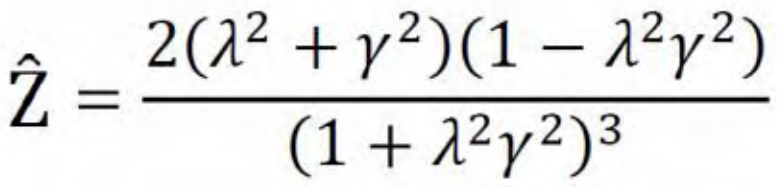

$$\hat{Z} = \frac{2(\lambda^2+\gamma^2)(1-\lambda^2\gamma^2)}{(1+\lambda^2\gamma^2)^3}$$

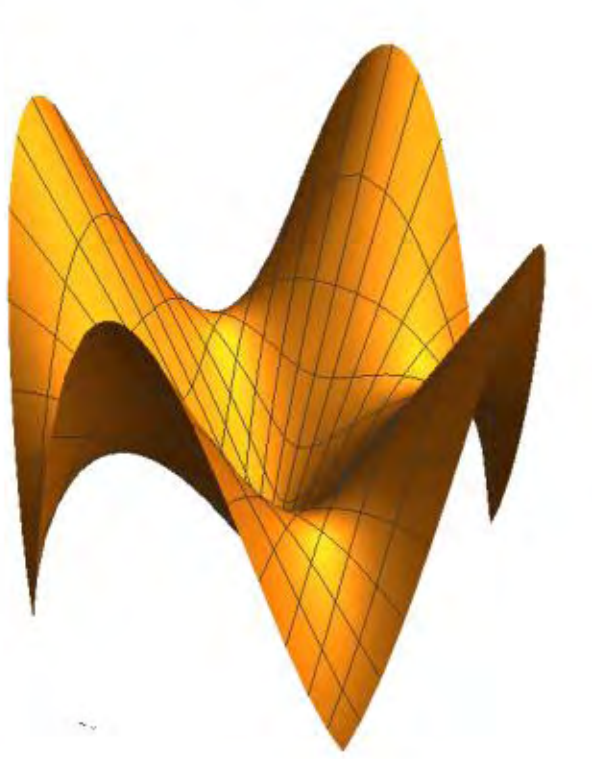

## Non-axial ½-Punch

$$\widehat{X} = \frac{\lambda}{(\lambda^2+\gamma^2)^{1/4}\,(1+\lambda^2\gamma^2)^{1/4}}$$

$$\hat{Y} = \frac{\gamma}{(\lambda^2+\gamma^2)^{1/4}\,(1+\lambda^2\gamma^2)^{1/4}}$$

$$\hat{Z} = \frac{1-\lambda^2\gamma^2}{2\,(1+\lambda^2\gamma^2)^{3/2}\sqrt{\lambda^2+\gamma^2}}$$

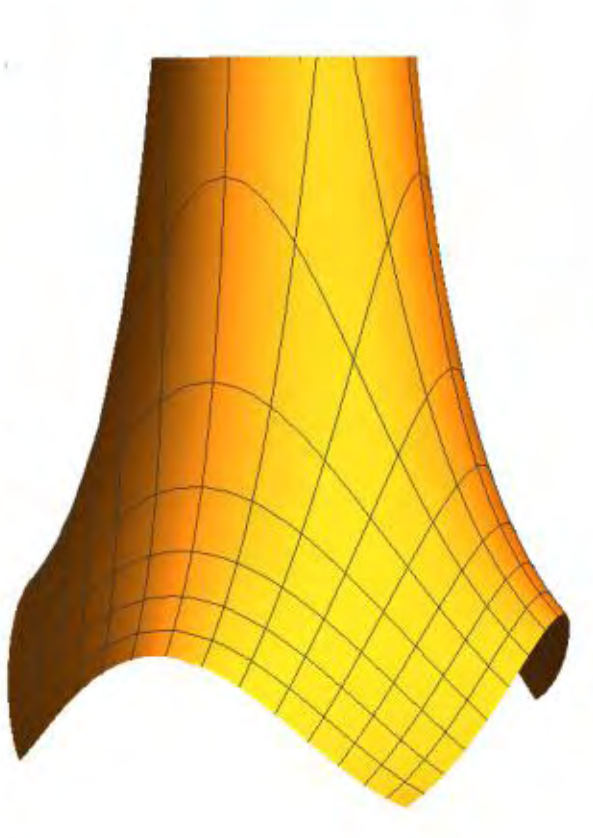

## Vertical Squelch

$$\widehat{X} = \lambda$$

$$\hat{Y} = \gamma\sqrt{1-\lambda^2}$$

$$\hat{Z} = \sqrt{1-\lambda^2}$$

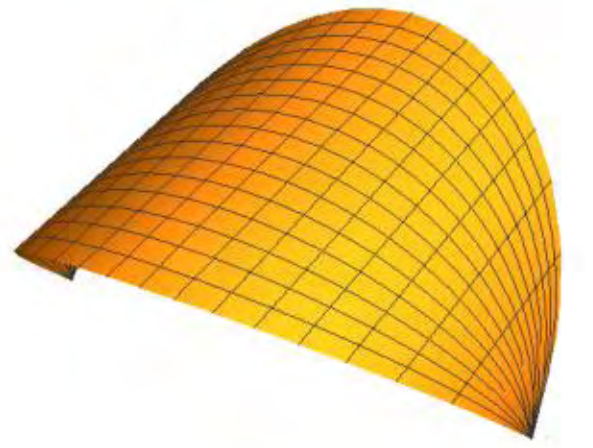

## Horizontal Squelch

$$\widehat{X} = \lambda\sqrt{1-\gamma^2}$$

$$\hat{Y} = \gamma$$

$$\hat{Z} = \sqrt{1-\gamma^2}$$

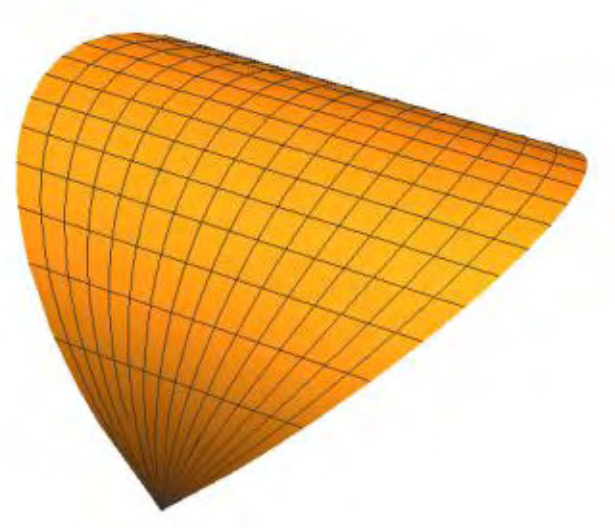

Figure 16: Different detJ surfaces for the mappings

$$\hat{X} = \lambda \sqrt{\frac{\lambda^2 + \gamma^2 - 2\lambda^2\gamma^2}{(\lambda^2 + \gamma^2)(1 - \lambda^2\gamma^2)}}$$

$$\hat{Y} = \gamma \sqrt{\frac{\lambda^2 + \gamma^2 - 2\lambda^2\gamma^2}{(\lambda^2 + \gamma^2)(1 - \lambda^2\gamma^2)}}$$

$$\hat{Z} = \frac{\lambda^2 + \gamma^2 + \lambda^4\gamma^2 - 4\lambda^2\gamma^2 + \lambda^2\gamma^4}{(\lambda^2 + \gamma^2)(1 - \lambda^2\gamma^2)^2}$$

## Cornerific Tapered2

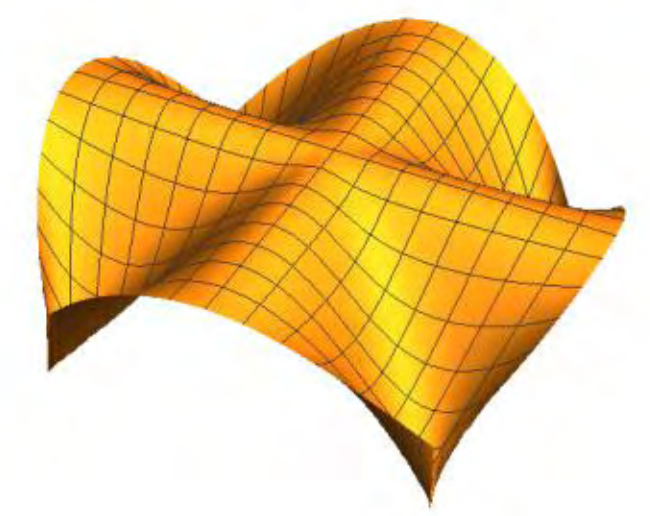

$$\hat{X} = \frac{sgn(\lambda\gamma)}{\gamma} \sqrt{\frac{-1 + \sqrt{1 + 4\lambda^4\gamma^2 + 4\lambda^2\gamma^4}}{2(\lambda^2 + \gamma^2)}}$$

$$\hat{Y} = \frac{sgn(\lambda\gamma)}{\lambda} \sqrt{\frac{-1 + \sqrt{1 + 4\lambda^4\gamma^2 + 4\lambda^2\gamma^4}}{2(\lambda^2 + \gamma^2)}}$$

$$\hat{Z} = \frac{-1 - \lambda^4\gamma^2 - \lambda^2\gamma^4 + \sqrt{1 + 4\lambda^4\gamma^2 + 4\lambda^2\gamma^4}}{\lambda^2\gamma^2(\lambda^2 + \gamma^2)\sqrt{1 + 4\lambda^4\gamma^2 + 4\lambda^2\gamma^4}}$$

## 3-Squircular

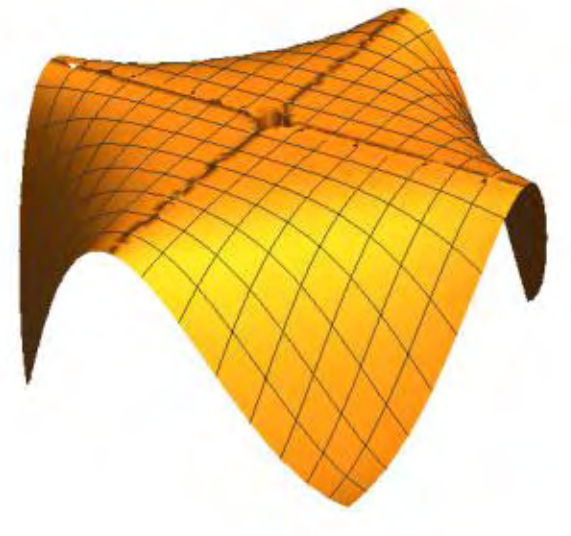

## Schwarz-Christoffel

$$\hat{X} = Re\left(\frac{1 - i}{\sqrt{2}}\, cn\left(K_e \frac{1 + i}{2}(\lambda + \gamma\, i) - K_e \;, \frac{1}{\sqrt{2}}\right)\right)$$

$$\hat{Y} = Im\left(\frac{1 - i}{\sqrt{2}}\, cn\left(K_e \frac{1 + i}{2}\,(\lambda + \gamma\, i) - K_e \;, \frac{1}{\sqrt{2}}\right)\right)$$

$$\hat{Z} = \frac{K_e^2}{2} \left| dn\left(K_e \frac{1 + i}{2}\,(\lambda + \gamma\, i) - K_e, \frac{1}{\sqrt{2}}\right)\; sn\left(K_e \frac{1 + i}{2}\,(\lambda + \gamma\, i) - K_e, \frac{1}{\sqrt{2}}\right) \right|^2$$

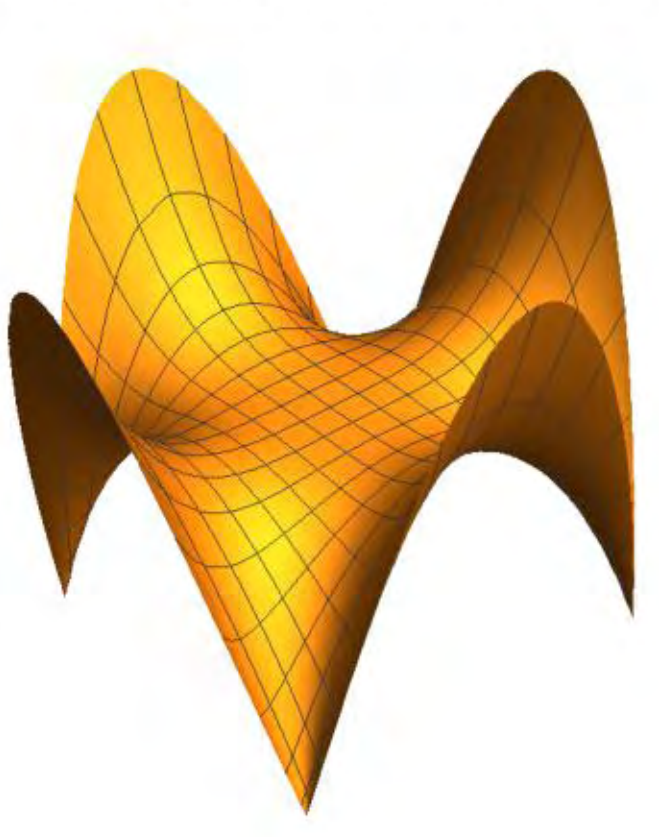

Figure 16 (continued)

## 9.3 The Sham Hemi-Ellipsoid

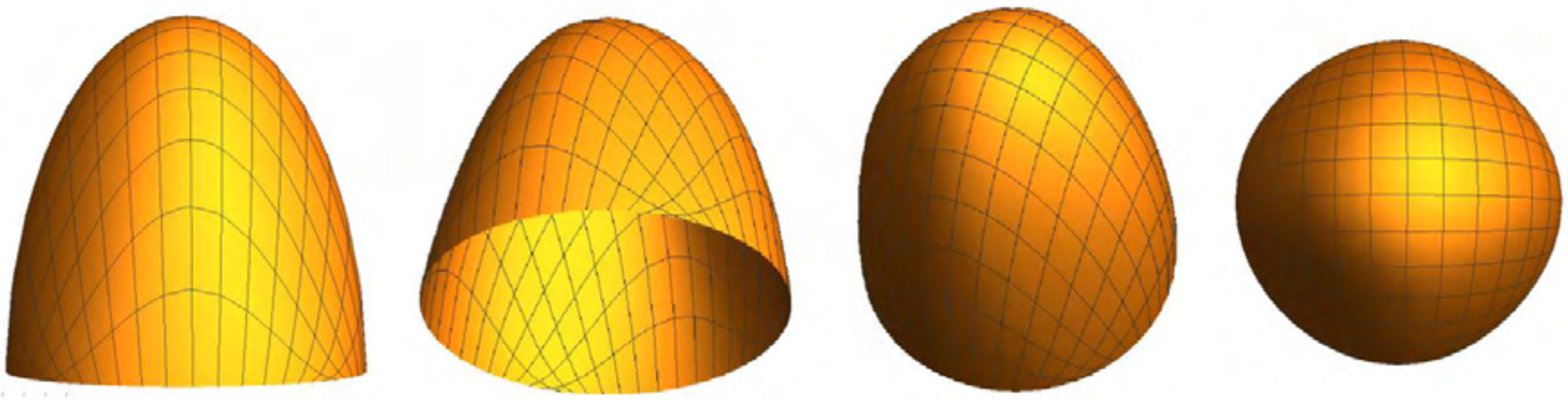

Figure 17: Different perspective views of the Sham Hemi-Ellipsoid

The Sham Quartic mapping produces the most interesting detJ surface among all the mappings discussed in this paper. We have deliberately set it apart in its own separate section to discuss it in detail. Figure 17 shows the detJ surface for the Sham Quartic mapping from different perspective views. We shall refer to this specific detJ surface as the *Sham Hemi-Ellipsoid.* The parametric equations for this surface are

$$\hat{X} = \frac{\lambda\sqrt{2 + 2\lambda^2\gamma^2 - \lambda^2 - \gamma^2}}{1 + \lambda^2\gamma^2}$$

$$\hat{Y} = \frac{\gamma\sqrt{2 + 2\lambda^2\gamma^2 - \lambda^2 - \gamma^2}}{1 + \lambda^2\gamma^2}$$

$$\hat{Z} = \frac{2(1 - \lambda^2)(1 - \gamma^2)(1 - \lambda^2\gamma^2)}{(1 + \lambda^2\gamma^2)^3}$$

This detJ surface appears very much like half of an ellipsoid, but it only approximates an ellipsoid. This is the reason why it is called Sham Hemi-Ellipsoid. Nevertheless, it comes very close to approximating the top half of the ellipsoid with the equation $X^2+Y^2+Z^2/4 = 1$. In fact, it has an identical base circular cross-section as the ellipsoid when cut by the xy-plane. Also, the Sham Hemi-Ellipsoid has a peak height of 2 at $(\lambda, \gamma)=(0,0)$.

## 9.4 The Coerced Hemisphere

This parametric equation for the Sham Hemi-Ellipsoid can easily be modified to produce parametric equations for the hemisphere. This can be done by using the equation for the unit sphere: $X^2 + Y^2 + Z^2 = 1$. By specifically setting $\hat{Z} = \sqrt{1 - \hat{X}^2 - \hat{Y}^2}$ in the parametric equations, we can effectively force the surface to be a hemisphere.

$$\hat{X} = \frac{\lambda\sqrt{2 + 2\lambda^2\gamma^2 - \lambda^2 - \gamma^2}}{1 + \lambda^2\gamma^2}$$

$$\hat{Y} = \frac{\gamma\sqrt{2 + 2\lambda^2\gamma^2 - \lambda^2 - \gamma^2}}{1 + \lambda^2\gamma^2}$$

$$\hat{Z} = \frac{(1 - \lambda^2)(1 - \gamma^2)}{1 + \lambda^2\gamma^2}$$

The derivation of the expression for $\hat{Z}$ is relatively simple but requires a bit of algebra. Start with

$$\hat{Z} = \sqrt{1 - \hat{X}^2 - \hat{Y}^2} = \sqrt{1 - \left(\frac{\lambda\sqrt{2 + 2\lambda^2\gamma^2 - \lambda^2 - \gamma^2}}{1 + \lambda^2\gamma^2}\right)^2 - \left(\frac{\gamma\sqrt{2 + 2\lambda^2\gamma^2 - \lambda^2 - \gamma^2}}{1 + \lambda^2\gamma^2}\right)^2}$$

$$= \sqrt{\frac{(1 + \lambda^2\gamma^2)^2 - \lambda^2(2 + 2\lambda^2\gamma^2 - \lambda^2 - \gamma^2) - \gamma^2(2 + 2\lambda^2\gamma^2 - \lambda^2 - \gamma^2)}{(1 + \lambda^2\gamma^2)^2}}$$

$$= \sqrt{\frac{1 + 2\lambda^2\gamma^2 + \lambda^4\gamma^4 - 2\lambda^2 - 2\lambda^4\gamma^2 + \lambda^4 + \lambda^2\gamma^2 - 2\gamma^2 - 2\lambda^2\gamma^4 + \lambda^2\gamma^2 + \gamma^4}{(1 + \lambda^2\gamma^2)^2}}$$

$$= \frac{\sqrt{(1 - \lambda^2 - \gamma^2 + \lambda^2\gamma^2)^2}}{1 + \lambda^2\gamma^2} = \frac{\sqrt{(1 - \lambda^2)^2(1 - \gamma^2)^2}}{1 + \lambda^2\gamma^2} = \frac{(1 - \lambda^2)(1 - \gamma^2)}{1 + \lambda^2\gamma^2}$$

Note that a coerced hemi-ellipsoid can be produced by multiplying this $\hat{Z}$ with a factor 2. Also note that the full sphere can be represented with these parametric equations by taking $\pm\hat{Z}$.

## 9.5 Mapping the Square to the Hemisphere

Recall that $(\lambda, \gamma)$ represents points inside the square and that our equation for the coerced hemisphere is parameterized by $(\lambda, \gamma)$. This means that we can use these parametric equations to effectively map the square to the hemisphere. An example of this application is shown in Figure 18.

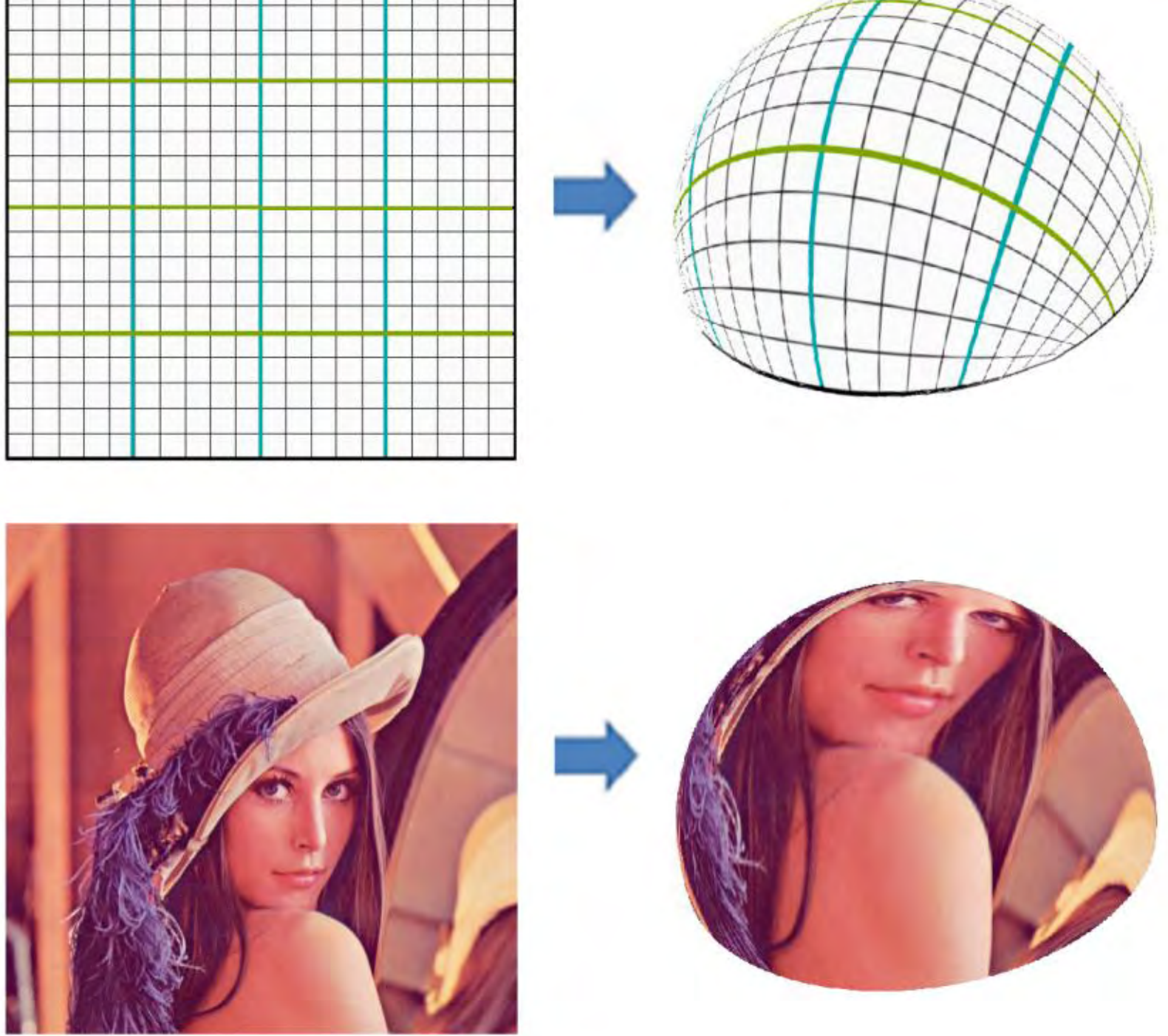

Figure 18: Lena square image mapped to a hemisphere

# 10 Blended Generalizations of the Mappings

In this section, we will introduce more blended mappings along the lines of the Blended Elliptical Grid mapping introduced in Section 7.8 of this paper. These blended mappings can be considered as a generalization of several explicit mappings already mentioned in this paper. Furthermore, all of these blended mappings will have explicit inverse equations.

These blended mappings have a free parameter $\beta$ that can have any value between 0 and 1, inclusive. In some sense, these blended mappings can be considered as an infinite family of mappings based on the adjustable blend parameter. This is a powerful generalization of the mappings previously covered.

## 10.1 Biased Squelch Blended Mapping

The Vertical and Horizontal Squelch open mappings can be blended together to produce an interesting family of mappings. The square-to-disc equations for the blended mapping are

$$\boldsymbol{u = x\sqrt{1-\beta y^2}}$$

$$\boldsymbol{v = y\sqrt{1-(1-\beta)x^2}}$$

When $\beta = 0$, these equations simplify to the Vertical Squelch mapping. When $\beta = 1$, these equations simplify to the Horizontal Squelch mapping. Moreover, when $\beta = ½$, these equations reduce to the Elliptical Grid mapping! In a way, the Elliptical Grid mapping can be considered as simply a midway mapping between the Vertical and Horizontal Squelch mappings. Here is a table summarizing the blending characteristic of this mapping.

| Mapping | blend value | equation for *u* | equation for *v* |
|---|---|---|---|
| Vertical Squelch | 0 | $u = x$ | $v = y\sqrt{1-x^2}$ |
| Horizontal Squelch | 1 | $u = x\sqrt{1-y^2}$ | $v = y$ |
| Blended | $\beta$ | $u = x\sqrt{1-\beta y^2}$ | $v = y\sqrt{1-(1-\beta)x^2}$ |
| Elliptical Grid | ½ | $u = x\sqrt{1-\frac{1}{2}y^2}$ | $v = y\sqrt{1-\frac{1}{2}x^2}$ |

Figure 18 shows the results of mapping a square grid to a disc for different values of the blending parameter $\beta$. When $\beta \neq 0$ and $\beta \neq 1$, this blended mapping is a closed mapping whose domain and range include the bounding rims of the circular disc and the square. The disc-to-square equations for this blended mapping are

$$\boldsymbol{x = \frac{sgn(u)}{\sqrt{2(1-\beta)}}\sqrt{1+(1-\beta)u^2-\beta v^2-\sqrt{(1+(1-\beta)u^2-\beta v^2)^2-4(1-\beta)u^2}}}$$

$$\boldsymbol{y = \frac{sgn(v)}{\sqrt{2\beta}}\sqrt{1-(1-\beta)u^2+\beta v^2-\sqrt{(1-(1-\beta)u^2+\beta v^2)^2-4\beta v^2}}}$$

In Figure 19, one can observe the grid lines enclosed inside each of the circular discs. For the Vertical Squelch mapping at $\beta$=0, the vertically-oriented grid lines are all straight and the horizontally-oriented grid lines are curved. In stark contrast, for the Horizontal Squelch mapping at $\beta$=1, the horizontally-oriented grid lines are all straight and the vertically-oriented grid lines are curved. In a way, one can say that the Vertical Squelch mapping is biased towards vertical lines and that the Horizontal Squelch mapping is biased towards horizontal lines.

This bias carries over for $\beta$ values in between 0 and 1. For $\beta$ < ½, vertically-oriented grid lines tend to have less curvature than horizontally-oriented grid lines. For $\beta$ > ½, the opposite is true. For $\beta$ = ½ with the Elliptical Grid mapping, there is no vertical or horizontal bias. In fact, the grid lines in the Elliptical Grid mapping are symmetric with respect to the two main diagonals. This symmetry about the diagonal axes does not carry over for other $\beta$ values in the Biased Squelch blended mapping.

This bias towards vertical or horizontal lines in the Biased Squelch blended mapping is the reason why the mapping is named that way. This blended mapping encompasses a spectrum of mappings that are either vertically-biased or horizontally-biased. The only exception to this bias is midway with the Elliptical Grid mapping.

Just like the Vertical and Horizontal Squelch mappings, this blended mapping maps circular contours inside the disc to Fernandez-Guasti squircles inside the square. In other words, this blended mapping is also rooted to the Fernandez-Guasti squircle.

This property can be shown by using the equations for *u* and *v* to evaluate

$$\begin{aligned} u^2 + v^2 &= x^2(1 - \beta y^2) + y^2(1 - (1 - \beta)x^2) \\ &= x^2 - \beta x^2 y^2 + y^2 - (1 - \beta)x^2 y^2 \\ &= x^2 + y^2 - x^2 y^2 - \beta x^2 y^2 + \beta x^2 y^2 \\ &= x^2 + y^2 - x^2 y^2 \end{aligned}$$

This is just the equation for the *linear squircular continuum*.

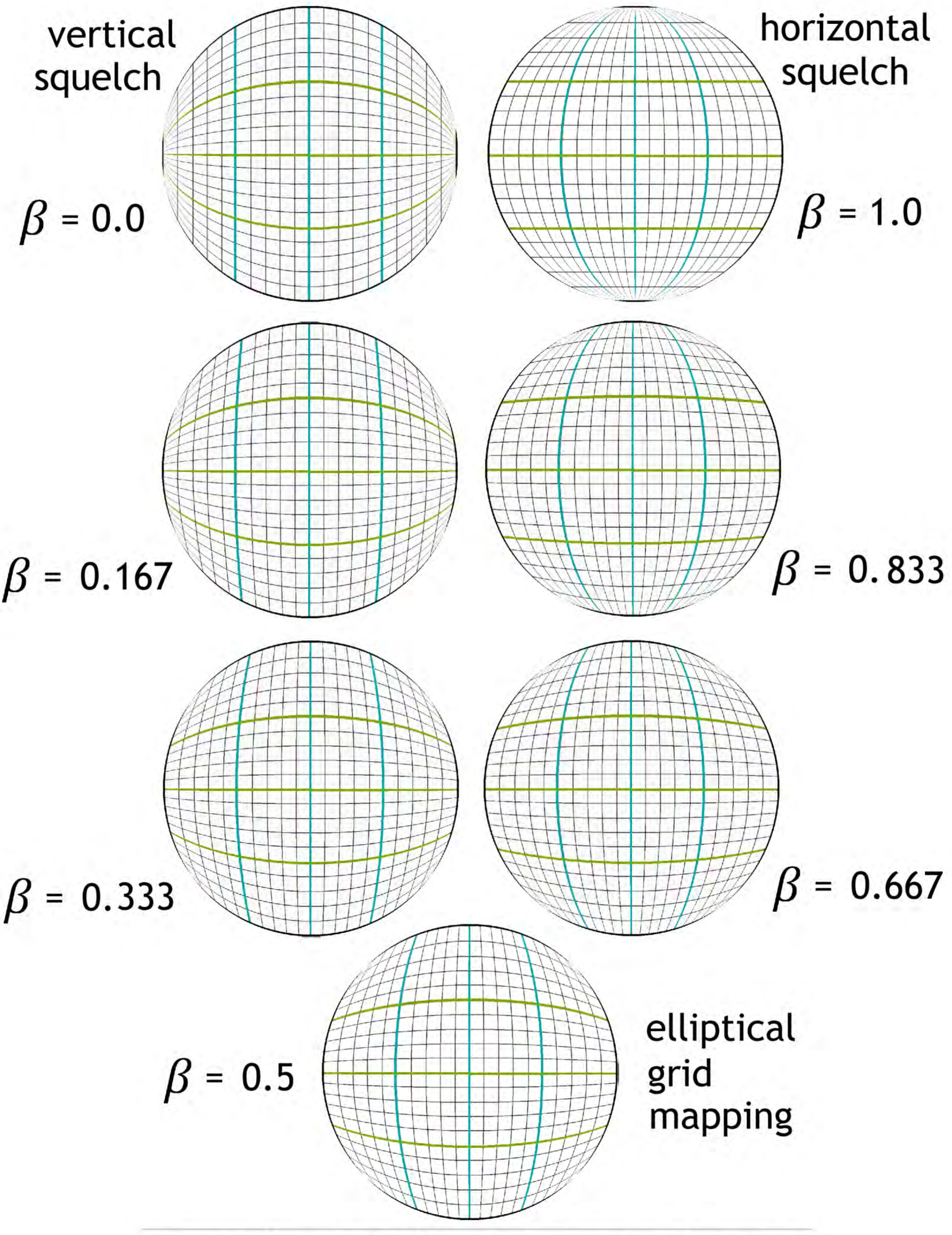

Figure 19: Biased Squelch Blended Mapping

## 10.2 Power2 Blend

The 2-Squircular and Cornerific Tapered2 mappings can be blended to produce a parameterized family of mappings. The derivation comes from introducing a blend parameter $\beta$ in the nonlinear relationship between the radius of the squircle and its squareness. Specifically, we can use this relation between $s$ and $t$

$$s = t\sqrt{2\beta + (1 - 2\beta)t^2}$$

When $\beta = 0$, the blended mapping reduces to the 2-Squircular mapping. When $\beta = 1$, the blended mapping reduces to the Cornerific Tapered2 mapping. Moreover, when $\beta = ½$, the blended mapping reduces to the FG-Squircular mapping. Here is a table summarizing this blended family of mappings.

| mapping | blend value | equation for *s* | equation for *t* |
|---|---|---|---|
| 2-Squircular | 0 | $s = t^2$ | $t = \sqrt{\dfrac{x^2 + y^2}{1 + x^2y^2}}$ |
| Cornerific Tapered2 | 1 | $s = t\sqrt{2 - t^2}$ | $t = \sqrt{\dfrac{x^2 + y^2 - 2x^2y^2}{1 - x^2y^2}}$ |
| blended | $\beta$ | $s = t\sqrt{2\beta + (1 - 2\beta)t^2}$ | $t = \sqrt{\dfrac{x^2 + y^2 - 2\beta x^2y^2}{1 + x^2y^2 - 2\beta x^2y^2}}$ |
| FG-Squircular | ½ | $s = t$ | $t = \sqrt{x^2 + y^2 - x^2y^2}$ |

The forward and inverse equations for this blended mapping are shown in vector format below.

Disc to Square Mapping:

$$\begin{bmatrix} x \\ y \end{bmatrix} = sgn(uv)\sqrt{\frac{u^2 + v^2 - \sqrt{(u^2 + v^2)[u^2 + v^2 - 4u^2v^2(u^2 + v^2 + 2\beta(1 - u^2 - v^2))]}}{2(u^2 + v^2 + 2\beta(1 - u^2 - v^2))}} \begin{bmatrix} \frac{1}{v} \\ \frac{1}{u} \end{bmatrix}$$

Square to Disc Mapping:

$$\begin{bmatrix} u \\ v \end{bmatrix} = \sqrt{\frac{x^2 + y^2 - 2\beta x^2y^2}{(x^2 + y^2)(1 + x^2y^2 - 2\beta x^2y^2)}} \begin{bmatrix} x \\ y \end{bmatrix}$$

## 10.3 Power3 Blend

The 3-Squircular and Tapered4 mappings can be blended to produce a parameterized family of mappings. The derivation comes from introducing a blend parameter $\beta$ in the nonlinear relationship between the radius of the squircle and its squareness. Specifically, we can use this relation between *s* and *t*

$$s = t\sqrt{\frac{3}{2}\beta + \left(1 - \frac{3}{2}\beta\right)t^4}$$

When $\beta = 0$, the blended mapping reduces to the 3-Squircular mapping. When $\beta = 1$, the blended mapping reduces to the Tapered4 mapping. Moreover, when $\beta = \frac{2}{3}$, the blended mapping reduces to the FG-Squircular mapping. Here is a table summarizing this blended family of mappings.

| mapping | blend value | equation for *s* | equation for *t* |
|---|---|---|---|
| 3-Squircular | 0 | $s = t^3$ | $t = \sqrt{\dfrac{-1 + \sqrt{4x^4y^2 + 4x^2y^4 + 1}}{2x^2y^2}}$ |
| Tapered4 | 1 | $s = t\sqrt{\frac{3}{2} - \frac{1}{2}t^4}$ | $t = \sqrt{\dfrac{1 - \sqrt{1 - 2x^4y^2 - 2x^2y^4 + 3x^4y^4}}{x^2y^2}}$ |
| blended | $\beta$ | $s = t\sqrt{\frac{3}{2}\beta + \left(1 - \frac{3}{2}\beta\right)t^4}$ | $t = \sqrt{\dfrac{1 - \sqrt{1 - 2(3\beta-2)x^4y^2 - 2(3\beta-2)x^2y^4 + 3\beta(3\beta-2)x^4y^4}}{(3\beta-2)x^2y^2}}$ |
| FG-Squircular | $\frac{2}{3}$ | $s = t$ | $t = \sqrt{x^2 + y^2 - x^2y^2}$ |

The forward and inverse equations for this blended mapping are shown in vector format below.

Disc to Square Mapping:

$$\begin{bmatrix} x \\ y \end{bmatrix} = sgn(uv)\sqrt{\frac{u^2 + v^2 - \sqrt{(u^2+v^2)[u^2 + v^2 - 2u^2v^2[3\beta - (3\beta-2)u^4 - 2(3\beta-2)u^2v^2 - (3\beta-2)v^4]]}}{3\beta - (3\beta-2)u^4 - 2(3\beta-2)u^2v^2 - (3\beta-2)v^4}}\begin{bmatrix} \frac{1}{v} \\ \frac{1}{u} \end{bmatrix}$$

Square to Disc Mapping:

$$\begin{bmatrix} u \\ v \end{bmatrix} = \frac{sgn(xy)}{\sqrt{x^2+y^2}}\sqrt{\frac{1 - \sqrt{1 - 2(3\beta-2)x^4y^2 - 2(3\beta-2)x^2y^4 + 3\beta(3\beta-2)x^4y^4}}{3\beta - 2}}\begin{bmatrix} \frac{1}{y} \\ \frac{1}{x} \end{bmatrix}$$

## 10.4 Blended Sham

The Sham Quartic and the Non-axial 2-Pinch mappings can be blended to produce a parameterized family of mappings. The derivation comes from introducing a blend parameter $\beta$ to the nonlinear modulator function

$$m(t) = t\sqrt{2\beta + (1-2\beta)t^2}$$

Using the same approach as the one outlined in Section 8.3, we can derive a square-to-disc equation for this parameterized blended mapping as

$$\begin{bmatrix} u \\ v \end{bmatrix} = \frac{\sqrt{x^2 + y^2 + 2\beta(1 + x^2y^2 - x^2 - y^2)}}{1 + x^2y^2} \begin{bmatrix} x \\ y \end{bmatrix}$$

When $\beta = 0$, the blended mapping reduces to the Non-axial 2-Pinch mapping. When $\beta = 1$, the blended mapping reduces to the Sham Quartic mapping. Moreover, when $\beta = ½$, the blended mapping reduces to the 2-Squircular mapping. Here is a table summarizing this blended family of mappings.

| mapping | blend value | equation for *m(t)* | square-to-disc equation |
|---|---|---|---|
| Non-axial 2-Pinch | 0 | $m(t) = t^2$ | $\begin{bmatrix} u \\ v \end{bmatrix} = \frac{\sqrt{x^2 + y^2}}{1 + x^2y^2} \begin{bmatrix} x \\ y \end{bmatrix}$ |
| Sham Quartic | 1 | $m(t) = t\sqrt{2 - t^2}$ | $\begin{bmatrix} u \\ v \end{bmatrix} = \frac{\sqrt{2 + 2x^2y^2 - x^2 - y^2}}{1 + x^2y^2} \begin{bmatrix} x \\ y \end{bmatrix}$ |
| blended | $\beta$ | $m(t) = t\sqrt{2\beta + (1-2\beta)t^2}$ | $\begin{bmatrix} u \\ v \end{bmatrix} = \frac{\sqrt{x^2 + y^2 + 2\beta(1 + x^2y^2 - x^2 - y^2)}}{1 + x^2y^2} \begin{bmatrix} x \\ y \end{bmatrix}$ |
| 2-Squircular | ½ | $m(t) = t$ | $\begin{bmatrix} u \\ v \end{bmatrix} = \frac{1}{\sqrt{1 + x^2y^2}} \begin{bmatrix} x \\ y \end{bmatrix}$ |

Just like the Sham Quartic mapping, the inverse equations require the quartic formula. To be more specific, we need the roots of this quartic polynomial equation

$$v^4\boldsymbol{q}^4 - 2\beta v^2\boldsymbol{q}^3 + (2\beta u^2 + 2\beta v^2 + 2u^2v^2 - u^2 - v^2)\boldsymbol{q}^2 - 2\beta u^2\boldsymbol{q} + u^4 = 0$$

The coefficients to be passed to the quartic formula are

$$a_4 = v^4 \qquad a_1 = -2\beta u^2$$

$$a_3 = -2\beta v^2 \qquad a_0 = u^4$$

$$a_2 = 2\beta(u^2 + v^2) + 2u^2v^2 - u^2 - v^2$$

Let $q_0$ be the smallest non-negative root of the given quartic equation. The formulas for x and y are

$$\boldsymbol{x = sgn(u)\sqrt{q_0}}$$

$$\boldsymbol{y = sgn(v)\left|\frac{v}{u}\right|\sqrt{q_0}}$$

Just like the other mappings discussed in this paper, we can produce a detJ surface associated with this blended mapping. This detJ surface will have an adjustable parameter $\beta$ that determines its shape. When $\beta = 1$, the surface becomes the Sham Hemi-Ellipsoid. The parametric equations for this detJ surface are

$$\hat{X} = \frac{\lambda\sqrt{\lambda^2 + \gamma^2 + 2\beta(1 + \lambda^2\gamma^2 - \lambda^2 - \gamma^2)}}{1 + \lambda^2\gamma^2}$$

$$\hat{Y} = \frac{\gamma\sqrt{\lambda^2 + \gamma^2 + 2\beta(1 + \lambda^2\gamma^2 - \lambda^2 - \gamma^2)}}{1 + \lambda^2\gamma^2}$$

$$\hat{Z} = \frac{2(1 - \lambda^2\gamma^2)(\lambda^2 + \gamma^2 + \beta(1 + \lambda^2\gamma^2 - 2\lambda^2 - 2\gamma^2))}{(1 + \lambda^2\gamma^2)^3}$$

Also, we can force the blended surface to always be a hemisphere by setting $\hat{Z} = \sqrt{1 - \hat{X}^2 - \hat{Y}^2}$. The equations for the coerced hemisphere at different $\beta$ values are given in the table below.

| mapping | blend value | coerced hemisphere equation for $\hat{Z}$ |
|---|---|---|
| Non-axial 2-Pinch | 0 | $\hat{Z} = \sqrt{\frac{(1-\lambda^2)(1-\gamma^2)}{1+\lambda^2\gamma^2}}$ |
| Sham Quartic | 1 | $\hat{Z} = \frac{(1-\lambda^2)(1-\gamma^2)}{1+\lambda^2\gamma^2}$ |
| blended | $\beta$ | $\hat{Z} = \frac{\sqrt{(1-\lambda^2)(1-\gamma^2)(1+\lambda^2+\gamma^2+\lambda^2\gamma^2-2\beta\left(\lambda^2+\gamma^2\right))}}{1+\lambda^2\gamma^2}$ |
| 2-Squircular | ½ | $\hat{Z} = \frac{\sqrt{(1-\lambda^4)(1-\gamma^4)}}{1+\lambda^2\gamma^2}$ |

## 11 Summary

In this paper, we presented a methodology for generalizing square-to-disc mappings into rectangle-to-ellipse mappings. The key idea is to remove the eccentricity first and then reintroduce it back after a chosen square-to-disc mapping is performed. We also presented and derived several invertible square-to-disc mappings which are mostly variants to the FG-Squircular and Elliptical Grid mappings. In addition, we introduced the concept of a detJ surface associated with each mapping.

| mapping | key property | offshoot from | open or closed | notes/comments |
|---|---|---|---|---|
| 2-Squircular | simple equations | FG-Squircular with squircular nonlinearity | closed | Radial mapping which has considerable size distortion near the corners |
| 3-Squircular | quasi-symmetric | FG-Squircular with squircular nonlinearity | closed | Very similar to 2-Squircular mapping; Deemphasizes the four corners |
| Cornerific Tapered2 | good corner behavior | FG-Squircular with squircular nonlinearity | closed | Radial mapping which has less distortion in corners; Center area has undesirable curvy distortions. |
| Tapered4 Squircular | good corner behavior | FG-Squircular with squircular nonlinearity | closed | Very similar to Cornerific Tapered2 mapping; puts emphasis on the four corners |
| Squelched Grid | simple equations | Elliptical Grid | open mapping | Mapping is open-type and does not include the boundary rim of the disc and square |
| Vertical Squelch | vertical lines preserved | Elliptical Grid | open mapping | Vertical features remain upright; Vertically-biased mapping is an open-type |
| Horizontal Squelch | horizontal lines preserved | Elliptical Grid | open mapping | Horizontal features remain sideways; Horizontally-biased mapping is an open-type |
| Non-Axial 2-Pinch | pinching on the center of disc | FG-Squircular with axial nonlinearity | closed | $x=u^2$ and $y=v^2$ relationship along the horizontal and vertical Cartesian axes, respectively |
| Non-Axial ½-Punch | bulging on the center of disc | FG-Squircular with axial nonlinearity | closed | $x=\sqrt{u}$ and $y=\sqrt{v}$ relationship along the horizontal and vertical Cartesian axes, respectively |
| Sham Quartic | hemispherical appearance | FG-Squircular with axial nonlinearity | closed | Non-linear axial relationship provides an excellent mapping of the square to the hemisphere |
| Blended E-Grid | blended generalization | Elliptical Grid | open mapping | Has an adjustable parameter to blend the Elliptical Grid with the Squelched Grid mappings |
| Biased Squelch | blended generalization | Elliptical Grid | open mapping | Has an adjustable parameter to blend the Vertical and Horizontal Squelch mappings |
| Power2 Blend | blended generalization | FG-Squircular | closed | Has an adjustable parameter to blend the 2-Squircular with Cornerific Tapered2 mapping |
| Power3 Blend | blended generalization | FG-Squircular | closed | Has an adjustable parameter to blend the 3-Squircular with Tapered4 mapping |
| Blended Sham | blended generalization | FG-Squircular | closed | Has an adjustable parameter to blend the Sham Quartic with Non-axial 2-Pinch mapping |

## 12 Acknowledgement

We would like to thank Douglas Norton, Karl Kattchee, and Anil Venkatesh for organizing the SIGMAA-ARTS sessions at the Joint Mathematics Meetings 2019 in Baltimore.

# Appendix A: *Radial Mapping Based on the Lamé Curve*

Most of the mappings discussed in this paper are based on the Fernandez-Guasti squircle. However, it is possible to create disc-to-square mappings that are not rooted on the Fernandez-Guasti squircle. For example, we will present a mapping based on the Lamé curve [Weinstein 2017] here.

The Lamé curve is a shape very similar to the Fernandez-Guasti squircle. The equation for the Lamé curve with no eccentricity is

$$|x|^p + |y|^p = t^p$$

This closed curve has the characteristic similar to the Fernandez-Guasti squircle in that it blends the circle with the square. When $p=2$, the curve is a circle with radius $t$. As $p$ tends to infinity, the curve becomes more and more square-like. In theory, the square cannot be explicitly represented by the Lamé curve because this requires an infinite power value for $p$, but in practice, any $p>10$ can fairly approximate a square shape.

Using the *radial constraint linear parametric equation* in conjunction with the equation for the Lamé curve, we can come up with a square-to-disc mapping, that is valid for $0<x<1$ and $0<y<1$.

$$u = \frac{x}{\sqrt{x^2+y^2}} \left( |x|^{\frac{2}{(1-|x|)(1-|y|)}} + |y|^{\frac{2}{(1-|x|)(1-|y|)}} \right)^{\frac{1}{2}(1-|x|)(1-|y|)}$$

$$v = \frac{y}{\sqrt{x^2+y^2}} \left( |x|^{\frac{2}{(1-|x|)(1-|y|)}} + |y|^{\frac{2}{(1-|x|)(1-|y|)}} \right)^{\frac{1}{2}(1-|x|)(1-|y|)}$$

Note that in the interest of brevity, we have not singled-out the degenerate cases that occur when x or y have numerical values of 0 or 1 in the equations given above. These cases will result in unwanted infinities in the equations. For these degenerate cases, simply set u=x and v=y.

We were unable to find explicit equations for the inverse mapping. This highlights the difficulty of using the Lamé curve in lieu of the Fernandez-Guasti squircle. The Fernandez-Guasti squircle is a quartic curve that involves polynomials with no higher power than 4, whereas the Lamé curve needs an infinite power to express the square. We believe that there are no closed-form expressions for the inverse of this Lamé-based mapping.

# Appendix B: *Another Mapping Based on the Lamé Curve*

The Lamé curve has parametric equations:

$$x(t) = sgn(\cos t)\ \ |\cos t|^{2/p}$$
$$y(t) = sgn(\sin t)\ \ |\sin t|^{2/p}$$

where $p$ is the specific power from the equation of Lamé curve with no eccentricity

$$|x|^p + |y|^p = t^p$$

Using these equations, it is possible to come up with another disc-to-square mapping

$$x = sgn(u)\ |u|^{1-u^2-v^2}$$
$$y = sgn(v)\ |v|^{1-u^2-v^2}$$

Just as in Appendix A, we were unable to find explicit equations for the inverse mapping. In other words, we do not have expressions for $u$ & $v$ in terms of $x$ & $y$. Again, this highlights the difficulty of using the Lamé curve in lieu of the Fernandez-Guasti squircle. The Fernandez-Guasti squircle is a simple quartic curve of degree 4, whereas the Lamé curve has unbounded exponents. These unbounded exponents make the Lamé curve unwieldy and not amenable to algebraic manipulation.

We have included mapping diagrams for this mapping in the next page. Qualitatively, this mapping looks similar to the Squelched Grid mapping. However, this mapping is not an axial mapping.

# Lamé-based Mapping

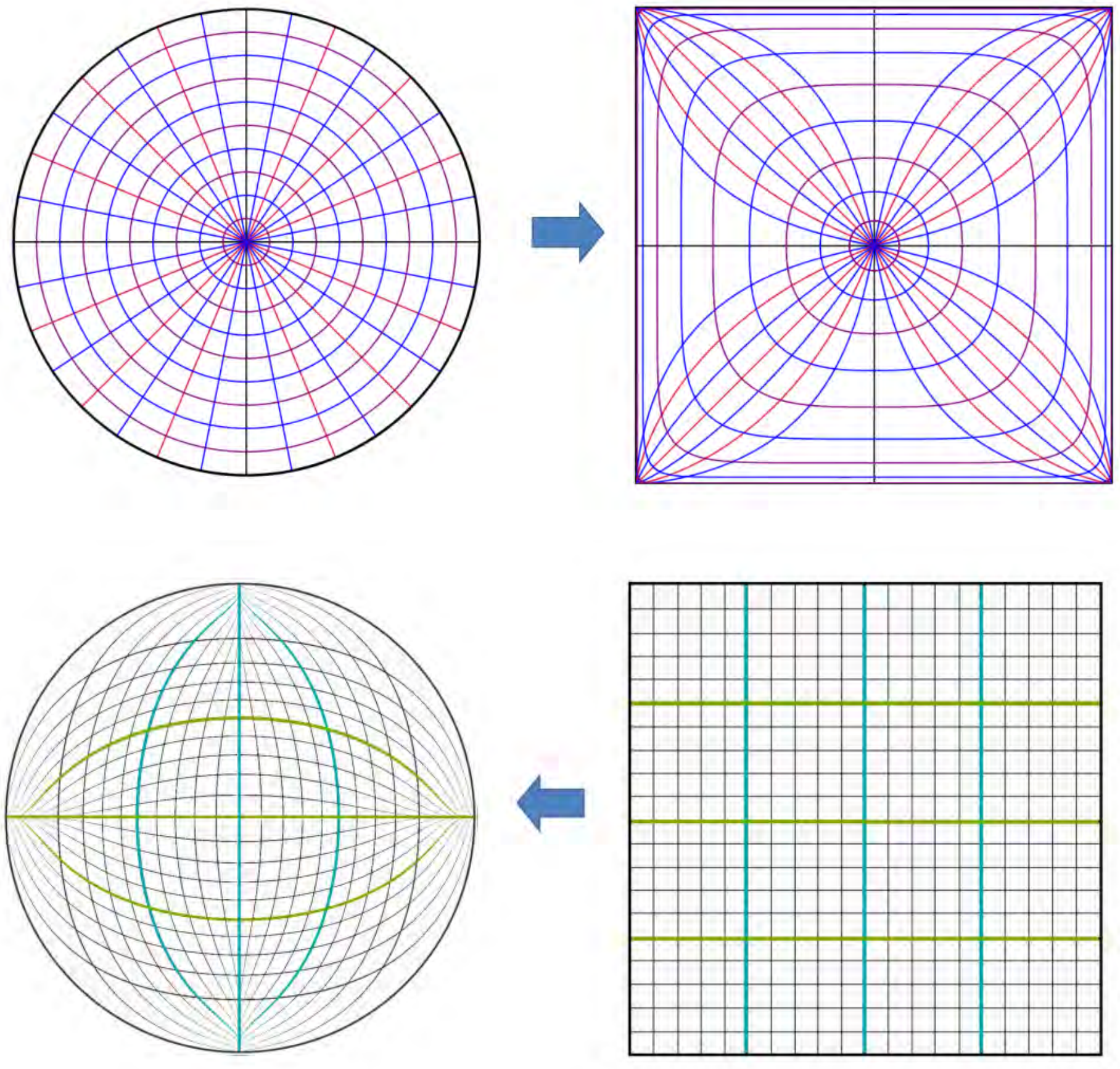

Disc to square mapping:

$$x = sgn(u)\ |u|^{1-u^2-v^2}$$
$$y = sgn(v)\ |v|^{1-u^2-v^2}$$

There are no known explicit inverse equations for this mapping. However, there are many techniques in numerical analysis for performing numerical inversion of these equations. This is how we produced the inverse diagram illustrated above.